\documentclass[12pt,a4paper]{article}

\usepackage[english]{babel}
\usepackage{longtable}
\usepackage[authoryear]{natbib}
\usepackage{bm}
\usepackage[table,xcdraw]{xcolor}
\usepackage[utf8]{inputenc}
\usepackage[official]{eurosym}
\usepackage{multirow}
\usepackage{threeparttable}
\usepackage{adjustbox}
\usepackage{booktabs}
\usepackage{amsmath, amsthm, amssymb, amsfonts, graphicx,tabularx,rotate}
\usepackage{longtable, lscape, adjustbox,empheq}
\usepackage[figuresright]{rotating}
\usepackage{multicol}
\usepackage{caption,rotfloat}
\usepackage{subcaption}
\usepackage{afterpage,url}
\usepackage[tmargin=2.5cm,bmargin=2.25cm,lmargin=2.25cm,rmargin=2.25cm]{geometry}
\usepackage{setspace}
\usepackage[dvipsnames]{xcolor}
\usepackage[hidelinks]{hyperref}
\usepackage{comment}
\usepackage{xcolor}
\usepackage[normalem]{ulem}
\usepackage{graphicx}  % Per inserire immagini
\usepackage{tikz}
\usepackage{caption}
\usepackage{adjustbox}

\usepackage{placeins}
\makeatletter
\AtBeginDocument{%
  \expandafter\renewcommand\expandafter\subsection\expandafter
    {\expandafter\@fb@secFB\subsection}%
  \newcommand\@fb@secFB{\FloatBarrier
    \gdef\@fb@afterHHook{\@fb@topbarrier \gdef\@fb@afterHHook{}}}%
  \g@addto@macro\@afterheading{\@fb@afterHHook}%
  \gdef\@fb@afterHHook{}%
}
\makeatother
\begin{document}

%%%%%%%%%%%%%%%%%%%%%%%%%%%%%%%%%%%%%%%%%%%%%%%%%%%%%%%%%%%%%%%%%%%%%%%%%%
% Title PAGE
%%%%%%%%%%%%%%%%%%%%%%%%%%%%%%%%%%%%%%%%%%%%%%%%%%%%%%%%%%%%%%%%%%%%%%%%%%
\setcounter{page}{0}
\pagenumbering{Alph}

\doublespacing
\begin{center}
\Large{\textbf{A Real-Time Framework for Forecasting Metal Prices}}\\[.25cm]
\large{%
\begin{tabular}{ccc}
Andrea Bastianin$^{a,b}$
& %
Luca Rossini$^{a,b}$
& % 
Lorenzo Tonni$^{a,\ast}$
\end{tabular}
}
\\
\vspace{1cm}
\normalsize{\today}\\
\vfill
\end{center}
\footnotesize{\textbf{Abstract.} This paper develops a real-time forecasting framework for monthly real prices of four key industrial metals -- aluminum, copper, nickel, and zinc -- whose demand is rising due to their widespread use in manufacturing and low-carbon technologies. To replicate the information set available to forecasters in real time, we construct a new dataset combining daily financial variables with first-release macroeconomic indicators and use nowcasting techniques to address publication lags. Within this real-time environment, we evaluate the predictive accuracy of a broad set of univariate, multivariate, and factor-augmented models, comparing their performance with two industry benchmarks: survey expectations and futures-spot spread models. Results show that although short-run metal price movements remain difficult to predict, medium-term horizons display substantial forecastability. Indicators of manufacturing activity tied to primary metals—such as new orders and capacity utilization—significantly improve forecasting accuracy for aluminum and copper, with more moderate gains for zinc and limited improvements for nickel. Futures and survey forecasts generally underperform the real-time econometric models. These findings highlight the value of incorporating timely macroeconomic information into forecasting frameworks for industrial metal markets.}
\vfill
\small{\noindent\textbf{Key Words:} First-Release Data; Energy Transition; Forecasting; Metals; Critical Raw materials.}\\\vspace{\fill}

\footnotesize{%
\noindent $^{(a)}$ Department of Economics, Management, and Quantitative Methods, University of Milan, Milan, Italy.\\
\noindent $^{(b)}$ Fondazione Eni Enrico Mattei, Milan, Italy.\\
\noindent $^{(\ast)}$ \textit{Corresponding author}: Lorenzo Tonni, Department of Economics, Management, and Quantitative Methods, University of Milan, Via Conservatorio, 7, 20122, Milan, IT. Email: \url{lorenzo.tonni@unimi.it}.}
\vfill
\scriptsize{
\noindent\textit{Acknowledgments:} 
The authors gratefully acknowledge H. Bjornland, F. Ravazzolo, A. Viselli and the participants at 10th Annual Conference of the Society for Economic Measurement (SEM), held at the Athens University of Economics and Business, and the 2025 International Association for Applied Econometrics (IAAE) Conference, held in Turin for their useful feedback.
The authors acknowledge financial support from the “Fund for Departments of Excellence” provided by the Ministero dell’Università e della Ricerca (MUR), established by the Stability Law (“Legge di Stabilità n. 232/2016, 2017”), within the projects of the Department of Economics, Management, and Quantitative Methods (University of Milan). 
}

\thispagestyle{empty}
\newpage
\setcounter{footnote}{0}
\pagenumbering{arabic}
\doublespacing
\normalsize

\section{Introduction}
\begin{figure}[ht]
\centering
\caption{Monthly Price and yearly world consumption of copper, aluminum, nickel, and zinc from January 1995 to December 2024.}
\label{fig:prices_ETM}
    \includegraphics[width=0.495\textwidth]{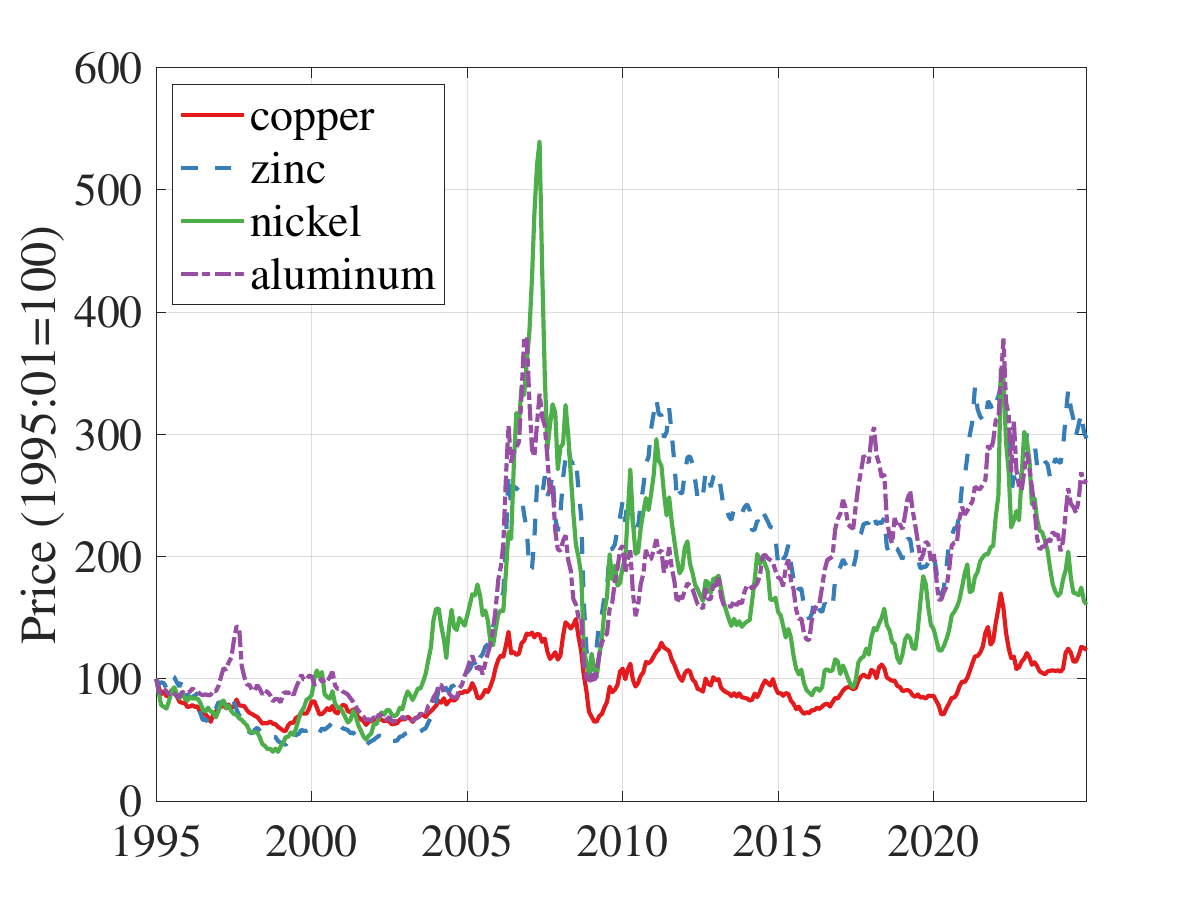} \includegraphics[width=0.495\textwidth]{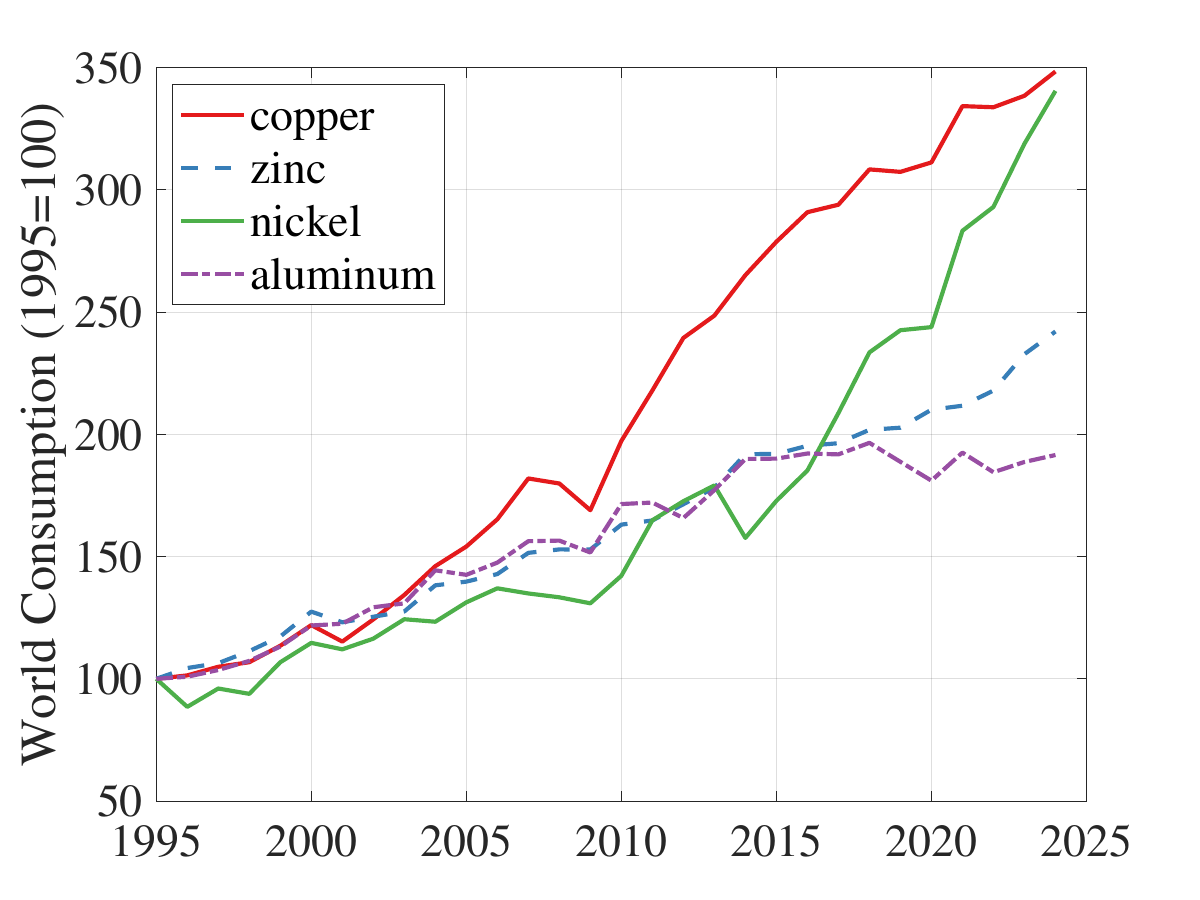} 
\caption*{\scriptsize\textit{Notes}: Prices are described in Section \ref{sec:data}. World consumption data are sourced from the World Bureau of Metal Statistics.}
\end{figure}

Prices of several metals are under strong demand pressures due to their expanding use in electric mobility and renewable energy technologies \citep{IEACRM25}. 
As shown in Figure \ref{fig:prices_ETM}, global consumption of aluminum, copper, nickel, and zinc has doubled since 1995, while prices have remained highly volatile. 
This trend is expected to continue. Under the International Energy Agency’s (IEA) ``Net Zero Emissions'' Scenario, global aluminum production is projected to grow at a Compound Annual Growth Rate (CAGR) of 1.39\% from 2022 to 2030.%
\footnote{The compound annual growth rate is defined as $\left(V_t / V_0\right)^{{1}/{t}} - 1$, where \(V_0\) and \(V_t\) denote the initial and terminal values, respectively, and \(t\) is the number of years. The IEA’s Net Zero Emissions by 2050 Scenario outlines a pathway for the global energy sector to achieve net-zero $\text{CO}_2$ emissions by mid-century, implying rapid clean energy deployment, no new fossil fuel development, and systemic changes across all sectors. See: \url{https://www.iea.org/reports/global-energy-and-climate-model/net-zero-emissions-by-2050-scenario-nze}. For aluminum: IEA (2023), \textit{Global aluminum production in the Net Zero Scenario, 2010--2030}, IEA, Paris. \url{https://www.iea.org/data-and-statistics/charts/global-aluminium-production-in-the-net-zero-scenario-2010-2030-3}. For copper, nickel, and zinc: IEA, Critical Minerals Dataset, IEA, Paris, \url{https://www.iea.org/data-and-statistics/data-product/critical-minerals-dataset}. Aluminum projections refer to production; for copper, nickel, and zinc we use demand.}
Demand for other metals is even more dynamic in clean energy applications: zinc demand for clean technologies is projected to grow at 17.2\% per year (2024--2030), copper at 17.4\% (vs. 4.3\% for total demand), and nickel at 22.5\% (vs. 8.0\% for total demand). 
These discrepancies between total and clean-energy demand highlight the growing influence of low-emission technologies on metal markets and the rising policy concerns around ensuring sustainable and resilient supply chains. They also underscore the urgent need for reliable short- and medium-term real-time metal price forecasts to support effective market monitoring and scenario analysis.

Timely and reliable data collection is therefore a prerequisite for both credible price forecasting and sound policy design in critical mineral markets. As emphasized by \citet{IEAtraceability}, traceability systems and standardized data protocols are essential for integrating real-time information on production, trade flows, and sectoral demand, which are key inputs for anticipating market developments and identifying potential bottlenecks. Incomplete or outdated information undermines the accuracy of short- and medium-term price forecasts and limits the effectiveness of policy responses aimed at managing supply-chain risks, stabilizing markets, and guiding investment decisions \citep{Wittenstein}.

This paper addresses this policy issue by developing a comprehensive database that brings together a range of predictors for forecasting the real price of key metals: aluminum, copper, nickel, and zinc. The database is designed to replicate the information environment faced by policymakers, investors, and practitioners when producing real-time forecasts at monthly sampling frequency. To this end, we rely on first-release values of macroeconomic indicators and employ nowcasting techniques to bridge data gaps due to publication lags. Within this framework, we evaluate the forecasting performance of several models relative to standard industry benchmarks, including survey-based forecasts and futures--spot spread models.

Modelling and forecasting commodity prices has a long tradition in economics. Within the forecasting literature, a vast body of research has systematically examined crude oil prices \citep[see, e.g.,][for a survey]{alquist2013forecasting} and, more broadly, energy prices \citep{ferrari2021forecasting}, covering refined oil products \citep{bastianin2014forecasting,baumeister2017inside}, natural gas \citep{Bjornland2025fore,baumeisterGas}, and electricity \citep{foroni2023low,gianfreda2020comparing,hauzenberger2025sparse}. In contrast, the literature on metal markets and critical raw materials remains notably fragmented -- despite their growing strategic importance in the energy transition. An early survey on nonferrous metals is provided by \citet{watkins2004econometric}.

Although this paper belongs to the broader literature on commodity price forecasting, its contribution is particularly aligned with research that employs real-time data to forecast commodity prices and with studies modeling mineral prices and their role in the energy and digital transitions. Within the strand of research using real-time data for commodity price forecasting, crude oil continues to dominate the field \citep[see][]{baumeister2012real,baumeister2015forecasting,RavazzoloRothman}, with \citet{baumeisterGas} representing a notable exception for natural gas. In contrast, research on forecasting metal prices remains limited, and none of the existing studies, to our knowledge, relies on real-time data. \citet{pierdzioch2013forecasting} analyze survey-based forecasts, while other works emphasize the role of macro-financial and structural drivers as predictors \citep{duarte2021commodity,gargano2014forecasting,issler2014using,stuermer2017industrialization,west2014factor}. Moreover, a growing number of recent studies have examined the role of metals in the ecological transition and the determinants of their prices using structural vector autoregressive (VAR) models \citep[see, e.g.,][]{BCG2023,BaumeisterSpecialFocus,boer2024energy,CONSIDINE}.

Compared with the existing literature, this paper makes three main contributions. 
First, it develops a real-time forecasting framework for the monthly real prices of aluminum, copper, nickel, and zinc, supported by a newly constructed dataset that combines first-release macroeconomic indicators with nowcasting techniques. This approach reproduces the information actually available to forecasters at each point in time -- a perspective absent from previous studies on metal price forecasting and crucial for policymakers and market participants operating under real-time uncertainty. 
Second, the paper provides a comprehensive evaluation of forecasting models, ranging from univariate benchmarks to multivariate specifications with macro-financial predictors, and compares their performance with industry standards such as survey expectations and futures--spot spread models. This comparison highlights the value of macroeconomic information relative to widely used market-based forecasts. 
Third, the analysis shows that the predictability of metal prices is highly horizon-dependent: short-term movements are largely erratic, while medium-term dynamics are systematically linked to manufacturing activity in sectors closely tied to primary metals. Taken together, these findings demonstrate that incorporating timely macroeconomic indicators substantially improves medium-term forecasts, while also clarifying the limits of predictability at shorter horizons. Our paper therefore extends real-time forecasting methods to an important class of industrial commodities and provides new evidence on the macroeconomic forces shaping base metal prices.

The remainder of the paper is structured as follows. Section \ref{sec:database} describes the real-time database and compares nowcasting approaches. Section \ref{sec:models} presents the models and the forecasting experiment. Section \ref{sec:forecasting} presents the main results, while Section \ref{sec:mod_comparison_&_pooling} focuses on the time-varying performance of forecasts and assesses the benefits of forecast pooling. Section \ref{sec:conclusion} concludes.

\section{Real-Time Dataset Construction}\label{sec:database}
This section describes the construction of the real-time dataset. We combine daily financial data with the first-release vintages of monthly macroeconomic predictors. These monthly series exhibit publication lags of varying length, so we employ nowcasting to make them usable in real-time as shown in Section \ref{sec:nowcast}.

\subsection{Data}\label{sec:data}
\noindent\textit{Daily data.} We collect daily nominal spot prices for aluminum, copper, nickel, and zinc traded on the London Metal Exchange (LME) from the London Stock Exchange Group (LSEG). Since aluminum, copper, nickel, and zinc are storable commodities, we also source daily data on visible inventories from LSEG, representing the volume of stocks held in LME-approved warehouses.\footnote{Inventories are termed \textit{visible} because they are publicly reported and correspond to the portion of global stocks available for delivery against LME contracts.} Inventories proxy for market tightness and expected scarcity -- the key state variable linking current market conditions to expectations about future supply and demand in the competitive storage model \citep{deaton1992behaviour,deaton1996competitive}. Because they capture physical imbalances that drive short- and medium-term real price movements, inventories are routinely included in empirical models of commodity markets \citep{Bjornland2025fore,alquist2020commodity,baumeisterGas,kilianmurphy2014}.

Both nominal price and inventory data are available in real-time and are not subject to revisions. In the forecasting exercise, daily observations are aggregated by taking calendar-month averages.

\bigskip

\noindent\textit{Monthly data.} Real-time data for selected monthly macroeconomic variables were retrieved from ALFRED (Archival Federal Reserve Economic Data), maintained by the Federal Reserve Bank of St. Louis. See Table \ref{tab::predictors_list} for the full list of variables. As shown in Table \ref{tab::predictors_list}, we include five broad groups of predictors. Producer price indices (PPI) capture inflationary trends and cost pressures in metal-intensive industries: PPI series for primary and battery manufacturing reflect upstream price dynamics.

Since our focus is on forecasting real metal prices, nominal spot prices are deflated using the CPI vintage series of the corresponding period, with the two most recent missing observations nowcasted as described in the Section \ref{sec:nowcast}. 

Indicators of industrial production, new orders, shipments, and manufacturing hours track cyclical demand for metals, while capacity utilization in the primary and semifinished metal sectors quantifies supply tightness. To account for structural shifts associated with the energy and digital transitions, we use shipments and producer prices in the battery-manufacturing sector as proxies for rising demand for copper, nickel, and aluminum in electrification and storage technologies.

Finally, real broad effective exchange-rate indices are included for Australia, Chile, China, Indonesia, Peru, the Philippines, and Russia -- countries accounting for a large share of global production of aluminum, copper, nickel, and zinc. As shown by \citet{chen2010can}, exchange rates of “commodity-currency” countries have robust predictive power for global commodity prices, since they anticipate future export earnings and terms-of-trade shifts.

With the exception of exchange rates, the remaining macroeconomic variables refer to the United States. This choice is motivated not only by the availability and ease of access to granular real-time data through ALFRED, but also by the prominent role of the U.S. economy in global mineral markets through its positions as a major consumer, refiner, and financial centre. Moreover, metal prices are quoted in U.S. dollars and respond strongly to U.S. macroeconomic and financial conditions.

\begin{table}[ht]
\caption{Real-time database structure: macroeconomic and financial variables.}
\newcolumntype{L}[1]{>{\scriptsize\hsize=#1\hsize\raggedright\arraybackslash}X}
\newcolumntype{C}[1]{>{\scriptsize\hsize=#1\hsize\centering\arraybackslash}X}
\begin{tabularx}{\textwidth}{C{.3}|C{.4}|L{2.9}|C{.7}|C{.6}}
\hline
Group & ID & Description & Transformation & Missing Obs\\
\hline
\multirow{2}{*}{\rotatebox[origin=c]{90}{\scriptsize Prices}} & CPI   & Consumer Price Index for All Urban Consumers: All Items & $\Delta^2\log(x_t)$ & 2\\
 & PPI-M & Producer Price Index: Primary Metal Manufacturing & $\Delta\log(x_t)$ & 2\\
\hline
\multirow{4}{*}{\rotatebox[origin=c]{90}{\scriptsize Ec. Act.}} 
 & IP    & US Industrial Production & $\Delta\log(x_t)$ & 2\\
 & HEM   & Hours of Employees, Manufacturing & $\log(x_t)$ & 2\\
 & NO\!-M & New Orders: Primary Metals & $\log(x_t)$ & 2\\
 & NO\!-A & New Orders: Aluminum and Nonferrous Metal Products & $\log(x_t)$ & 3\\\hline
 \multirow{2}{*}{\rotatebox[origin=c]{90}{\scriptsize CU}} 
 & CU\!-P & Capacity Utilization: Primary \& Semifinished Processing & $\log(x_t)$ & 2\\
 & CU\!-M & Capacity Utilization: Primary Metal & $\log(x_t)$ & 2\\\hline
\multirow{2}{*}{\rotatebox[origin=c]{90}{\scriptsize ET}} 
 & PPI\!-B & Producer Price Index: Battery Manufacturing & $\Delta\log(x_t)$ & 2\\
 & MVS   & Manufacturers’ Value of Shipments: Battery Manufacturing & $\log(x_t)$ & 3\\
\hline
\multirow{7}{*}{\rotatebox[origin=c]{90}{\scriptsize Exchange Rates}} 
 & AUS   & Real Broad Effective Exchange Rate for Australia & $\Delta\log(x_t)$ & 2\\
 & CHL   & Real Broad Effective Exchange Rate for Chile & $\Delta\log(x_t)$ & 2\\
 & CHN   & Real Broad Effective Exchange Rate for China & $\Delta\log(x_t)$ & 2\\
 & IDN   & Real Broad Effective Exchange Rate for Indonesia & $\Delta\log(x_t)$ & 2\\
 & PER   & Real Broad Effective Exchange Rate for Peru & $\Delta\log(x_t)$ & 2\\
 & PHL   & Real Broad Effective Exchange Rate for Philippines & $\Delta\log(x_t)$ & 2\\
 & RUS   & Real Broad Effective Exchange Rate for Russia & $\Delta\log(x_t)$ & 2\\
\hline
\multirow{4}{*}{\rotatebox[origin=c]{90}{\scriptsize Inventories}} 
 & ALU\!-V & Aluminum Inventory Volume & $\Delta\log(x_t)$ & --\\
 & COP\!-V & Copper Inventory Volume & $\Delta\log(x_t)$ & --\\
 & NIC\!-V & Nickel Inventory Volume & $\Delta\log(x_t)$ & --\\
 & ZNC\!-V & Zinc Inventory Volume & $\Delta\log(x_t)$ & --\\
\hline
\end{tabularx}
\caption*{\scriptsize\textit{Notes}: In column 1, ``EC. Act.'' refers to Economic Activity; ``CU'' stands for Capacity Utilization, while ``ET'' refers to predictors related to the increasing use of metals in technologies involved in the energy and digital transitions.}
\label{tab::predictors_list}
\end{table}

\bigskip

\noindent\textit{Real-time data issues.} As shown in Table~\ref{tab::predictors_list}, all macroeconomic series are released with publication lags of two up to three months. Consequently, the dataset exhibits a ragged-edge structure \citep{wallis1986forecasting}, as observations for different variables are available at different times. We address this issue by nowcasting the missing observations each time a real-time forecast is produced. Conducting a real-time forecasting exercise requires using data vintages exactly as they were available at the time each forecast was made.

\begin{figure}[ht]
    \centering
    \caption{Effects of revisions of Manufacturers' New Orders for Primary Metals (black line) from February 2011 (red dot) to January 2025 (blue dot) in millions of dollars.} 
    \includegraphics[width=0.6\textwidth]{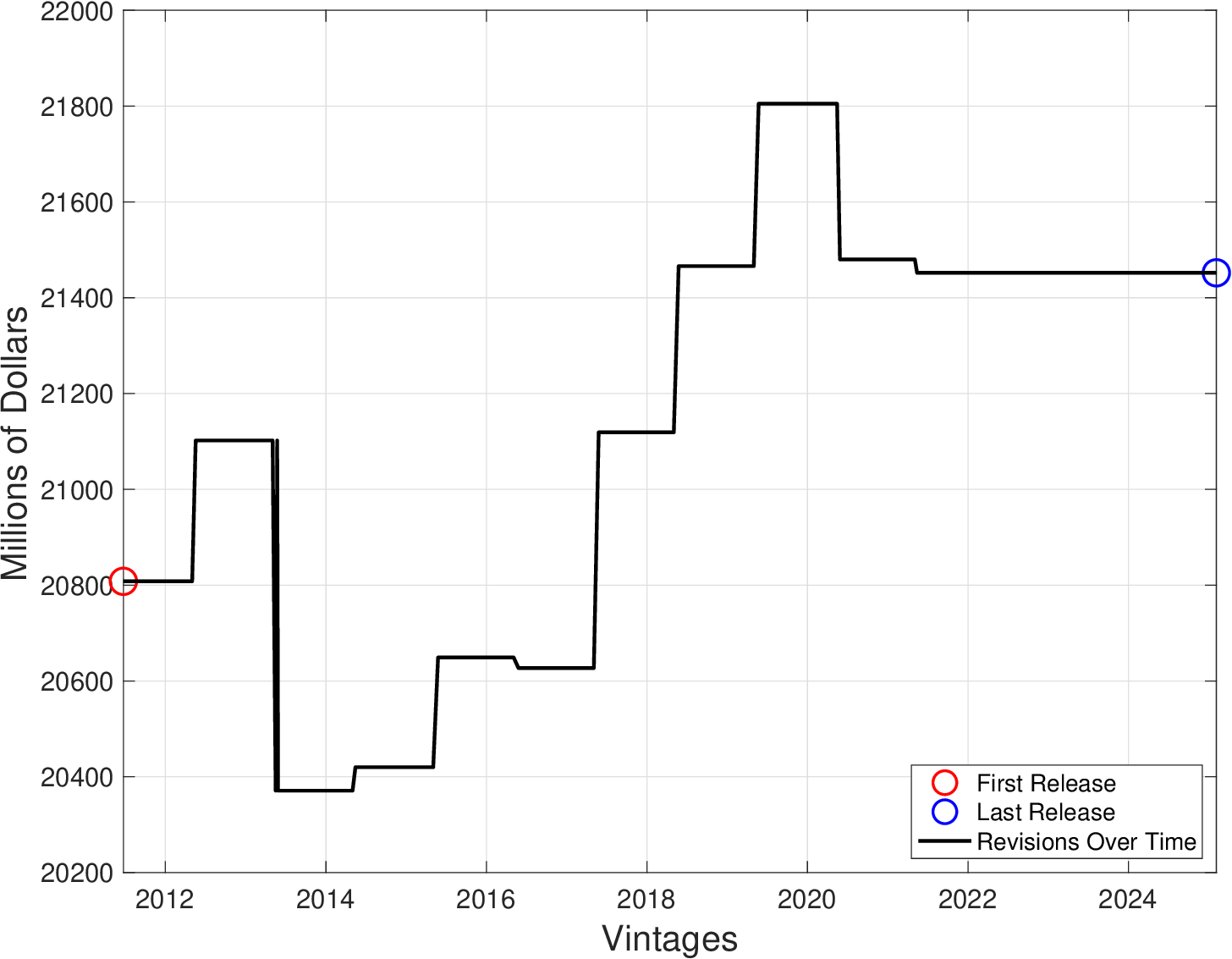} % Dimensione più appropriata
    \label{fig:A31SNO_revisions} % Label per riferimenti incrociati
\end{figure}

As an illustration of the magnitude of data revisions, Figure~\ref{fig:A31SNO_revisions} shows the history of the Manufacturers’ New Orders for Primary Metals series between February 2011 and January 2025. The initial release in 2011 reported a value of 20,800 million dollars (red dot). Over subsequent years, revisions alternated between downward and upward adjustments--rising to 20,400 million in 2013 and peaking at 21,800 million in 2020. In recent years, revisions have become smaller, stabilizing around 21,400 million dollars. This pattern highlights how successive data updates can alter both the level and trajectory of a series, reinforcing the importance of using first-release information in real-time forecasting.

To replicate the informational constraints faced by real-world forecasters, we collect the first-release value of each observation from ALFRED, constructing a dataset in which each vintage expands only through the addition of newly available observations (see Supplement for details on dataset construction). Table~\ref{tab::predictors_list} also reports the transformations applied to each predictor as it enters forecasting models, as well as the number of missing observations requiring nowcasting.\footnote{Transformations follow, as closely as possible, those adopted by \citet{mccracken2016fred}, with deviations made when comparable series were unavailable. Additional information on data sources and series identifiers is provided in the Supplement.}

\subsection{Nowcasting and backcasting}\label{sec:nowcast}
As shown in Table \ref{tab::predictors_list}, most predictors are observed with a two-month lag relative to the forecast period, except for the New Orders for Aluminum products (NO-A) and Manufacturers' Shipments (MVS) series, which exhibits three missing observations. 
To forecast the prices of aluminum, copper, nickel, and zinc, we first address missing values in the series by nowcasting and backcasting them using alternative time series models. To bridge these data gaps and replace missing observations with model-based estimates, we conduct a “horse race” among univariate time series models, estimated under both frequentist and Bayesian frameworks. This exercise is essential because, unlike for widely studied macroeconomic indicators such as CPI or industrial production, there is limited empirical evidence to guide the selection of appropriate nowcasting models for many of the commodity price series considered here.

Our benchmark model is the Random Walk with drift (RW-D), a common reference in forecasting due to its simplicity and robust empirical performance. As frequentist alternatives, we consider a simple autoregressive model with one lag (AR(1)) and an autoregressive model with a lag length selected each period by the Akaike Information Criterion (AR(AIC)), allowing for time-varying dynamics.

On the Bayesian side, we start with a Bayesian AR(1) with constant volatility (BAR(1)), and then extend it to account for stochastic volatility (BAR(1)-SV), following evidence that time-varying uncertainty improves macroeconomic forecasts \citep{Carriero2019, Chan2020}. Finally, since our sample covers the COVID-19 period, we also consider a Bayesian AR(1) model with stochastic volatility and outliers (BAR(1)-SVo), as proposed by \cite{chan2023comparing} and \cite{Mertens2024}, to better capture the presence of extreme shocks.

For each data vintage and nowcast horizon, we estimate the model parameters and project the process \textit{h}-steps ahead, where $h$ ranges from 1 to 3 depending on the publication lag of the series.
The Bayesian and frequentist estimation follows a rolling window approach: as we move to the next data vintage, we shift the in-sample window forward by one period, incorporating a newly available first-release observation and repeating the process until the latest vintage, November 2023. 

To assess the goodness of the forecasts and compare the model performance, we compute the Root Mean Square Forecasting Errors (RMSFEs) using forecast errors from all six models, measured in the original units of each variable. 
We report the RMSFEs for the baseline RW-D model, whereas for the other autoregressive models, we report the ratios between the computed metric of the current model over the computed metric of the baseline model. 
Then, entries of less than $1$ indicate that the given current model yields nowcasts more accurate than those provided by the baseline random walk model.
In line with \cite{Carriero2015} and \cite{Clark2015}, rely the Diebold-Mariano test \citep{diebold1995comparing} to evaluate the forecast accuracy against the RWD model. The differences in accuracy that are statistically different from zero are denoted with one, two, or three asterisks, corresponding to significance level of $10\%$, $5\%$, and $1\%$, respectively.

Table \ref{tab:nowcast_rmspe} presents the RMSFE along with the ratios and the associated Diebold-Mariano test for each variable and nowcasting horizon, with the lowest ratios highlighted in bold. 
All autoregressive models show similar nowcasting performances across variables and nowcast horizon. 
As the gains from more sophisticated models appear marginal and considering that a real-time forecaster cannot know in advance which models will perform best over the entire sample, we adopt the simple RW-D model for subsequent analysis. 
We notice that when the advantages of using sophisticated models is higher in terms of the ratio, the Diebold-Mariano test does not provide evidence of their significance with respect to the benchmark (e.g. PPI-M rows).

\begin{table}[!ht]
  \centering
  \caption{Nowcasting of different predictors for each horizon and models.}
  \begin{threeparttable}
  \resizebox{6.5in}{!}{
% Table generated by Excel2LaTeX from sheet 'Nowcasting'
    \begin{tabular}{c|cllllll|c|rrrrrrr}
    \bottomrule
    \multicolumn{1}{c}{ID} & Horizon & \multicolumn{1}{c}{RW-D} & \multicolumn{1}{c}{AR(1)} & \multicolumn{1}{c}{AR(AIC)} & \multicolumn{1}{c}{BAR(1)} & \multicolumn{1}{c}{BAR(1) - SV} & \multicolumn{1}{c|}{BAR(1) - SVo} & \multicolumn{1}{c}{ID} & \multicolumn{1}{c}{Horizon} & \multicolumn{1}{c}{RW-D} & \multicolumn{1}{c}{AR(1)} & \multicolumn{1}{c}{AR(AIC)} & \multicolumn{1}{c}{BAR(1)} & \multicolumn{1}{c}{BAR(1) - SV} & \multicolumn{1}{c}{BAR(1) - SVo} \\
    \midrule
    \midrule
    \multirow{3}[2]{*}{IP} & 1     & 2.50  & 1.17*** & 1.23*** & 1.14*** & 1.07*** & 1.06*** & \multirow{3}[2]{*}{PPI-B} & \multicolumn{1}{c}{1} & \multicolumn{1}{l}{1.80} & \multicolumn{1}{l}{1.02***} & \multicolumn{1}{l}{\textbf{0.98}} & \multicolumn{1}{l}{1.02***} & \multicolumn{1}{l}{\textbf{0.99**}} & \multicolumn{1}{l}{1**} \\
          & 2     & 4.18  & 1.05*** & 1.21*** & 1.04*** & 1.04*** & 1.05*** &       & \multicolumn{1}{c}{2} & \multicolumn{1}{l}{2.69} & \multicolumn{1}{l}{1.03***} & \multicolumn{1}{l}{\textbf{0.94}} & \multicolumn{1}{l}{1***} & \multicolumn{1}{l}{1**} & \multicolumn{1}{l}{1***} \\
          & 3     & -     & -     & -     & -     & -     & -     &       & \multicolumn{1}{c}{3} & \multicolumn{1}{l}{-} & \multicolumn{1}{l}{-} & \multicolumn{1}{l}{-} & \multicolumn{1}{l}{-} & \multicolumn{1}{l}{-} & \multicolumn{1}{l}{-} \\
    \midrule
    \multirow{3}[2]{*}{NO-M} & 1     & 1126.48 & \textbf{0.92} & \textbf{0.92} & \textbf{0.92} & \textbf{0.92} & \textbf{0.92} & \multirow{3}[2]{*}{CHL} & \multicolumn{1}{c}{1} & \multicolumn{1}{l}{1.98} & \multicolumn{1}{l}{1.04***} & \multicolumn{1}{l}{1.04***} & \multicolumn{1}{l}{1.04***} & \multicolumn{1}{l}{1.03***} & \multicolumn{1}{l}{1.03***} \\
          & 2     & 1529.06 & \textbf{0.94} & \textbf{0.92} & \textbf{0.94} & \textbf{0.95} & \textbf{0.95} &       & \multicolumn{1}{c}{2} & \multicolumn{1}{l}{2.94} & \multicolumn{1}{l}{1**} & \multicolumn{1}{l}{1***} & \multicolumn{1}{l}{1**} & \multicolumn{1}{l}{\textbf{0.99}} & \multicolumn{1}{l}{\textbf{0.99*}} \\
          & 3     & -     & -     & -     & -     & -     & -     &       & \multicolumn{1}{c}{3} & \multicolumn{1}{l}{-} & \multicolumn{1}{l}{-} & \multicolumn{1}{l}{-} & \multicolumn{1}{l}{-} & \multicolumn{1}{l}{-} & \multicolumn{1}{l}{-} \\
    \midrule
    \multirow{3}[2]{*}{NO-A} & 1     & 739.94 & 1.06*** & 1.06*** & 1.06*** & 1.07*** & 1.07*** & \multirow{3}[2]{*}{AUS} & \multicolumn{1}{c}{1} & \multicolumn{1}{l}{1.49} & \multicolumn{1}{l}{1.02***} & \multicolumn{1}{l}{1.01***} & \multicolumn{1}{l}{1.02***} & \multicolumn{1}{l}{1.01***} & \multicolumn{1}{l}{1***} \\
          & 2     & 1067.93 & 1.01*** & 1***  & 1.01*** & 1.03*** & 1.03*** &       & \multicolumn{1}{c}{2} & \multicolumn{1}{l}{2.15} & \multicolumn{1}{l}{1.04***} & \multicolumn{1}{l}{1.04***} & \multicolumn{1}{l}{1.03***} & \multicolumn{1}{l}{1***} & \multicolumn{1}{l}{1***} \\
          & 3     & 1336.36 & \textbf{0.98**} & \textbf{0.97} & \textbf{0.98*} & 1***  & 1***  &       & \multicolumn{1}{c}{3} & \multicolumn{1}{l}{-} & \multicolumn{1}{l}{-} & \multicolumn{1}{l}{-} & \multicolumn{1}{l}{-} & \multicolumn{1}{l}{-} & \multicolumn{1}{l}{-} \\
    \midrule
    \multirow{3}[2]{*}{CPI} & 1     & \textbf{0.00} & \textbf{0.97} & \textbf{0.87} & \textbf{0.97} & \textbf{0.97} & \textbf{0.98} & \multirow{3}[2]{*}{CHN} & \multicolumn{1}{c}{1} & \multicolumn{1}{l}{1.52} & \multicolumn{1}{l}{\textbf{0.98}} & \multicolumn{1}{l}{\textbf{0.98}} & \multicolumn{1}{l}{\textbf{0.98}} & \multicolumn{1}{l}{\textbf{0.98}} & \multicolumn{1}{l}{\textbf{0.98}} \\
          & 2     & \textbf{0.01} & 1.07*** & \textbf{0.91} & 1.08*** & 1.11*** & 1.11*** &       & \multicolumn{1}{c}{2} & \multicolumn{1}{l}{2.41} & \multicolumn{1}{l}{1**} & \multicolumn{1}{l}{1***} & \multicolumn{1}{l}{\textbf{0.99*}} & \multicolumn{1}{l}{1***} & \multicolumn{1}{l}{1**} \\
          & 3     & -     & -     & -     & -     & -     & -     &       & \multicolumn{1}{c}{3} & \multicolumn{1}{l}{-} & \multicolumn{1}{l}{-} & \multicolumn{1}{l}{-} & \multicolumn{1}{l}{-} & \multicolumn{1}{l}{-} & \multicolumn{1}{l}{-} \\
    \midrule
    \multirow{3}[2]{*}{HEM} & 1     & \textbf{0.32} & 1.01*** & 1.03*** & 1.02*** & 1***  & 1.01*** & \multirow{3}[2]{*}{IDN} & \multicolumn{1}{c}{1} & \multicolumn{1}{l}{1.51} & \multicolumn{1}{l}{1.02***} & \multicolumn{1}{l}{\textbf{0.96}} & \multicolumn{1}{l}{1.03***} & \multicolumn{1}{l}{1.02***} & \multicolumn{1}{l}{1.02***} \\
          & 2     & \textbf{0.44} & \textbf{0.97} & \textbf{0.97} & 1**   & \textbf{0.98*} & \textbf{0.98} &       & \multicolumn{1}{c}{2} & \multicolumn{1}{l}{2.31} & \multicolumn{1}{l}{1.01***} & \multicolumn{1}{l}{\textbf{0.97}} & \multicolumn{1}{l}{1.01***} & \multicolumn{1}{l}{1.01***} & \multicolumn{1}{l}{1.01***} \\
          & 3     & -     & -     & -     & -     & -     & -     &       & \multicolumn{1}{c}{3} & \multicolumn{1}{l}{-} & \multicolumn{1}{l}{-} & \multicolumn{1}{l}{-} & \multicolumn{1}{l}{-} & \multicolumn{1}{l}{-} & \multicolumn{1}{l}{-} \\
    \midrule
    \multirow{3}[2]{*}{MVS} & 1     & 50.74 & 1.05*** & 1.06*** & 1.05*** & 1.04*** & 1.04*** & \multirow{3}[2]{*}{PER} & \multicolumn{1}{c}{1} & \multicolumn{1}{l}{2.20} & \multicolumn{1}{l}{\textbf{0.97*}} & \multicolumn{1}{l}{\textbf{0.97*}} & \multicolumn{1}{l}{\textbf{0.97*}} & \multicolumn{1}{l}{\textbf{0.97*}} & \multicolumn{1}{l}{\textbf{0.97*}} \\
          & 2     & 68.92 & 1.01*** & 1.02*** & 1.01*** & 1***  & 1**   &       & \multicolumn{1}{c}{2} & \multicolumn{1}{l}{3.31} & \multicolumn{1}{l}{\textbf{0.99}} & \multicolumn{1}{l}{\textbf{0.99}} & \multicolumn{1}{l}{\textbf{0.98}} & \multicolumn{1}{l}{\textbf{0.98}} & \multicolumn{1}{l}{\textbf{0.98}} \\
          & 3     & 76.58 & 1.02*** & 1.06*** & 1.02*** & 1.01*** & 1***  &       & \multicolumn{1}{c}{3} & \multicolumn{1}{l}{-} & \multicolumn{1}{l}{-} & \multicolumn{1}{l}{-} & \multicolumn{1}{l}{-} & \multicolumn{1}{l}{-} & \multicolumn{1}{l}{-} \\
    \midrule
    \multirow{3}[2]{*}{CU-M} & 1     & 1.84  & 1.09*** & 1.06*** & 1.09*** & 1.09*** & 1.09*** & \multirow{3}[2]{*}{PHL} & \multicolumn{1}{c}{1} & \multicolumn{1}{l}{1.36} & \multicolumn{1}{l}{1.01***} & \multicolumn{1}{l}{1**} & \multicolumn{1}{l}{1.01***} & \multicolumn{1}{l}{1.02***} & \multicolumn{1}{l}{1.01***} \\
          & 2     & 3.17  & 1.01*** & 1***  & 1.01*** & 1.01*** & 1.01*** &       & \multicolumn{1}{c}{2} & \multicolumn{1}{l}{2.07} & \multicolumn{1}{l}{1.04***} & \multicolumn{1}{l}{1.04***} & \multicolumn{1}{l}{1.03***} & \multicolumn{1}{l}{1.04***} & \multicolumn{1}{l}{1.04***} \\
          & 3     & -     & -     & -     & -     & -     & -     &       & \multicolumn{1}{c}{3} & \multicolumn{1}{l}{-} & \multicolumn{1}{l}{-} & \multicolumn{1}{l}{-} & \multicolumn{1}{l}{-} & \multicolumn{1}{l}{-} & \multicolumn{1}{l}{-} \\
    \midrule
    \multirow{3}[2]{*}{CU-P} & 1     & 1.17  & 1.04*** & 1.15*** & 1.04*** & 1.04*** & 1.04*** & \multirow{3}[2]{*}{RUS} & \multicolumn{1}{c}{1} & \multicolumn{1}{l}{6.05} & \multicolumn{1}{l}{\textbf{0.78}} & \multicolumn{1}{l}{\textbf{0.8}} & \multicolumn{1}{l}{\textbf{0.78}} & \multicolumn{1}{l}{\textbf{0.75}} & \multicolumn{1}{l}{\textbf{0.75}} \\
          & 2     & 1.84  & 1.02*** & 1.24*** & 1.02*** & 1***  & 1.01*** &       & \multicolumn{1}{c}{2} & \multicolumn{1}{l}{9.18} & \multicolumn{1}{l}{\textbf{0.85}} & \multicolumn{1}{l}{\textbf{0.86}} & \multicolumn{1}{l}{\textbf{0.85}} & \multicolumn{1}{l}{\textbf{0.86}} & \multicolumn{1}{l}{\textbf{0.86}} \\
          & 3     & -     & -     & -     & -     & -     & -     &       & \multicolumn{1}{c}{3} & \multicolumn{1}{l}{-} & \multicolumn{1}{l}{-} & \multicolumn{1}{l}{-} & \multicolumn{1}{l}{-} & \multicolumn{1}{l}{-} & \multicolumn{1}{l}{-} \\
    \midrule
    \multirow{3}[2]{*}{PPI-M} & 1     & 6.61  & \textbf{0.76} & \textbf{0.76} & \textbf{0.76} & \textbf{0.75} & \textbf{0.75} & \multirow{3}[2]{*}{-} &       &       &       &       &       &       &  \\
          & 2     & 11.89 & \textbf{0.8} & \textbf{0.8} & \textbf{0.79} & \textbf{0.78} & \textbf{0.78} &       &       &       &       &       &       &       &  \\
          & 3     & -     & -     & -     & -     & -     & -     &       &       &       &       &       &       &       &  \\
    \bottomrule
    \bottomrule
    \end{tabular}%
    }
 \begin{scriptsize}
           \begin{tablenotes}
        \item \parbox[t]{6in}{%
          \textbf{Notes:} For the baseline (RW-D) model, we report the RMSFE, whereas for the other models, the ratio relative to the baseline model. 
          A ratio below one indicates that the model outperforms the RW-D model. $^{\ast \ast \ast}$, $^{\ast \ast}$, and $^{\ast}$ indicate that ratios are significantly different from $1$ at the 1\%, 5\%, and 10\% based on the \cite{diebold1995comparing} test. The best-performing model at each horizon is highlighted in bold.
        }
      \end{tablenotes}
    \end{scriptsize}
  \end{threeparttable}
  \label{tab:nowcast_rmspe}%
\end{table}

\section{Forecasting Models and evaluations}
\label{sec:models}

\subsection{Model Description}
We consider both univariate and multivariate models for the real price of aluminum, copper, nickel and zinc. Moreover, we include various predictors as outlined in Section~\ref{sec:data} and we consider forecast at different horizons. 
As a natural comparison for the univariate and multivariate models, we also rely on forecasts already available in the market, such as those implied by future contracts \citep{alquist2010futures} or produced by professional forecasters \citep{ang2007macro,croushore2010evaluation}.

\subsubsection{Univariate models}
As a baseline specification, we consider the RW-D model. In addition, we include two alternative univariate time series models: an AR(1) model and an AR model in which the number of lags is selected according to the AIC (AR(AIC)).

Moreover, to leverage information from the set of real-data predictors discussed in Section~\ref{sec:data}, we augment the AR model with exogenous variables. Thus, we consider Autoregressive Distributed Lags (ARDL(p,s)) models, where we include the $p$ lags of the dependent variable and $s$ lags of the predictor variables:
\begin{equation}\label{eq:ardl}
        \widehat{y}_{t+h} = \widehat{c}_h + \sum_{i=1}^{p} \widehat{a}_{i,h} y_{t-i+1} + \sum_{j=1}^{s}\mathbf{\widehat{\Gamma}}_{j,h} \mathbf{Z}_{t-j+1},
\end{equation}
where $\widehat{y}_{t+h}$ is the real metal prices forecasted $h$ steps ahead, $\widehat{c}_h$ is the estimated intercept, $\widehat{a}_{i,h}$  the estimated autoregressive coefficient for each lag $i = 1,..., p$. 
The term $\mathbf{Z}_{t-j+1}$ is a $K \times 1$ vector of exogenous predictors at lag $t-j+1$ and $\mathbf{\widehat{\Gamma}}_{j}$ is the corresponding $1 \times K$ vector of coefficients for lag $j=1,\ldots, s$.

As for the AR model, the ARDL model is estimated either with 1 lags for both autoregressive and predictor variables or with the number of lags selected using the AIC. 
Regarding the choice of the $K$ real-time predictors, we include a single predictor at time or we use all predictors within the same categories as outlined in Section~\ref{sec:data}.

Another widely used approach in the forecasting literature to exploit information from a dataset with a large cross-sectional dimension is to estimate factors that summarize the information contained in the real-time dataset.
Following \cite{stock2002forecasting}, we augment the AR model with one or two factors extracted from the set of real-time predictors and we define the Autoregressive Diffusion Index (ARDI) model. 

The specification is analogous to Equation~\ref{eq:ardl}, except that the vector of exogenous variables $\mathbf{Z}_{t-j-1}$ is replaced by an $r \times 1$ vector of estimated factors $\widehat{\mathbf{F}}_{t-j+1}$ at lag $j$ and the corresponding coefficients are collected in a $1 \times r$ vector $\mathbf{\widehat{\Gamma}}_j$, for $j = 1, \ldots, q$.
These factors are estimated with the PCA approach by \citet{stock2002forecasting}, where $r$ is the number of factors. 
The $r \times N$ matrix of factor loadings $\widehat{\bm\Lambda}$ is estimated as the product between $\sqrt N$ and the $r$ normalized eigenvectors corresponding to the $r$-largest eigenvalues of the sample covariance matrix of the standardized data $\tilde{\mathbf x}_t$.\footnote{The target metal price is excluded from $\tilde{\mathbf{x}}_t$, as it already enters the model via the dependent variable $y$.} Then, the factors, $\widehat{\mathbf F}_t$, are obtained by projecting the estimated loadings onto the data, i.e., $\widehat{\mathbf F}_t=\widehat{\bm\Lambda} \mathbf{x}_t$.

\subsubsection{Multivariate models}
As a further set of specification, we employ on a Vector Autoregressive (VAR) models of order $p$: 

\begin{equation}
\label{eq:var}
    \mathbf{\widehat{Y}}_{t+h} = \mathbf{\widehat{C}}_h + \sum_{i=1}^{p} \mathbf{\widehat{A}}_i \mathbf{Y}_{t-i+1}, 
\end{equation}
where $\mathbf{\widehat{Y}}_{t+h}$ is the $n \times 1$ estimated vector of endogenous variables at horizon $t+h$. This vector includes the price and inventory volume for the mineral of interest, along with metals' new orders in the manufacturing sector (NO-M), for a total of $n = 3$ endogenous variables. The aim is to jointly model a measure of the metal price, a measure of the metal stock quantity, and a proxy for metal demand. $\mathbf{\widehat{C}}_h$ is the $n \times 1$ estimated vector of constants, $\mathbf{Y}_{t-i+1}$ is the $n \times 1$ vector of endogenous variables at lag $t-i+1$ and $\mathbf{\widehat{A}}_i$ is the associated $n \times n$ matrix of autoregressive coefficients.

Additionally, we consider an alternative multivariate specification where we augment the VAR model in Eq.~\eqref{eq:var} with either one or two factors constructed as in the ARDI model, thereby estimating a Factor-Augmented VAR (FAVAR)\footnote{In this case, we exclude from $\tilde{x}_t$ the metal price, metal inventory volume, and metals' new orders in the manufacturing sector (NO-M), since these are already included in the vector of endogenous variables $\mathbf{Y}$.}. As in the ARDI framework, this approach has the advantage of automatically selecting the most relevant information from a large set of predictors by weighting them differently in the factor estimation. Consequently, it removes the burden of manually identifying the most informative predictors from the list in Table \ref{tab::predictors_list}. Furthermore, because the factors are re-estimated at each iteration, the associated factor loadings are allowed to evolve over time, thereby offering some flexibility to accommodate potential structural breaks in the sample.

\subsubsection{Model-free forecasts}
The model previously described are compared against a baseline model, but they could also be compared with the forecasts available in the market, such as those implied by the future contracts or produced by professional forecasters.

\bigskip

\noindent\textit{Forecasts based on futures prices.} Since future contracts specify a price agreed upon in the present for the exchange of a commodity at a future date, their prices inherently reflect the market's expectations regarding future spot prices \citep{alquist2010futures}. 
Leveraging this characteristics, we employ LSEG Data \& Analytics on future contracts with $3$- and $15$- month maturity\footnote{Other maturities lower than 24 months were not available.}. The market’s expectation for the metal spot real price at time $T + h$ is computed as
\begin{equation}
    \widehat{P}_{T+h|T} = \frac{F^{T+h}_{T}}{\widehat{p}_{T+h|T}},
\end{equation}
where $F^{T+h}_{T}$ denotes the futures price at time $T$ for a transaction scheduled at time $T+h$, and $\widehat{p}_{T+h|T}$ is the forecast at time $T$ for the CPI index at time $T+h$\footnote{We estimate $\widehat{p}_{T+h|T}$ as ${p}_{T}(1+\widehat{g}^h_p)$, where $\widehat{g}^h_p$ is the CPI sample growth rate over $h$ periods.}.

\bigskip

\noindent\textit{Professional forecasters.} Another model-free forecast is provided by projections produced by professional forecasters. We compute the average forecast across individual forecasters published in the Consensus Economics survey. 
In detail, each monthly survey contains forecasts for fixed calendar dates -- March, June, September, and December -- regardless of the survey month. As a result, the forecast horizons are not constant but vary depending on the survey date.
Since these forecasts follow a \textit{fixed-event} structure, $y^{FE}_{T+h|T}$, we convert them into \textit{fixed-horizon} forecasts, $y^{FH}_{T+h|T}$, using an interpolation method as in \citet{dovern2012disagreement}:
\begin{equation}\label{eq::dovern}
    y^{FH}_{T+h|T} = \frac{d - |h - h_1|}{d} y^{FE}_{T+h_1|T} + \frac{d - |h - h_2|}{d} y^{FE}_{T+h_2|T},
\end{equation}
where $d$ is the distance between two consecutive forecasted horizons in the survey and in our case is equal to $3$ since forecasts are issued for a single month every quarter. The variable $h$ denotes the fixed forecast horizon of interest, while $h_1$ and $h_2$ are the survey horizons immediately preceding and following $h$, respectively.

\subsection{Forecast evaluation methods}

Based on the model described above, the datasets cover the period from January 2000 to November 2023, where the macroeconomic real-time observations start in April 2015\footnote{As shown in the Supplement, the start of real-time recording differs across series. To construct a balanced panel, we set April 2015 as the initial vintage, corresponding to the beginning of real-time availability for the Producer Price Index variables, which are the latest to begin real-time recording.}.
To conduct the forecasting exercise, we set April 2015 as the last in-sample date for the first data vintage.
Then, we forecast the real metal prices for different horizons, ranging from $1$ to $24$ months to better describe the short and medium term. 

The metal prices are forecasted in growth rates and then we reconstruct the forecasts in the original measurement unit before computing the relative forecasting errors.
As we move to subsequent data vintages, we shift the in-sample window one period forward and we repeat the procedure until November 2023 by adopting a rolling window approach.
Since we are interested in forecasting up to $24$ horizons, we use the direct forecasting method for the exogenous variables \citep[see, e.g.][]{Marcellino2006}.

As stated in Section~\ref{sec:nowcast}, we compute the Root Mean Square Forecasting Error (RMSFE) to evaluate the model performance. We report the RMSFEs for the baseline RW-D model, while for the other models we report the ratios between the computed metric of the current model over the one of the baseline model. 
Moreover,  we apply the Diebold-Mariano $t$ test \citep{diebold1995comparing}, where we compare each model with respect to the benchmark model and we denote the difference in accuracy that are statistically different from zero with one, two, or three stars corresponding to 10\%, 5\%, and 1\%, respectively. 
Finally, to jointly compare different model specifications, we employ the Model Confidence Set procedure proposed by \cite{hansen2011model} for a fixed horizon. 
The differences have been tested separately for each horizon and the MCS has been used to compare the predictive power without disentangling between univariate, multivariate or factor models. 
Results are presented in cells highlighted in light blue with confidence level $\alpha$ fixed at $0.25$\footnote{Results for different values of $\alpha$ and for all models are available in the Supplement.}.

\section{Forecasting the real prices of raw materials}\label{sec:forecasting}
Our forecasting evaluation is based on horizons up to 24 months ahead and compares a wide range of univariate and multivariate models, from simple benchmarks to factor-augmented specifications. The main results for copper and aluminum are summarised in Table~\ref{tab:forecasting_best_models_cop_alu} for selected horizons, while the corresponding full set of results is reported in the Supplement. For zinc and nickel, the key models are shown in Table~\ref{tab:forecasting_best_models_nick_zinc}. In all cases, we assess whether including real-time macroeconomic predictors delivers statistically significant gains in forecast accuracy relative to a random walk with drift (RW-D) baseline. The tables report RMSPEs (for RW-D) and RMSPE ratios (for all other models), alongside Diebold--Mariano tests and the Set of Superior Models (SSM) obtained from the MCS procedure.

Two general features emerge. First, different models outperform the baseline across metals and horizons, and no single specification dominates. This is consistent with the model selection uncertainty highlighted by the MCS results. Second, forecastability is horizon- and metal-specific: short-horizon gains are modest, while medium-term horizons (6--18 months ahead) are where most of the economically relevant improvements arise. Below we discuss each metal in turn and then compare these model-based forecasts to model-free benchmarks.

\bigskip

\noindent\textit{Copper.} For copper, forecasting gains are particularly pronounced at medium-term horizons (6 to 18 months ahead), in line with the metal’s pivotal role for global industrial activity. As shown in the top panel of Table~\ref{tab:forecasting_best_models_cop_alu} and in the Supplement, models incorporating information from the U.S. manufacturing sector (NO-M) deliver the strongest improvements. In particular, the ARDL model using manufacturers’ new orders for primary metals as an exogenous regressor reduces the RMSPE by roughly 20\% at the 6-month horizon and by about 30\% at the 12-month horizon relative to RW-D, with both gains statistically significant and the model frequently included in the SSM.

Moving from univariate to multivariate specifications further improves performance, although the gains are somewhat smaller than for the best ARDL(NO-M) model. When new orders are included in a VAR alongside inventories and copper prices, medium-term horizons still exhibit RMSPE reductions of around 20\%. This pattern is consistent with the idea that inventories transmit information on current and expected market tightness, but that new orders remain the dominant driver of medium-term predictability.

Factor-augmented models (ARDI and FAVAR) confirm and refine this picture. Both one- and two-factor versions perform strongly and enter the SSM at several horizons. The first factor places high loadings on metals-related manufacturing activity, capacity utilisation and producer prices, capturing a broad demand-driven macroeconomic component. The second factor loads more heavily on inventories and other metals’ prices, reflecting a market-specific component related to co-movements across metals and short-run supply adjustments. The forecasting gains of these factor models therefore arise from their ability to separate copper price movements into a macro-demand factor and a metals-specific factor.

Over short (1-month) and very long (24-month) horizons, the improvements relative to RW-D are more modest--on the order of 6--10\% for the best models. By contrast, at medium-term horizons, the best-performing ARDL and factor-augmented models achieve RMSPE reductions in the range of 17--31\% (from horizon 9 to 18), indicating that copper prices are substantially more predictable over these horizons when macroeconomic information is exploited.

\begin{figure}[!ht]
    \centering
        \caption{First (left) and second (right) estimated factors from the FAVAR model for real copper price}
 \begin{tabular}{cc}
   \includegraphics[width=0.5\textwidth]{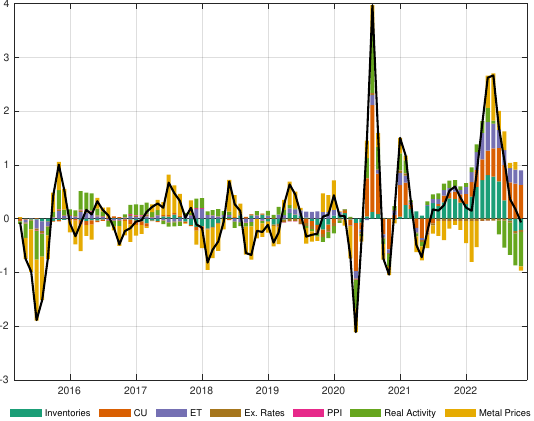}   &  
   \includegraphics[width=0.5\textwidth]{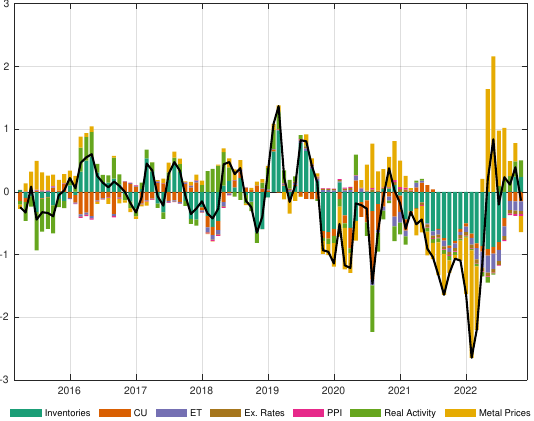}
 \end{tabular}

    \label{fig:factors_FAVAR_copper}
\end{figure}

To illustrate the underlying mechanisms, Figure~\ref{fig:factors_FAVAR_copper} plots the first two FAVAR factors for copper at the 12-month horizon.\footnote{The corresponding factors for the other commodities and for the ARDI models are reported in the Supplement.} The left panel shows a broad macroeconomic factor largely driven by real economic activity, capacity utilisation and producer prices, while the right panel highlights a metals-market factor dominated by inventories and other metals’ prices.

\begin{table}[!ht]
  \centering
  \caption{Copper and aluminum real prices point forecasts based on different models and horizons.}
  \begin{threeparttable}
  \resizebox{6.5in}{!}{
     \begin{tabular}{c|l|lllllllll}
    \toprule
    \multicolumn{1}{c}{Metal} & \multicolumn{1}{c}{Model} & \multicolumn{1}{c}{Horizon 1} & \multicolumn{1}{c}{Horizon 3} & \multicolumn{1}{c}{Horizon 6} & \multicolumn{1}{c}{Horizon 9} & \multicolumn{1}{c}{Horizon 12} & \multicolumn{1}{c}{Horizon 15} & \multicolumn{1}{c}{Horizon 18} & \multicolumn{1}{c}{Horizon 21} & \multicolumn{1}{c}{Horizon 24} \\
    \midrule
    \midrule
    \multirow{14}[4]{*}{Copper} & RW-D  & 292.26 & 610.34 & 918.10 & 1139.00 & 1363.14 & 1527.96 & 1628.31 & \cellcolor[rgb]{ .792,  .929,  .984}1634.84 & \cellcolor[rgb]{ .792,  .929,  .984}1625.21 \\
\cmidrule{2-11}          & Futures & \cellcolor[rgb]{ .792,  .929,  .984}- & 1.145*** & \cellcolor[rgb]{ .792,  .929,  .984}- & \cellcolor[rgb]{ .792,  .929,  .984}- & \cellcolor[rgb]{ .792,  .929,  .984}- & 1.004*** & \cellcolor[rgb]{ .792,  .929,  .984}- & \cellcolor[rgb]{ .792,  .929,  .984}- & \cellcolor[rgb]{ .792,  .929,  .984}- \\
          & Professional Forecasters & \cellcolor[rgb]{ .792,  .929,  .984}- & \cellcolor[rgb]{ .792,  .929,  .984}- & 1.069*** & 1.000*** & \textbf{0.942} & \cellcolor[rgb]{ .792,  .929,  .984}\textbf{0.898} & \cellcolor[rgb]{ .792,  .929,  .984}\textbf{0.857**} & \cellcolor[rgb]{ .792,  .929,  .984}- & \cellcolor[rgb]{ .792,  .929,  .984}- \\
          & ARDL(1) - Ec. Activity & 1.018*** & 1.029*** & 1.006*** & \textbf{0.941} & \cellcolor[rgb]{ .792,  .929,  .984}\textbf{0.860*} & \cellcolor[rgb]{ .792,  .929,  .984}\textbf{0.883} & \cellcolor[rgb]{ .792,  .929,  .984}\textbf{0.926} & \cellcolor[rgb]{ .792,  .929,  .984}\textbf{0.991*} & \cellcolor[rgb]{ .792,  .929,  .984}1.058*** \\
          & ARDL(1) - Ex. Rates & \textbf{0.975} & \textbf{0.980*} & 1.011*** & 1.045*** & 1.039*** & 1.052*** & 1.062*** & 1.069*** & 1.063*** \\
          & ARDL(1) - ETM Prices & \textbf{0.984} & 1.011*** & 1.171*** & \cellcolor[rgb]{ .792,  .929,  .984}1.197*** & 1.190*** & 1.141*** & \cellcolor[rgb]{ .792,  .929,  .984}1.105*** & 1.077*** & 1.099*** \\
          & ARDL(1) - ET & \textbf{0.973} & \textbf{0.926} & \textbf{0.931} & \textbf{0.933} & \textbf{0.929} & \textbf{0.950} & 1.000*** & 1.115*** & \cellcolor[rgb]{ .792,  .929,  .984}1.199*** \\
          & ARDL(1) - CU & 1.007*** & 1.068*** & 1.167*** & 1.199*** & 1.272*** & 1.280*** & 1.308*** & 1.445*** & 1.530*** \\
          & ARDL(1) - Inventories & \textbf{0.976} & 1.007*** & 1.079*** & 1.106*** & 1.085*** & 1.149*** & 1.136*** & \cellcolor[rgb]{ .792,  .929,  .984}1.167*** & 1.271*** \\
          & VAR(1) & \textbf{0.985} & \textbf{0.893*} & \cellcolor[rgb]{ .792,  .929,  .984}\textbf{0.796**} & \cellcolor[rgb]{ .792,  .929,  .984}\textbf{0.772**} & \cellcolor[rgb]{ .792,  .929,  .984}\textbf{0.750***} & \cellcolor[rgb]{ .792,  .929,  .984}\textbf{0.797***} & \cellcolor[rgb]{ .792,  .929,  .984}\textbf{0.855***} & \cellcolor[rgb]{ .792,  .929,  .984}\textbf{0.939*} & \cellcolor[rgb]{ .792,  .929,  .984}\textbf{0.997*} \\
          & ARDI(1) - 1 Factors & \cellcolor[rgb]{ .792,  .929,  .984}\textbf{0.942} & \cellcolor[rgb]{ .792,  .929,  .984}\textbf{0.856**} & \cellcolor[rgb]{ .792,  .929,  .984}\textbf{0.779**} & \cellcolor[rgb]{ .792,  .929,  .984}\textbf{0.717**} & \cellcolor[rgb]{ .792,  .929,  .984}\textbf{0.765**} & \cellcolor[rgb]{ .792,  .929,  .984}\textbf{0.832**} & \cellcolor[rgb]{ .792,  .929,  .984}\textbf{0.894**} & \cellcolor[rgb]{ .792,  .929,  .984}1.004*** & 1.122*** \\
          & ARDI(1) - 2 Factors & \cellcolor[rgb]{ .792,  .929,  .984}\textbf{0.943} & \cellcolor[rgb]{ .792,  .929,  .984}\textbf{0.861**} & \cellcolor[rgb]{ .792,  .929,  .984}\textbf{0.786**} & \cellcolor[rgb]{ .792,  .929,  .984}\textbf{0.691**} & \cellcolor[rgb]{ .792,  .929,  .984}\textbf{0.733**} & \cellcolor[rgb]{ .792,  .929,  .984}\textbf{0.815**} & \cellcolor[rgb]{ .792,  .929,  .984}\textbf{0.869**} & \cellcolor[rgb]{ .792,  .929,  .984}1.002*** & 1.152*** \\
          & FAVAR(1) - 1 Factors & \cellcolor[rgb]{ .792,  .929,  .984}\textbf{0.956} & \cellcolor[rgb]{ .792,  .929,  .984}\textbf{0.859**} & \cellcolor[rgb]{ .792,  .929,  .984}\textbf{0.784**} & \cellcolor[rgb]{ .792,  .929,  .984}\textbf{0.760**} & \cellcolor[rgb]{ .792,  .929,  .984}\textbf{0.743***} & \cellcolor[rgb]{ .792,  .929,  .984}\textbf{0.780***} & \cellcolor[rgb]{ .792,  .929,  .984}\textbf{0.841***} & \cellcolor[rgb]{ .792,  .929,  .984}\textbf{0.946*} & \cellcolor[rgb]{ .792,  .929,  .984}1.035*** \\
          & FAVAR(1) - 2 Factors & \textbf{0.957} & \cellcolor[rgb]{ .792,  .929,  .984}\textbf{0.868**} & \cellcolor[rgb]{ .792,  .929,  .984}\textbf{0.789**} & \cellcolor[rgb]{ .792,  .929,  .984}\textbf{0.734**} & \cellcolor[rgb]{ .792,  .929,  .984}\textbf{0.709***} & \cellcolor[rgb]{ .792,  .929,  .984}\textbf{0.766***} & \cellcolor[rgb]{ .792,  .929,  .984}\textbf{0.829***} & \cellcolor[rgb]{ .792,  .929,  .984}\textbf{0.965} & 1.097*** \\
    \midrule
    \multirow{14}[4]{*}{Aluminum} & RW-D  & 89.91 & \cellcolor[rgb]{ .792,  .929,  .984}192.38 & \cellcolor[rgb]{ .792,  .929,  .984}276.64 & 350.61 & 424.51 & 483.25 & 525.46 & \cellcolor[rgb]{ .792,  .929,  .984}529.15 & \cellcolor[rgb]{ .792,  .929,  .984}523.40 \\
\cmidrule{2-11}          & Futures & \cellcolor[rgb]{ .792,  .929,  .984}- & 1.803*** & \cellcolor[rgb]{ .792,  .929,  .984}- & \cellcolor[rgb]{ .792,  .929,  .984}- & \cellcolor[rgb]{ .792,  .929,  .984}- & \cellcolor[rgb]{ .792,  .929,  .984}1.093*** & \cellcolor[rgb]{ .792,  .929,  .984}- & \cellcolor[rgb]{ .792,  .929,  .984}- & \cellcolor[rgb]{ .792,  .929,  .984}- \\
          & Professional Forecasters & \cellcolor[rgb]{ .792,  .929,  .984}- & \cellcolor[rgb]{ .792,  .929,  .984}- & 1.151*** & 1.054*** & \textbf{0.978} & \cellcolor[rgb]{ .792,  .929,  .984}\textbf{0.918} & \cellcolor[rgb]{ .792,  .929,  .984}\textbf{0.848**} & \cellcolor[rgb]{ .792,  .929,  .984}- & \cellcolor[rgb]{ .792,  .929,  .984}- \\
          & ARDL(1) - Ec. Activity & \cellcolor[rgb]{ .792,  .929,  .984}\textbf{0.936} & \cellcolor[rgb]{ .792,  .929,  .984}1.020*** & \cellcolor[rgb]{ .792,  .929,  .984}\textbf{0.989*} & \cellcolor[rgb]{ .792,  .929,  .984}\textbf{0.996**} & \cellcolor[rgb]{ .792,  .929,  .984}\textbf{0.970} & \cellcolor[rgb]{ .792,  .929,  .984}\textbf{0.970} & \textbf{0.998**} & \cellcolor[rgb]{ .792,  .929,  .984}1.035*** & 1.090*** \\
          & ARDL(1) - Ex. Rates & \cellcolor[rgb]{ .792,  .929,  .984}\textbf{0.961*} & \cellcolor[rgb]{ .792,  .929,  .984}1.003*** & \cellcolor[rgb]{ .792,  .929,  .984}1.016*** & \cellcolor[rgb]{ .792,  .929,  .984}1.032*** & \cellcolor[rgb]{ .792,  .929,  .984}1.052*** & \cellcolor[rgb]{ .792,  .929,  .984}1.043*** & \cellcolor[rgb]{ .792,  .929,  .984}1.036*** & \cellcolor[rgb]{ .792,  .929,  .984}1.022*** & \cellcolor[rgb]{ .792,  .929,  .984}1.013*** \\
          & ARDL(1) - ETM Prices & \cellcolor[rgb]{ .792,  .929,  .984}\textbf{0.952} & \cellcolor[rgb]{ .792,  .929,  .984}1.051*** & \cellcolor[rgb]{ .792,  .929,  .984}1.143*** & \cellcolor[rgb]{ .792,  .929,  .984}1.131*** & \cellcolor[rgb]{ .792,  .929,  .984}1.130*** & \cellcolor[rgb]{ .792,  .929,  .984}1.084*** & \cellcolor[rgb]{ .792,  .929,  .984}1.053*** & \cellcolor[rgb]{ .792,  .929,  .984}1.029*** & \cellcolor[rgb]{ .792,  .929,  .984}1.040*** \\
          & ARDL(1) - ET & \cellcolor[rgb]{ .792,  .929,  .984}\textbf{0.919} & \cellcolor[rgb]{ .792,  .929,  .984}\textbf{0.992} & \cellcolor[rgb]{ .792,  .929,  .984}\textbf{0.955} & \textbf{0.935} & \textbf{0.924} & \textbf{0.918} & \textbf{0.955} & 1.013*** & 1.042*** \\
          & ARDL(1) - CU & \cellcolor[rgb]{ .792,  .929,  .984}\textbf{0.926} & \cellcolor[rgb]{ .792,  .929,  .984}1.015*** & \cellcolor[rgb]{ .792,  .929,  .984}1.046*** & \textbf{0.995**} & \textbf{0.954} & \textbf{0.938} & \cellcolor[rgb]{ .792,  .929,  .984}\textbf{0.955} & \cellcolor[rgb]{ .792,  .929,  .984}1.012*** & \cellcolor[rgb]{ .792,  .929,  .984}1.085*** \\
          & ARDL(1) - Inventories & \textbf{0.951} & \cellcolor[rgb]{ .792,  .929,  .984}1.030*** & 1.017*** & 1.024*** & 1.022*** & 1.010*** & \textbf{0.996*} & \cellcolor[rgb]{ .792,  .929,  .984}1.001*** & \textbf{0.990} \\
          & VAR(1) & \cellcolor[rgb]{ .792,  .929,  .984}\textbf{0.927} & \cellcolor[rgb]{ .792,  .929,  .984}1.012*** & \cellcolor[rgb]{ .792,  .929,  .984}\textbf{0.939*} & \cellcolor[rgb]{ .792,  .929,  .984}\textbf{0.882**} & \cellcolor[rgb]{ .792,  .929,  .984}\textbf{0.863**} & \cellcolor[rgb]{ .792,  .929,  .984}\textbf{0.858***} & \cellcolor[rgb]{ .792,  .929,  .984}\textbf{0.884**} & \cellcolor[rgb]{ .792,  .929,  .984}\textbf{0.942} & \cellcolor[rgb]{ .792,  .929,  .984}\textbf{0.971} \\
          & ARDI(1) - 1 Factors & \cellcolor[rgb]{ .792,  .929,  .984}\textbf{0.915*} & \cellcolor[rgb]{ .792,  .929,  .984}\textbf{0.948} & \cellcolor[rgb]{ .792,  .929,  .984}\textbf{0.906*} & \cellcolor[rgb]{ .792,  .929,  .984}\textbf{0.860*} & \cellcolor[rgb]{ .792,  .929,  .984}\textbf{0.843*} & \cellcolor[rgb]{ .792,  .929,  .984}\textbf{0.852} & \cellcolor[rgb]{ .792,  .929,  .984}\textbf{0.884} & \cellcolor[rgb]{ .792,  .929,  .984}\textbf{0.926} & \cellcolor[rgb]{ .792,  .929,  .984}\textbf{0.999***} \\
          & ARDI(1) - 2 Factors & \textbf{0.982} & \cellcolor[rgb]{ .792,  .929,  .984}\textbf{0.993} & \cellcolor[rgb]{ .792,  .929,  .984}\textbf{0.942} & \cellcolor[rgb]{ .792,  .929,  .984}\textbf{0.871*} & \cellcolor[rgb]{ .792,  .929,  .984}\textbf{0.847*} & \cellcolor[rgb]{ .792,  .929,  .984}\textbf{0.830*} & \cellcolor[rgb]{ .792,  .929,  .984}\textbf{0.858} & \cellcolor[rgb]{ .792,  .929,  .984}\textbf{0.920} & \cellcolor[rgb]{ .792,  .929,  .984}1.002*** \\
          & FAVAR(1) - 1 Factors & \textbf{0.977} & \cellcolor[rgb]{ .792,  .929,  .984}1.046*** & \cellcolor[rgb]{ .792,  .929,  .984}\textbf{0.997**} & \cellcolor[rgb]{ .792,  .929,  .984}\textbf{0.923} & \cellcolor[rgb]{ .792,  .929,  .984}\textbf{0.899*} & \cellcolor[rgb]{ .792,  .929,  .984}\textbf{0.885**} & \cellcolor[rgb]{ .792,  .929,  .984}\textbf{0.894*} & \cellcolor[rgb]{ .792,  .929,  .984}\textbf{0.939} & \cellcolor[rgb]{ .792,  .929,  .984}\textbf{0.974} \\
          & FAVAR(1) - 2 Factors & \textbf{0.991} & \cellcolor[rgb]{ .792,  .929,  .984}1.048*** & \cellcolor[rgb]{ .792,  .929,  .984}1.001*** & \cellcolor[rgb]{ .792,  .929,  .984}\textbf{0.944} & \cellcolor[rgb]{ .792,  .929,  .984}\textbf{0.935*} & \cellcolor[rgb]{ .792,  .929,  .984}\textbf{0.885**} & \cellcolor[rgb]{ .792,  .929,  .984}\textbf{0.879} & \cellcolor[rgb]{ .792,  .929,  .984}\textbf{0.957} & \cellcolor[rgb]{ .792,  .929,  .984}1.006*** \\
    \midrule
    \end{tabular}%
    }
 \begin{scriptsize}
      \begin{tablenotes}
        \item \parbox[t]{6in}{%
          \textbf{\textit{Notes:}} For the RW-D model, we report the RMSPE, while for all other models, we present the RMSPE ratio relative to the RW-D model. RMSPE refers to the real price in USD per metric ton, with February 2015 CPI as base year. A ratio below one (in bold) indicates that the model outperforms the RW-D model. Asterisks denote statistical significance at the 1\% (***), 5\% (**) and 10\% (*) based on the \cite{diebold1995comparing} test. Light blue cells indicate models included in the Set of Superior Models (SSM) at $\alpha = 0.25$ with the MCS procedure.
        }
      \end{tablenotes}
    \end{scriptsize}
  \end{threeparttable}
  \label{tab:forecasting_best_models_cop_alu}%
\end{table}

\bigskip

\noindent\textit{Aluminum.} Turning to aluminum (bottom panel of Table~\ref{tab:forecasting_best_models_cop_alu}), forecasting gains are again concentrated at medium-term horizons (roughly 6 to 15 months ahead). The largest improvements, of about 19--22\%, are obtained from univariate ARDL models that incorporate manufacturing-sector indicators, especially capacity utilisation and new orders for primary metals. These results reflect the tight link between aluminum demand and industrial production, with capacity utilisation in metals-related manufacturing acting as a particularly informative proxy.

As with copper, factor-augmented models improve on the benchmark and enter the SSM at several horizons. However, for aluminum the FAVAR specifications tend to underperform the best ARDI and simple VAR models. The VAR with inventories consistently beats RW-D and often ranks among the better performers, although it does not dominate across all horizons. The first factor in the ARDI and FAVAR models loads strongly on aggregate manufacturing activity (industrial production, manufacturing hours and producer prices), summarising a demand-side component that aligns with the univariate results. The second factor assigns more weight to inventory volumes and co-movements with other base metals, capturing a supply-side or market-structure element. 

Quantitatively, at short horizons (up to 3 months) and at the longest horizons (21--24 months), most models deliver only limited gains relative to RW-D, typically on the order of 8--10\%. At medium-term horizons, however, aluminum price forecasts benefit more clearly from univariate models linked to capacity utilisation and from factor models that combine industrial demand and inventory information.

\begin{table}[htbp!]
  \centering
  \caption{Zinc and nickel real prices point forecasts based on different models and horizons.}
  \begin{threeparttable}
  \resizebox{6.5in}{!}{
     \begin{tabular}{c|l|lllllllll}
    \toprule
    \multicolumn{1}{c}{Metal} & \multicolumn{1}{c}{Model} & \multicolumn{1}{c}{Horizon 1} & \multicolumn{1}{c}{Horizon 3} & \multicolumn{1}{c}{Horizon 6} & \multicolumn{1}{c}{Horizon 9} & \multicolumn{1}{c}{Horizon 12} & \multicolumn{1}{c}{Horizon 15} & \multicolumn{1}{c}{Horizon 18} & \multicolumn{1}{c}{Horizon 21} & \multicolumn{1}{c}{Horizon 24} \\
    \midrule
    \midrule
    \multirow{14}[4]{*}{Zinc} & RW-D  & \cellcolor[rgb]{ .792,  .929,  .984}153.58 & \cellcolor[rgb]{ .792,  .929,  .984}297.88 & \cellcolor[rgb]{ .792,  .929,  .984}448.55 & 548.23 & 633.02 & \cellcolor[rgb]{ .792,  .929,  .984}714.20 & \cellcolor[rgb]{ .792,  .929,  .984}782.88 & 817.70 & 850.19 \\
\cmidrule{2-11}          & Futures & \cellcolor[rgb]{ .792,  .929,  .984}- & \cellcolor[rgb]{ .792,  .929,  .984}\textbf{0.973} & \cellcolor[rgb]{ .792,  .929,  .984}- & \cellcolor[rgb]{ .792,  .929,  .984}- & \cellcolor[rgb]{ .792,  .929,  .984}- & \cellcolor[rgb]{ .792,  .929,  .984}\textbf{0.863} & \cellcolor[rgb]{ .792,  .929,  .984}- & \cellcolor[rgb]{ .792,  .929,  .984}- & \cellcolor[rgb]{ .792,  .929,  .984}- \\
          & Professional Forecasters & \cellcolor[rgb]{ .792,  .929,  .984}- & \cellcolor[rgb]{ .792,  .929,  .984}- & \cellcolor[rgb]{ .792,  .929,  .984}1.017*** & \cellcolor[rgb]{ .792,  .929,  .984}\textbf{0.917} & \cellcolor[rgb]{ .792,  .929,  .984}\textbf{0.850} & \cellcolor[rgb]{ .792,  .929,  .984}\textbf{0.800*} & \cellcolor[rgb]{ .792,  .929,  .984}\textbf{0.769**} & \cellcolor[rgb]{ .792,  .929,  .984}- & \cellcolor[rgb]{ .792,  .929,  .984}- \\
          & ARDL(1) - Ec. Activity & \cellcolor[rgb]{ .792,  .929,  .984}1.037*** & \cellcolor[rgb]{ .792,  .929,  .984}1.052*** & \cellcolor[rgb]{ .792,  .929,  .984}1.049*** & \cellcolor[rgb]{ .792,  .929,  .984}1.076*** & 1.110*** & 1.119*** & 1.128*** & 1.149*** & 1.104*** \\
          & ARDL(1) - Ex. Rates & \cellcolor[rgb]{ .792,  .929,  .984}\textbf{0.993} & \cellcolor[rgb]{ .792,  .929,  .984}1.003*** & 1.027*** & 1.045*** & 1.059*** & \cellcolor[rgb]{ .792,  .929,  .984}1.046*** & \cellcolor[rgb]{ .792,  .929,  .984}1.022*** & \cellcolor[rgb]{ .792,  .929,  .984}\textbf{0.977} & \textbf{0.939} \\
          & ARDL(1) - ETM Prices & \cellcolor[rgb]{ .792,  .929,  .984}1.007*** & \cellcolor[rgb]{ .792,  .929,  .984}1.026*** & \cellcolor[rgb]{ .792,  .929,  .984}1.134*** & \cellcolor[rgb]{ .792,  .929,  .984}1.145*** & \cellcolor[rgb]{ .792,  .929,  .984}1.212*** & \cellcolor[rgb]{ .792,  .929,  .984}1.147*** & \cellcolor[rgb]{ .792,  .929,  .984}1.054*** & \cellcolor[rgb]{ .792,  .929,  .984}\textbf{0.954} & \textbf{0.919} \\
          & ARDL(1) - ET & \cellcolor[rgb]{ .792,  .929,  .984}\textbf{0.986} & \cellcolor[rgb]{ .792,  .929,  .984}\textbf{0.969} & \cellcolor[rgb]{ .792,  .929,  .984}\textbf{0.925*} & \textbf{0.897} & \textbf{0.883} & \textbf{0.889} & \cellcolor[rgb]{ .792,  .929,  .984}\textbf{0.909} & \cellcolor[rgb]{ .792,  .929,  .984}\textbf{0.940} & \textbf{0.950} \\
          & ARDL(1) - CU & \cellcolor[rgb]{ .792,  .929,  .984}1.014*** & \cellcolor[rgb]{ .792,  .929,  .984}1.071*** & 1.190*** & 1.226*** & \cellcolor[rgb]{ .792,  .929,  .984}1.219*** & \cellcolor[rgb]{ .792,  .929,  .984}1.135*** & \cellcolor[rgb]{ .792,  .929,  .984}1.079*** & \cellcolor[rgb]{ .792,  .929,  .984}1.033*** & 1.011*** \\
          & ARDL(1) - Inventories & \cellcolor[rgb]{ .792,  .929,  .984}1.002*** & \cellcolor[rgb]{ .792,  .929,  .984}1.038*** & 1.081*** & 1.045*** & 1.008*** & \textbf{0.983} & \cellcolor[rgb]{ .792,  .929,  .984}\textbf{0.958} & \cellcolor[rgb]{ .792,  .929,  .984}1.046*** & 1.058*** \\
          & VAR(1) & \cellcolor[rgb]{ .792,  .929,  .984}1.005*** & \cellcolor[rgb]{ .792,  .929,  .984}\textbf{0.995} & \cellcolor[rgb]{ .792,  .929,  .984}\textbf{0.987} & \cellcolor[rgb]{ .792,  .929,  .984}\textbf{0.963} & \cellcolor[rgb]{ .792,  .929,  .984}\textbf{0.923**} & \cellcolor[rgb]{ .792,  .929,  .984}\textbf{0.895***} & \cellcolor[rgb]{ .792,  .929,  .984}\textbf{0.913**} & \cellcolor[rgb]{ .792,  .929,  .984}1.012*** & 1.094*** \\
          & ARDI(1) - 1 Factors & \cellcolor[rgb]{ .792,  .929,  .984}\textbf{0.973} & \cellcolor[rgb]{ .792,  .929,  .984}\textbf{0.955*} & \cellcolor[rgb]{ .792,  .929,  .984}\textbf{0.951} & \cellcolor[rgb]{ .792,  .929,  .984}\textbf{0.946*} & \cellcolor[rgb]{ .792,  .929,  .984}\textbf{0.925**} & \cellcolor[rgb]{ .792,  .929,  .984}\textbf{0.917**} & \cellcolor[rgb]{ .792,  .929,  .984}\textbf{0.935} & \cellcolor[rgb]{ .792,  .929,  .984}\textbf{0.959} & \textbf{0.955} \\
          & ARDI(1) - 2 Factors & \cellcolor[rgb]{ .792,  .929,  .984}\textbf{0.955*} & \cellcolor[rgb]{ .792,  .929,  .984}\textbf{0.951*} & \cellcolor[rgb]{ .792,  .929,  .984}\textbf{0.966} & \cellcolor[rgb]{ .792,  .929,  .984}\textbf{0.950} & \cellcolor[rgb]{ .792,  .929,  .984}\textbf{0.913**} & \cellcolor[rgb]{ .792,  .929,  .984}\textbf{0.899**} & \cellcolor[rgb]{ .792,  .929,  .984}\textbf{0.910} & \cellcolor[rgb]{ .792,  .929,  .984}\textbf{0.925} & \textbf{0.937} \\
          & FAVAR(1) - 1 Factors & \cellcolor[rgb]{ .792,  .929,  .984}\textbf{0.960} & \cellcolor[rgb]{ .792,  .929,  .984}\textbf{0.962} & \cellcolor[rgb]{ .792,  .929,  .984}\textbf{0.991} & \cellcolor[rgb]{ .792,  .929,  .984}\textbf{0.988} & \cellcolor[rgb]{ .792,  .929,  .984}\textbf{0.960} & \cellcolor[rgb]{ .792,  .929,  .984}\textbf{0.939} & \cellcolor[rgb]{ .792,  .929,  .984}\textbf{0.955} & \cellcolor[rgb]{ .792,  .929,  .984}1.058*** & 1.148*** \\
          & FAVAR(1) - 2 Factors & \cellcolor[rgb]{ .792,  .929,  .984}\textbf{0.957} & \cellcolor[rgb]{ .792,  .929,  .984}\textbf{0.965} & \cellcolor[rgb]{ .792,  .929,  .984}1.000*** & \cellcolor[rgb]{ .792,  .929,  .984}\textbf{0.982} & \cellcolor[rgb]{ .792,  .929,  .984}\textbf{0.941} & \cellcolor[rgb]{ .792,  .929,  .984}\textbf{0.922} & \cellcolor[rgb]{ .792,  .929,  .984}\textbf{0.955} & \cellcolor[rgb]{ .792,  .929,  .984}1.064*** & 1.183*** \\
    \midrule
    \multirow{14}[4]{*}{Nickel} & RW-D  & \cellcolor[rgb]{ .792,  .929,  .984}1557.23 & \cellcolor[rgb]{ .792,  .929,  .984}2913.44 & \cellcolor[rgb]{ .792,  .929,  .984}3526.04 & \cellcolor[rgb]{ .792,  .929,  .984}3610.10 & \cellcolor[rgb]{ .792,  .929,  .984}3936.27 & \cellcolor[rgb]{ .792,  .929,  .984}4100.56 & \cellcolor[rgb]{ .792,  .929,  .984}4675.75 & \cellcolor[rgb]{ .792,  .929,  .984}4858.20 & \cellcolor[rgb]{ .792,  .929,  .984}5283.20 \\
\cmidrule{2-11}          & Futures & \cellcolor[rgb]{ .792,  .929,  .984}- & \cellcolor[rgb]{ .792,  .929,  .984}\textbf{0.929} & \cellcolor[rgb]{ .792,  .929,  .984}- & \cellcolor[rgb]{ .792,  .929,  .984}- & \cellcolor[rgb]{ .792,  .929,  .984}- & \cellcolor[rgb]{ .792,  .929,  .984}\textbf{0.960} & \cellcolor[rgb]{ .792,  .929,  .984}- & \cellcolor[rgb]{ .792,  .929,  .984}- & \cellcolor[rgb]{ .792,  .929,  .984}- \\
          & Professional Forecasters & \cellcolor[rgb]{ .792,  .929,  .984}- & \cellcolor[rgb]{ .792,  .929,  .984}- & \cellcolor[rgb]{ .792,  .929,  .984}\textbf{0.961} & \cellcolor[rgb]{ .792,  .929,  .984}1.004*** & \cellcolor[rgb]{ .792,  .929,  .984}\textbf{0.955} & \cellcolor[rgb]{ .792,  .929,  .984}\textbf{0.968} & \cellcolor[rgb]{ .792,  .929,  .984}\textbf{0.927} & \cellcolor[rgb]{ .792,  .929,  .984}- & \cellcolor[rgb]{ .792,  .929,  .984}- \\
          & ARDL(1) - Ec. Activity & \cellcolor[rgb]{ .792,  .929,  .984}1.077*** & \cellcolor[rgb]{ .792,  .929,  .984}1.047*** & \cellcolor[rgb]{ .792,  .929,  .984}1.115*** & \cellcolor[rgb]{ .792,  .929,  .984}1.195*** & 1.149*** & 1.156*** & 1.129*** & 1.141*** & 1.135*** \\
          & ARDL(1) - Ex. Rates & \cellcolor[rgb]{ .792,  .929,  .984}1.051*** & \cellcolor[rgb]{ .792,  .929,  .984}1.019*** & \cellcolor[rgb]{ .792,  .929,  .984}1.096*** & \cellcolor[rgb]{ .792,  .929,  .984}1.067*** & \cellcolor[rgb]{ .792,  .929,  .984}1.069*** & \cellcolor[rgb]{ .792,  .929,  .984}1.091*** & \cellcolor[rgb]{ .792,  .929,  .984}1.094*** & \cellcolor[rgb]{ .792,  .929,  .984}1.066*** & \cellcolor[rgb]{ .792,  .929,  .984}1.074*** \\
          & ARDL(1) - ETM Prices & \cellcolor[rgb]{ .792,  .929,  .984}1.054*** & \cellcolor[rgb]{ .792,  .929,  .984}1.031*** & \cellcolor[rgb]{ .792,  .929,  .984}1.187*** & \cellcolor[rgb]{ .792,  .929,  .984}1.211*** & \cellcolor[rgb]{ .792,  .929,  .984}1.281*** & \cellcolor[rgb]{ .792,  .929,  .984}1.283*** & 1.226*** & 1.119*** & 1.129*** \\
          & ARDL(1) - ET & \cellcolor[rgb]{ .792,  .929,  .984}1.068*** & \cellcolor[rgb]{ .792,  .929,  .984}1.012*** & \cellcolor[rgb]{ .792,  .929,  .984}1.075*** & \cellcolor[rgb]{ .792,  .929,  .984}1.010*** & 1.029*** & \cellcolor[rgb]{ .792,  .929,  .984}1.045*** & 1.052*** & \cellcolor[rgb]{ .792,  .929,  .984}1.092*** & 1.089*** \\
          & ARDL(1) - CU & \cellcolor[rgb]{ .792,  .929,  .984}1.065*** & 1.035*** & 1.127*** & 1.126*** & 1.118*** & 1.153*** & \textbf{0.972} & \textbf{0.845} & 1.053*** \\
          & ARDL(1) - Inventories & \cellcolor[rgb]{ .792,  .929,  .984}1.074*** & \cellcolor[rgb]{ .792,  .929,  .984}1.066*** & \cellcolor[rgb]{ .792,  .929,  .984}1.234*** & 1.249*** & 1.251*** & 1.254*** & 1.209*** & 1.145*** & 1.157*** \\
          & VAR(1) & \cellcolor[rgb]{ .792,  .929,  .984}1.061*** & \cellcolor[rgb]{ .792,  .929,  .984}\textbf{0.995} & \cellcolor[rgb]{ .792,  .929,  .984}1.079*** & \cellcolor[rgb]{ .792,  .929,  .984}1.067*** & \cellcolor[rgb]{ .792,  .929,  .984}\textbf{0.984} & \cellcolor[rgb]{ .792,  .929,  .984}\textbf{0.965} & \cellcolor[rgb]{ .792,  .929,  .984}\textbf{0.972} & \cellcolor[rgb]{ .792,  .929,  .984}\textbf{0.945} & \cellcolor[rgb]{ .792,  .929,  .984}\textbf{0.970} \\
          & ARDI(1) - 1 Factors & \cellcolor[rgb]{ .792,  .929,  .984}1.040*** & \cellcolor[rgb]{ .792,  .929,  .984}\textbf{0.978} & \cellcolor[rgb]{ .792,  .929,  .984}1.052*** & \cellcolor[rgb]{ .792,  .929,  .984}1.033*** & \cellcolor[rgb]{ .792,  .929,  .984}\textbf{0.940} & \cellcolor[rgb]{ .792,  .929,  .984}\textbf{0.926} & \cellcolor[rgb]{ .792,  .929,  .984}\textbf{0.889} & \cellcolor[rgb]{ .792,  .929,  .984}\textbf{0.861} & \cellcolor[rgb]{ .792,  .929,  .984}1.007*** \\
          & ARDI(1) - 2 Factors & \cellcolor[rgb]{ .792,  .929,  .984}1.072*** & \cellcolor[rgb]{ .792,  .929,  .984}1.032*** & \cellcolor[rgb]{ .792,  .929,  .984}1.139*** & \cellcolor[rgb]{ .792,  .929,  .984}1.076*** & \cellcolor[rgb]{ .792,  .929,  .984}1.081*** & \cellcolor[rgb]{ .792,  .929,  .984}1.105*** & \cellcolor[rgb]{ .792,  .929,  .984}1.103*** & \cellcolor[rgb]{ .792,  .929,  .984}1.075*** & \cellcolor[rgb]{ .792,  .929,  .984}1.084*** \\
          & FAVAR(1) - 1 Factors & \cellcolor[rgb]{ .792,  .929,  .984}1.038*** & \cellcolor[rgb]{ .792,  .929,  .984}\textbf{0.987} & \cellcolor[rgb]{ .792,  .929,  .984}1.081*** & \cellcolor[rgb]{ .792,  .929,  .984}1.078*** & \cellcolor[rgb]{ .792,  .929,  .984}\textbf{0.998**} & \cellcolor[rgb]{ .792,  .929,  .984}\textbf{0.973} & \cellcolor[rgb]{ .792,  .929,  .984}\textbf{0.981} & \cellcolor[rgb]{ .792,  .929,  .984}\textbf{0.940} & \cellcolor[rgb]{ .792,  .929,  .984}\textbf{0.996*} \\
          & FAVAR(1) - 2 Factors & \cellcolor[rgb]{ .792,  .929,  .984}1.041*** & \cellcolor[rgb]{ .792,  .929,  .984}\textbf{0.990} & \cellcolor[rgb]{ .792,  .929,  .984}1.093*** & \cellcolor[rgb]{ .792,  .929,  .984}1.074*** & \cellcolor[rgb]{ .792,  .929,  .984}\textbf{0.987} & \cellcolor[rgb]{ .792,  .929,  .984}\textbf{0.947} & \cellcolor[rgb]{ .792,  .929,  .984}\textbf{0.959} & \cellcolor[rgb]{ .792,  .929,  .984}\textbf{0.903} & \cellcolor[rgb]{ .792,  .929,  .984}1.001*** \\
    \midrule
    \bottomrule
    \end{tabular}%
    }
 \begin{scriptsize}
      \begin{tablenotes}
        \item See notes in Table~\ref{tab:forecasting_best_models_cop_alu}
      \end{tablenotes}
    \end{scriptsize}
  \end{threeparttable}
  \label{tab:forecasting_best_models_nick_zinc}%
\end{table}

\bigskip

\noindent\textit{Zinc.} The picture is somewhat different for zinc, as shown in the top panel of Table~\ref{tab:forecasting_best_models_nick_zinc}. Here, model-free forecasts (futures and professional forecasts) perform relatively well at intermediate horizons and are selected into the SSM for horizons between 12 and 18 months. However, models that exploit macroeconomic information still yield meaningful gains, especially at medium and long horizons.

In particular, zinc forecasts benefit from models incorporating capacity utilisation, new orders and inventory dynamics, with RMSPE reductions of roughly 12\% at the 12-month horizon and up to about 21\% at the 24-month horizon. By contrast, at short horizons (1--6 months ahead), the more complex specifications typically perform close to the RW-D benchmark. Capacity utilisation and new orders help anticipate zinc demand pressures and thus provide early signals about zinc consumption. Unlike for copper and aluminum, zinc inventories play an especially prominent role, reflecting their tight link to global galvanising cycles (construction, autos and infrastructure). Inventory behaviour therefore acts as an early indicator of impending supply tightness, with shorter and more frequent inventory cycles than for copper and aluminum.

Energy-transition (ET) indicators alone generate smaller gains than when they are incorporated into a broader factor structure. Including battery manufacturing producer prices and shipments is, however, useful to capture the growing importance of electrification and storage technologies. As with copper and aluminum, factor models (ARDI and FAVAR) deliver additional improvements, particularly in the ARDI specification with two factors. The first factor mirrors demand-driven components and loads on industrial production, new orders and producer price indices, while the second factor is dominated by inventories and co-movements with copper and aluminum. This confirms the value of combining standard macroeconomic indicators with metals-specific variables.

\bigskip

\noindent\textit{Nickel.} Nickel behaves quite differently from the other three metals (bottom panel of Table~\ref{tab:forecasting_best_models_nick_zinc}). At short horizons, virtually no model systematically outperforms the RW-D benchmark, and model-free forecasts are often competitive or slightly better. Forecastability improves somewhat at medium horizons (12--18 months ahead), but only a limited set of models achieves statistically significant gains, and the magnitude of these gains remains modest compared with those observed for copper, aluminum and zinc.

The best-performing models for nickel are factor-augmented specifications, which deliver RMSPE reductions of roughly 12--15\% at selected horizons. The first factor appears to capture aggregate macroeconomic activity but displays a relatively noisy pattern, making it difficult to associate it with a specific sector. The second factor relies heavily on inventory series and metals price co-movements. Even so, the overall evidence points to nickel being intrinsically harder to forecast, likely due to its geographically concentrated and technologically fragmented production, together with heightened exposure to geopolitical shocks. These structural features are only imperfectly captured by the macroeconomic and market indicators in our dataset.

\bigskip

\noindent\textit{Cross-metal comparison and model-free benchmarks.} Taken together, the results suggest that univariate models based on a single, well-chosen predictor often deliver the largest gains relative to RW-D, particularly for copper and aluminum. Factor models that summarise information from multiple macroeconomic series further improve forecast accuracy for some metals and horizons, especially when they combine demand-side indicators with inventories and other metals’ prices. For zinc, inventories and co-movements with other base metals are crucial, while for nickel the scope for improving on the random walk is limited.

Comparing our model-based forecasts with model-free benchmarks highlights three additional points. First, futures prices are generally not reliable predictors of future spot prices, and their RMSPE is often higher than that of even simple macro-based models. This suggests that market participants do not fully incorporate available information -- such as professional forecasts and macroeconomic signals -- into futures prices. Second, across most horizons and metals, our models outperform both futures-based forecasts and those issued by professional forecasters, especially at medium-term horizons where macroeconomic predictors are most informative. Third, nickel remains particularly challenging: both our models and professional forecasters deliver relatively weak and often statistically insignificant gains over the RW-D benchmark, underscoring the intrinsic difficulty of forecasting nickel prices.

\section{Model Comparison and Forecast Pooling}\label{sec:mod_comparison_&_pooling}
\subsection{Model Confidence Set Information}

Given the large number of competing models in our analysis, we use the Model
Confidence Set (MCS) procedure of \citet{hansen2011model} to evaluate their
relative performance. The MCS iteratively removes the worst-performing
specifications until a set of models remains that cannot be statistically
distinguished from the best one. The Supplement reports MCS $p$-values across
horizons, where one and two asterisks denote retention at the $\alpha = 0.10$
and $\alpha = 0.25$ levels, respectively.

A key result is that, for most metals and horizons, the SSM is large and
typically contains many models. This indicates substantial model uncertainty:
the real-time data do not provide sharp statistical discrimination among
specifications. Only for nickel at the 24-month horizon does the SSM shrink
meaningfully, reflecting clearer performance differences at very long horizons.

Although the MCS rarely identifies a small set of superior models, the
associated $p$-values allow us to rank specifications. For copper, factor-augmented
models (ARDI and FAVAR) consistently receive the highest rankings up to
18 months ahead, reinforcing the importance of broad macroeconomic information.
At longer horizons, simpler models based on new orders or producer price indices
often perform best. For aluminum, the top-ranked specifications are almost
exclusively univariate ARDL models using capacity utilisation and metals-sector
new orders, with producer price indices becoming more relevant at longer
horizons.

Zinc displays greater heterogeneity: factor models rank highly at very short
horizons, ARDI models with energy-transition variables perform well up to one
year, and model-free forecasts (futures and surveys) enter the SSM for several
medium-term horizons. At longer horizons, producer price indices and metals-sector
indicators reappear among the best-performing models. Nickel remains the most
difficult metal to forecast: factor models, capacity-utilisation specifications,
and model-free forecasts achieve similar MCS rankings at short and medium
horizons, while models including metals-sector new orders are favored at very
long horizons.

Overall, the MCS results confirm the main finding of Section~\ref{sec:forecasting}:
no single model dominates across metals or horizons, but certain predictors—
especially manufacturing new orders and capacity utilisation—repeatedly appear
in the best-performing specifications.

\subsection{Forecast pooling}

It is well known in the forecasting literature that pooling forecasts from multiple models can improve predictive accuracy compared to relying on a single model \citep{bates1969combination}.
Forecast combination leverages the strengths of different models, mitigating the risk of selecting a single, potentially misspecified, model. 
By averaging or weighting forecasts based on their past performance, combination methods can enhance robustness and reduce forecast uncertainty. 
This is particularly relevant and useful in our analysis, as the MCS procedure in the previous section revealed that it is difficult to identify a set of models that are clearly superior -- particularly for nickel and zinc. As a result, combining forecasts from multiple models may represent an effective strategy.

In this section, we perform model selection in real time using the MCS procedure and then combine the forecasts of the corresponding SSM at a confidence level equal to $\alpha = 0.25$. 
Specifically, during the first year of the evaluation period, we average the forecasts of all models considered in Section \ref{sec:forecasting}. 
Starting from the thirteenth month, we apply the MCS procedure to the past 12 observations and then construct forecasts using two different approaches: (i) averaging the forecasts of all selected models and (ii) averaging the forecasts of the two best-performing models, as determined by their MCS p-values.

As shown in Table \ref{tab:forecast_pooling}, pooling forecasts using both methods consistently yields greater forecast accuracy compared to the RW-D model across all horizons and metals, except for nickel, where the random-walk outperform the pooling forecast up to 10\% in the first 6 horizons.
As highlighted by the gray cells, we notice that the forecasting pooling provides more accurate predictions than any individual model over the horizon ranging from 9 to 21 horizons depending from the metal.

For copper, however, individual models based on Manufacturers' New Orders for Primary Metals (NO-M) -- such as the corresponding ARDL, VAR, and FAVAR specifications -- continue to outperform forecast combination strategies in the short and medium term. 
This result is likely due to the exceptionally strong predictive power of new orders for copper prices relative to other metals, combined with the weak forecasting performance of other macroeconomic predictors. 
Indeed, the gap in forecasting accuracy between models based on new orders and those based on other predictors is particularly pronounced for copper compared to the other three metals. 
As a result, pooling forecasts in the case of copper may mitigate the signal provided by a key indicator of planned production by incorporating information from variables with limited predictive content, thus introducing additional noise.
Indeed, it is important in the long-term (from 21 horizons), when doing forecasting combination outperform the best individual models. 
To wrap up, we notice that in the long-term adding a combination tool leads to better results with respect to single models thus highlighting the importance of using different variables but also different models in the forecasting exercise when less information are available.

% Table generated by Excel2LaTeX from sheet 'Forecast Pooling'
\begin{table}[!ht]
  \centering
  \caption{Point forecasting measures by combining the forecasting based on MCS procedures.
  }
  \begin{threeparttable}
    \resizebox{6.5in}{!}{
    \begin{tabular}{l|lllllllll}
    \toprule
    \multicolumn{1}{l}{Model} & \multicolumn{1}{c}{Horizon 1} & \multicolumn{1}{c}{Horizon 3} & \multicolumn{1}{c}{Horizon 6} & \multicolumn{1}{c}{Horizon 9} & \multicolumn{1}{c}{Horizon 12} & \multicolumn{1}{c}{Horizon 15} & \multicolumn{1}{c}{Horizon 18} & \multicolumn{1}{c}{Horizon 21} & \multicolumn{1}{c}{Horizon 24} \\
    \midrule
    \midrule
    \multicolumn{10}{c}{Copper} \\
    \midrule
    \multicolumn{1}{l}{RW-D} & 292   & 610   & 918   & 1139  & 1363  & 1528  & 1628  & 1635  & 1625 \\
    \midrule
    SSM25 & \textbf{0.972} & \textbf{0.923**} & \textbf{0.845**} & \textbf{0.753**} & \textbf{0.744***} & \textbf{0.813***} & \cellcolor[rgb]{ .816,  .816,  .816}\textbf{0.823***} & \cellcolor[rgb]{ .816,  .816,  .816}\textbf{0.863***} & \cellcolor[rgb]{ .816,  .816,  .816}\textbf{0.882**} \\
    Top 2 & \textbf{0.974} & \textbf{0.897**} & \textbf{0.816**} & \textbf{0.742**} & \textbf{0.746***} & \textbf{0.824***} & \textbf{0.833***} & \cellcolor[rgb]{ .816,  .816,  .816}\textbf{0.831***} & \cellcolor[rgb]{ .816,  .816,  .816}\textbf{0.852**} \\
    \midrule
    \multicolumn{10}{c}{Aluminum} \\
    \midrule
    \multicolumn{1}{l}{RW-D} & 90    & 192   & 277   & 351   & 425   & 483   & 525   & 529   & 523 \\
    \midrule
    SSM25 & \textbf{0.939} & 1.008*** & \textbf{0.922*} & \textbf{0.829***} & \textbf{0.819**} & \cellcolor[rgb]{ .816,  .816,  .816}\textbf{0.762**} & \cellcolor[rgb]{ .816,  .816,  .816}\textbf{0.772**} & \cellcolor[rgb]{ .816,  .816,  .816}\textbf{0.845***} & \textbf{0.906**} \\
    Top 2 & \textbf{0.938} & {1.015***} & \textbf{0.906*} & \cellcolor[rgb]{ .816,  .816,  .816}\textbf{0.805**} & \cellcolor[rgb]{ .816,  .816,  .816}\textbf{0.769**} & \cellcolor[rgb]{ .816,  .816,  .816}\textbf{0.761**} & \cellcolor[rgb]{ .816,  .816,  .816}\textbf{0.762**} & \cellcolor[rgb]{ .816,  .816,  .816}\textbf{0.800**} & \textbf{0.914**} \\
    \midrule
    \multicolumn{10}{c}{Nickel} \\
    \midrule
    \multicolumn{1}{l}{RW-D} & 1557  & 2913  & 3526  & 3610  & 3936  & 4101  & 4676  & 4858  & 5283 \\
    \midrule
    SSM25 & 1.077*** & 1.039*** & 1.090*** & 1.035*** & \textbf{0.922*} & \textbf{0.958} & \textbf{0.814} & \textbf{0.759} & \cellcolor[rgb]{ .816,  .816,  .816}\textbf{0.926} \\
    Top 2 & 1.075*** & 1.038*** & 1.005*** & \textbf{0.976} & \cellcolor[rgb]{ .816,  .816,  .816}\textbf{0.889*} & \textbf{0.928} & \textbf{0.798*} & \cellcolor[rgb]{ .816,  .816,  .816}\textbf{0.685} & \textbf{0.961} \\
    \midrule
    \multicolumn{10}{c}{Zinc} \\
    \midrule
    \multicolumn{1}{l}{RW-D} & 154   & 298   & 449   & 548   & 633   & 714   & 783   & 818   & 850 \\
    \midrule
    SSM25 & \textbf{0.989} & \textbf{0.988} & \textbf{0.981} & \textbf{0.909**} & \cellcolor[rgb]{ .816,  .816,  .816}\textbf{0.816***} & \cellcolor[rgb]{ .816,  .816,  .816}\textbf{0.787***} & \cellcolor[rgb]{ .816,  .816,  .816}\textbf{0.758***} & \cellcolor[rgb]{ .816,  .816,  .816}\textbf{0.779**} & \textbf{0.803*} \\
    Top 2 & \textbf{0.993} & \textbf{0.976} & \textbf{0.936*} & \cellcolor[rgb]{ .816,  .816,  .816}\textbf{0.864***} & \cellcolor[rgb]{ .816,  .816,  .816}\textbf{0.836**} & \cellcolor[rgb]{ .816,  .816,  .816}\textbf{0.811***} & \cellcolor[rgb]{ .816,  .816,  .816}\textbf{0.775***} & \cellcolor[rgb]{ .816,  .816,  .816}\textbf{0.747***} & \cellcolor[rgb]{ .816,  .816,  .816}\textbf{0.761*} \\
    \bottomrule
    \end{tabular}%
    }
 \begin{footnotesize}
      \begin{tablenotes}
        \item \parbox[t]{6in}{%
          \textbf{Note:} For the RW-D model, we report the RMSPE, while for all other models, we present the RMSPE ratio relative to the RW-D model. RMSPE refers to the real price in USD per metric ton, with February 2015 CPI as base year. A ratio below one indicates that the model outperforms the RW-D model. Asterisks denote statistical significance at the 1\% (***), 5\% (**) and 10\% (*) based on the \cite{diebold1995comparing} test. Forecast combinations that outperform the RW-D model are highlighted in bold. Grey cells indicate forecast combinations that outperform the best individual model at the corresponding horizon.
        }
      \end{tablenotes}
    \end{footnotesize}
  \end{threeparttable}
  \label{tab:forecast_pooling}%
\end{table}

\subsection{Performance over time}
 
\begin{figure}[!ht]
\centering
\setlength{\tabcolsep}{.008\textwidth}
\caption*{}
\caption{Cumulative RMSPE ratios for different models (colored lines) and metals (panel) at horizon 6.}
\begin{tabular}{cc}
\begin{footnotesize} {Copper}\end{footnotesize} &\begin{footnotesize} {Aluminum}\end{footnotesize}\\
\includegraphics[trim=0cm 0cm 0cm 0cm, clip, width=0.48\textwidth]{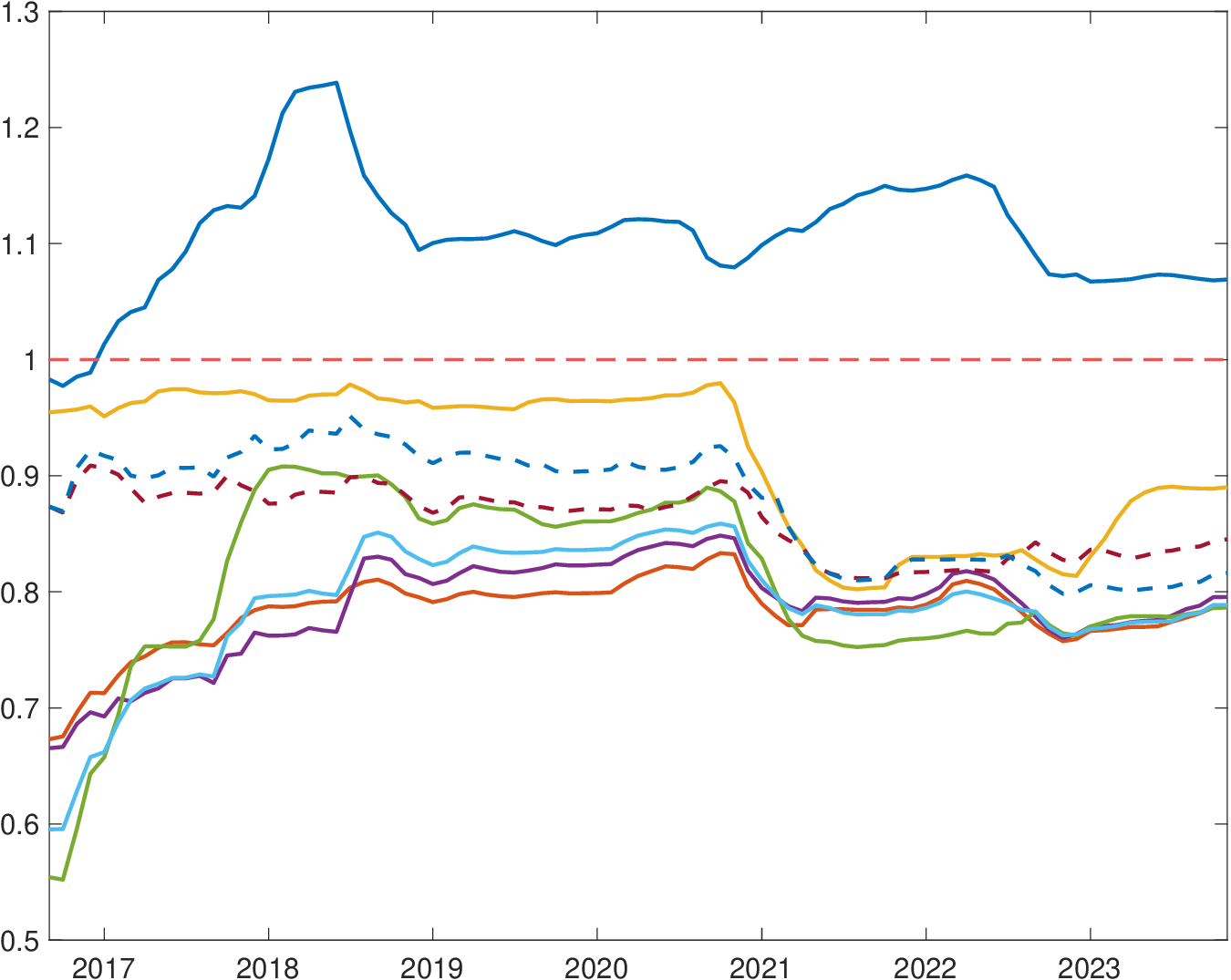} &
\includegraphics[trim=0cm 0cm 0cm 0cm, clip, width=0.48\textwidth]{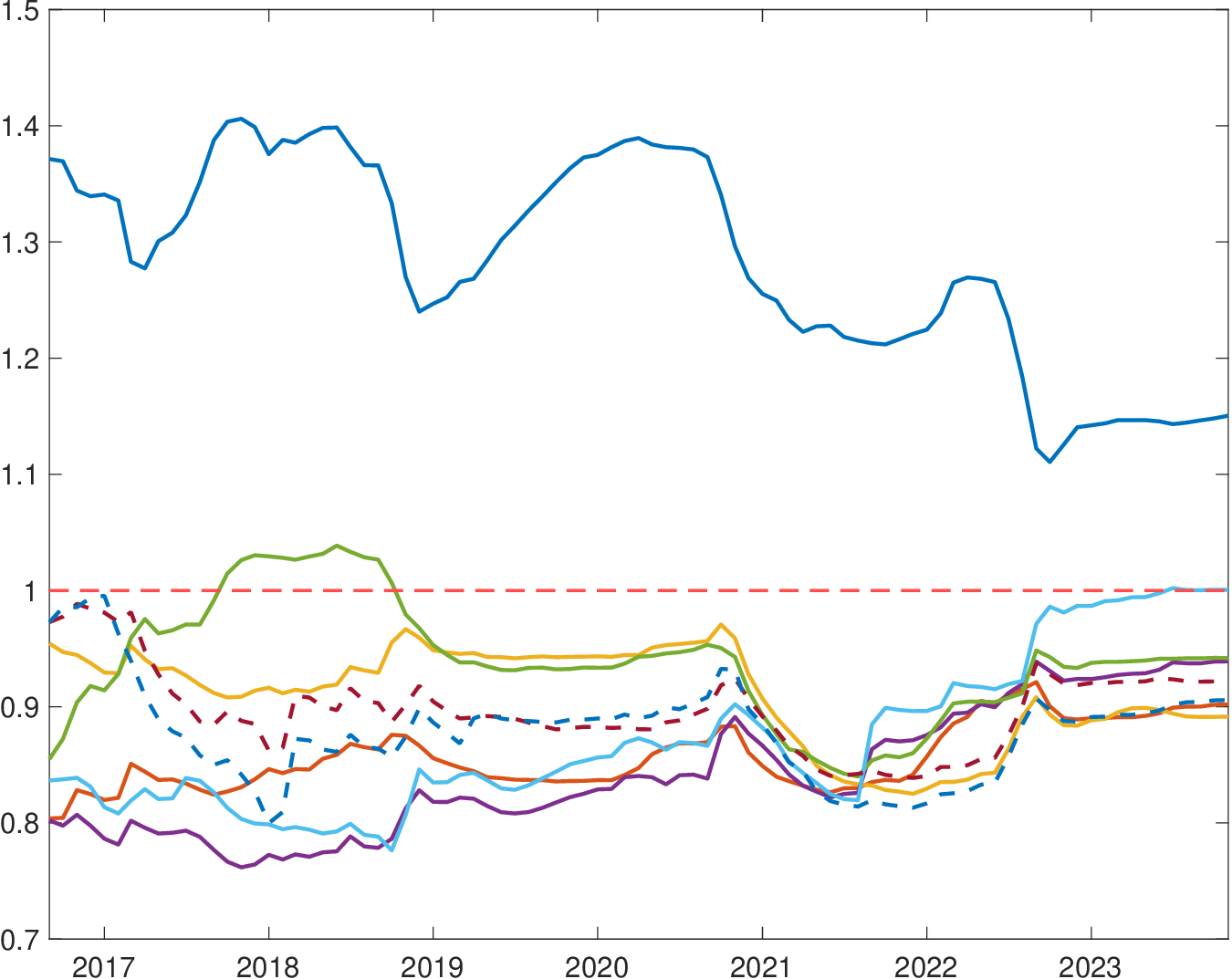}\\
\end{tabular}
\begin{tabular}{cc}\begin{footnotesize}\centering{ {Nickel}}\end{footnotesize} &\begin{footnotesize} {Zinc}\end{footnotesize}\\
\includegraphics[trim=0cm 0cm 0cm 0cm, clip, width=0.48\textwidth]{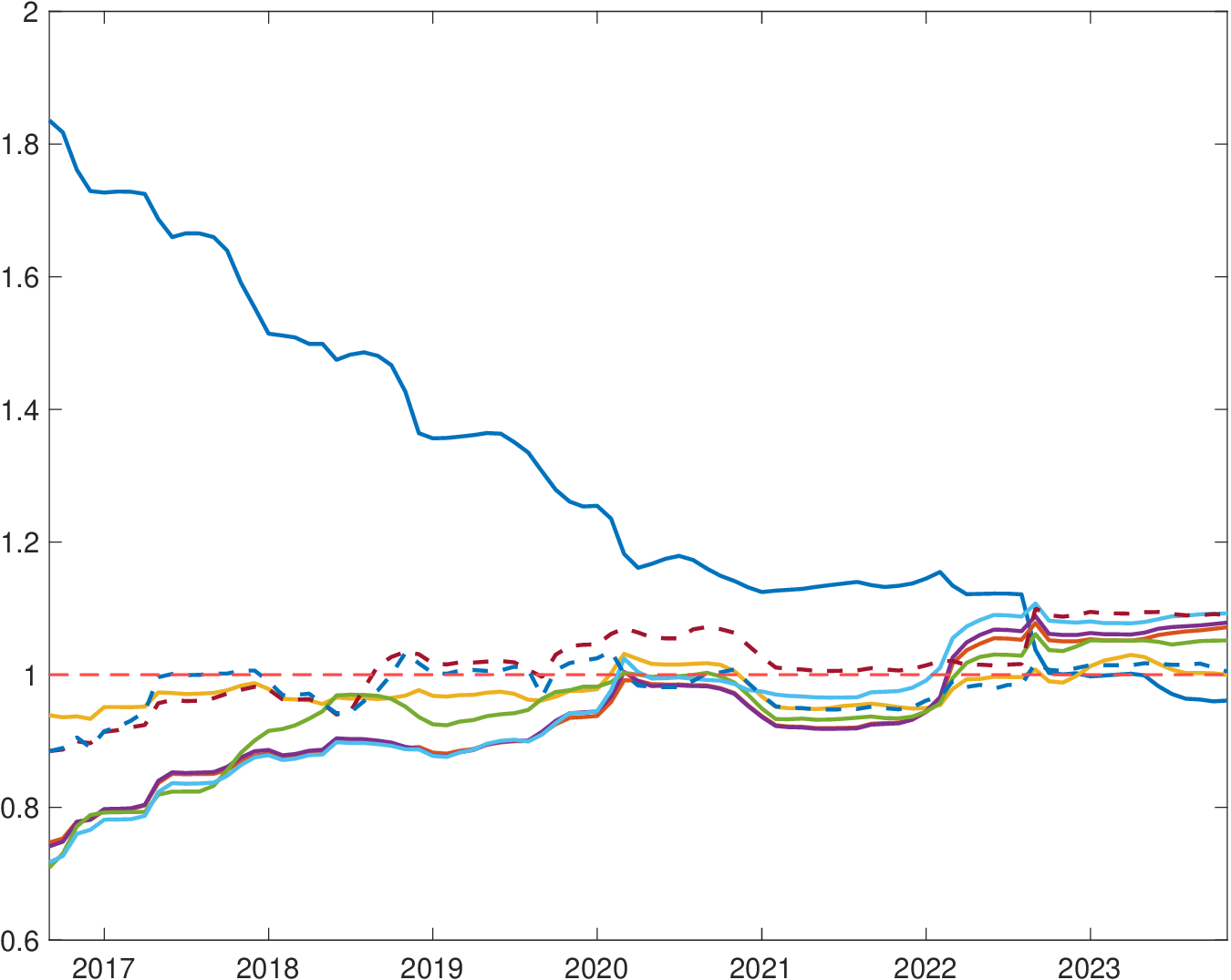} &
\includegraphics[trim=0cm 0cm 0cm 0cm, clip, width=0.48\textwidth]{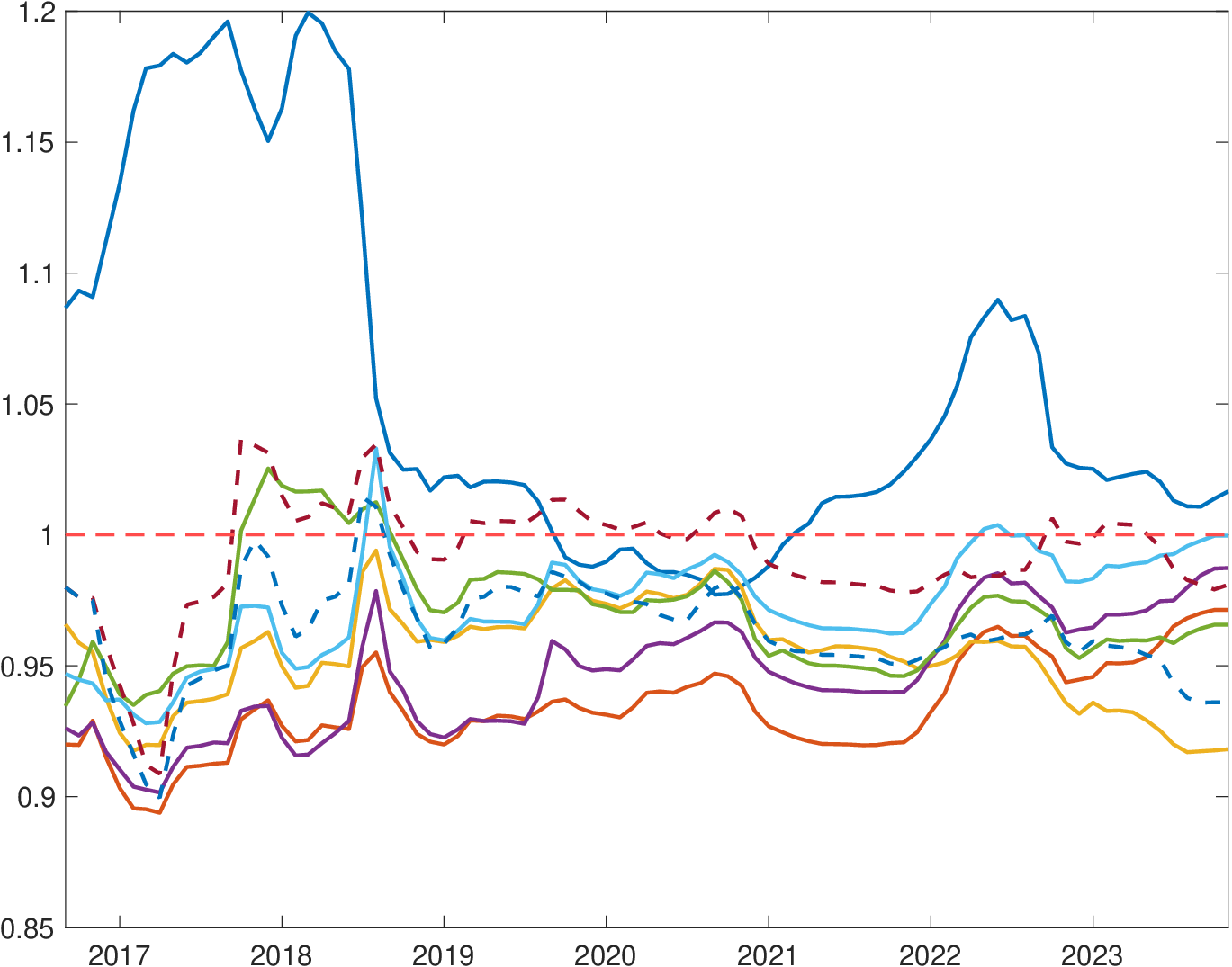}\\
\end{tabular}
\begin{tabular}{c}
\includegraphics[trim=2cm 2cm 2cm 2cm, clip, width=1\textwidth]{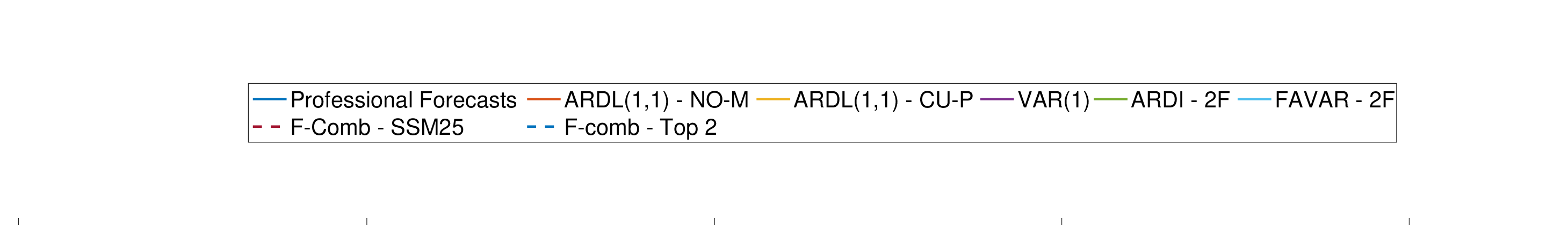}
\end{tabular}

\label{fig:RMSPE_cum_6month}
\end{figure}

\begin{figure}[!ht]
\centering
\setlength{\tabcolsep}{.008\textwidth}
\caption*{}
\caption{Cumulative RMSPE ratios for different models (colored lines) and metals (panel) at horizon 12.}
\begin{tabular}{cc}
\begin{footnotesize} {Copper}\end{footnotesize} &\begin{footnotesize} {Aluminum}\end{footnotesize}\\
\includegraphics[trim=0cm 0cm 0cm 0cm, clip, width=0.48\textwidth]{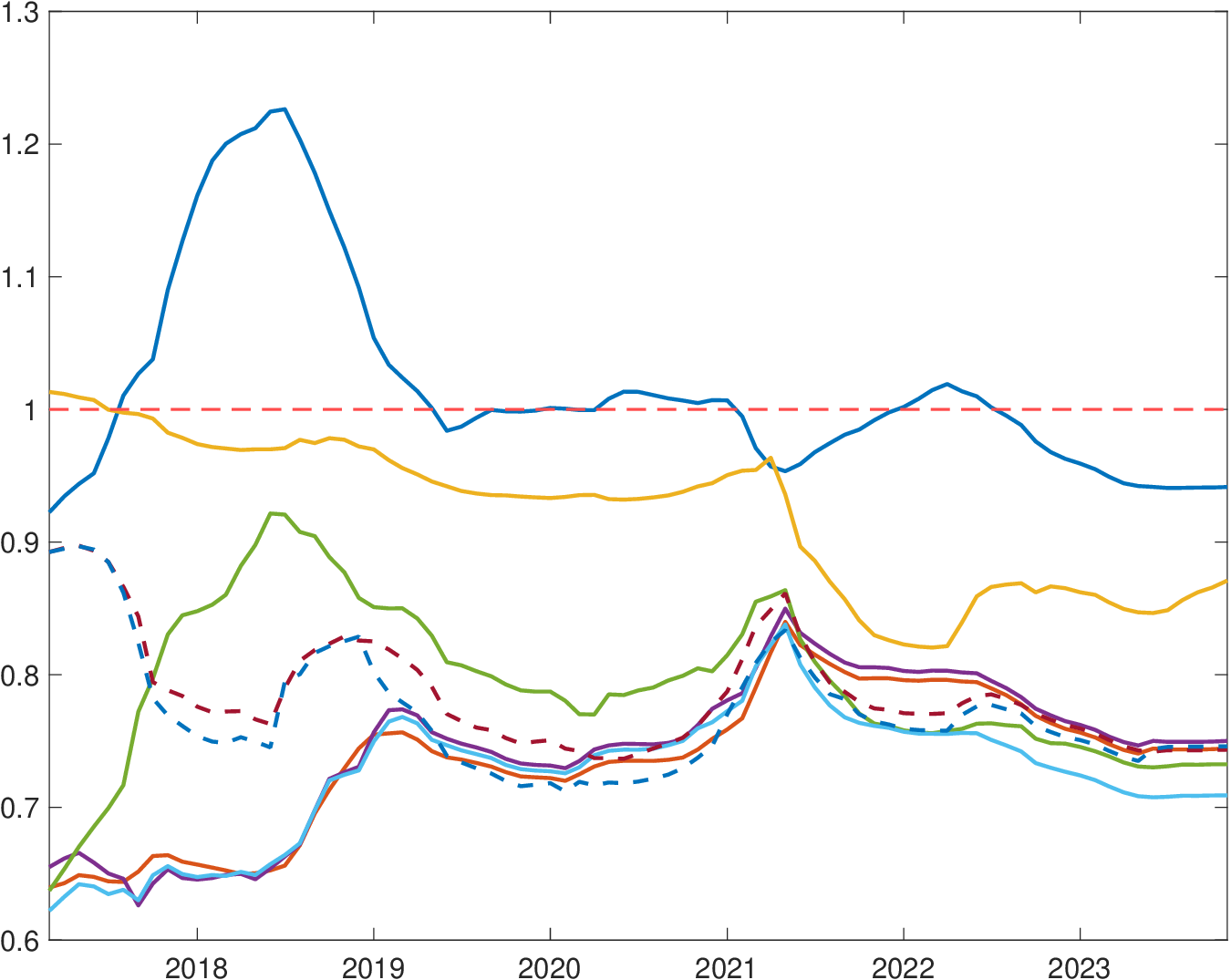} &
\includegraphics[trim=0cm 0cm 0cm 0cm, clip, width=0.48\textwidth]{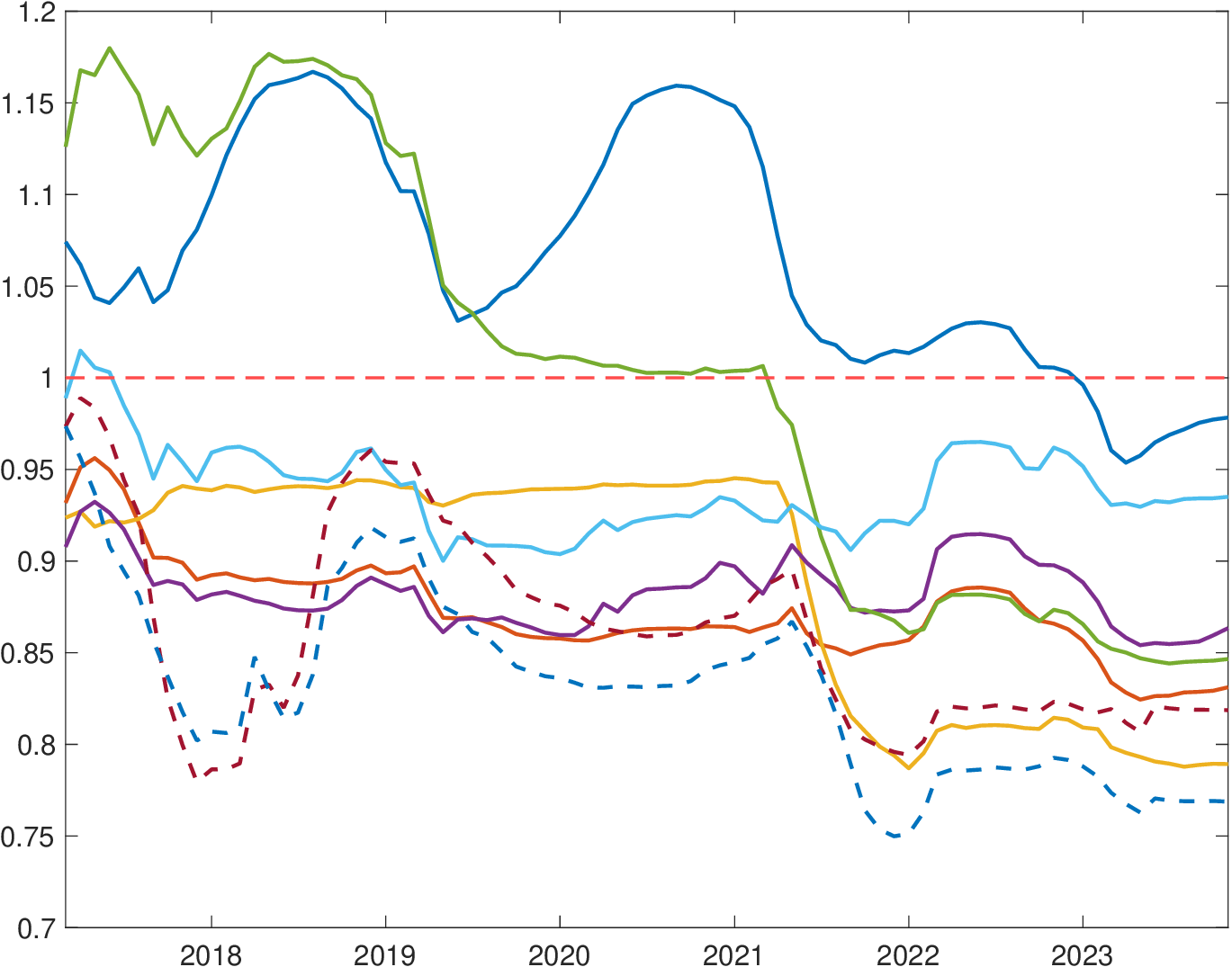}\\
\end{tabular}
\begin{tabular}{cc}\begin{footnotesize}\centering{ {Nickel}}\end{footnotesize} &\begin{footnotesize} {Zinc}\end{footnotesize}\\
\includegraphics[trim=0cm 0cm 0cm 0cm, clip, width=0.48\textwidth]{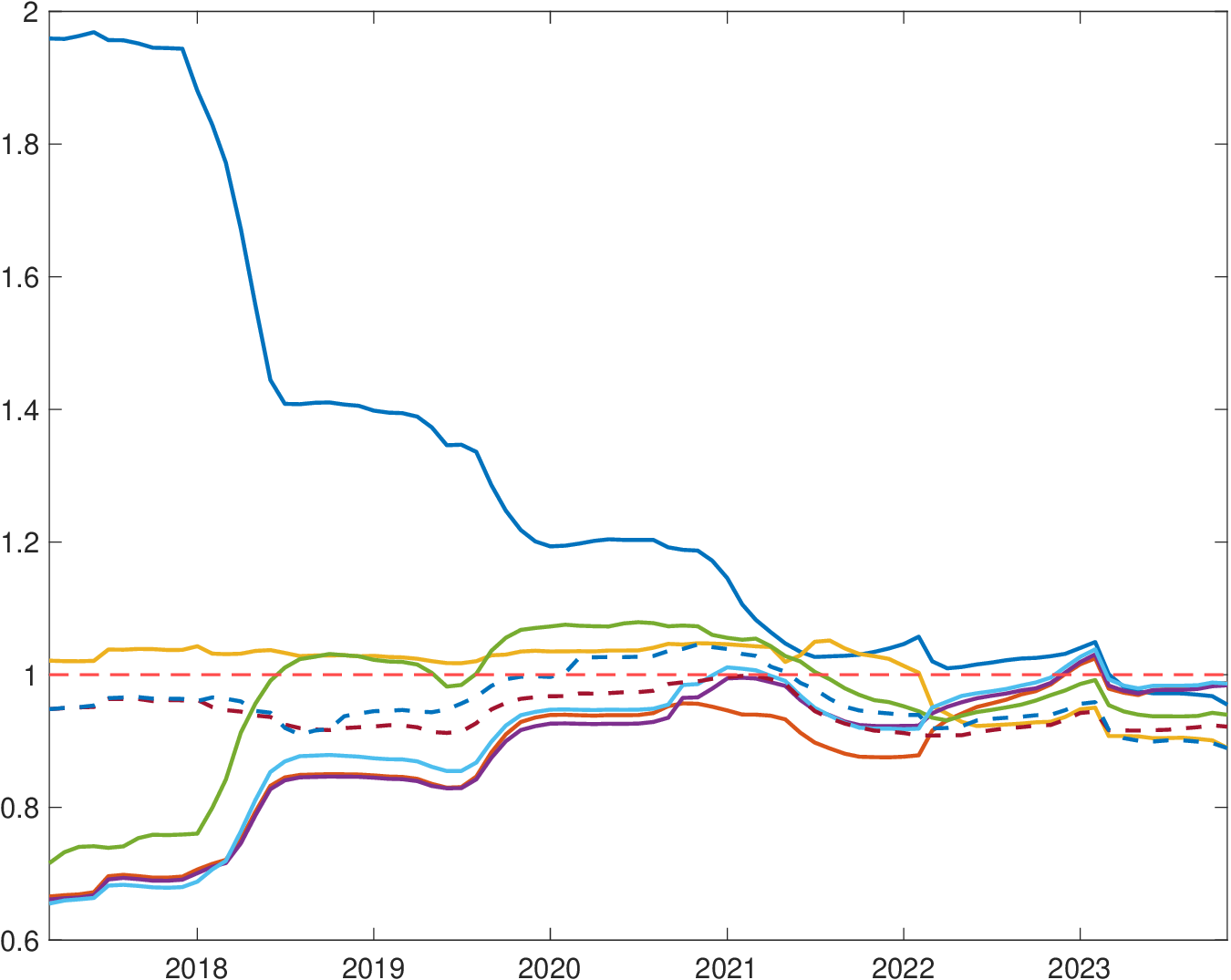} &
\includegraphics[trim=0cm 0cm 0cm 0cm, clip, width=0.48\textwidth]{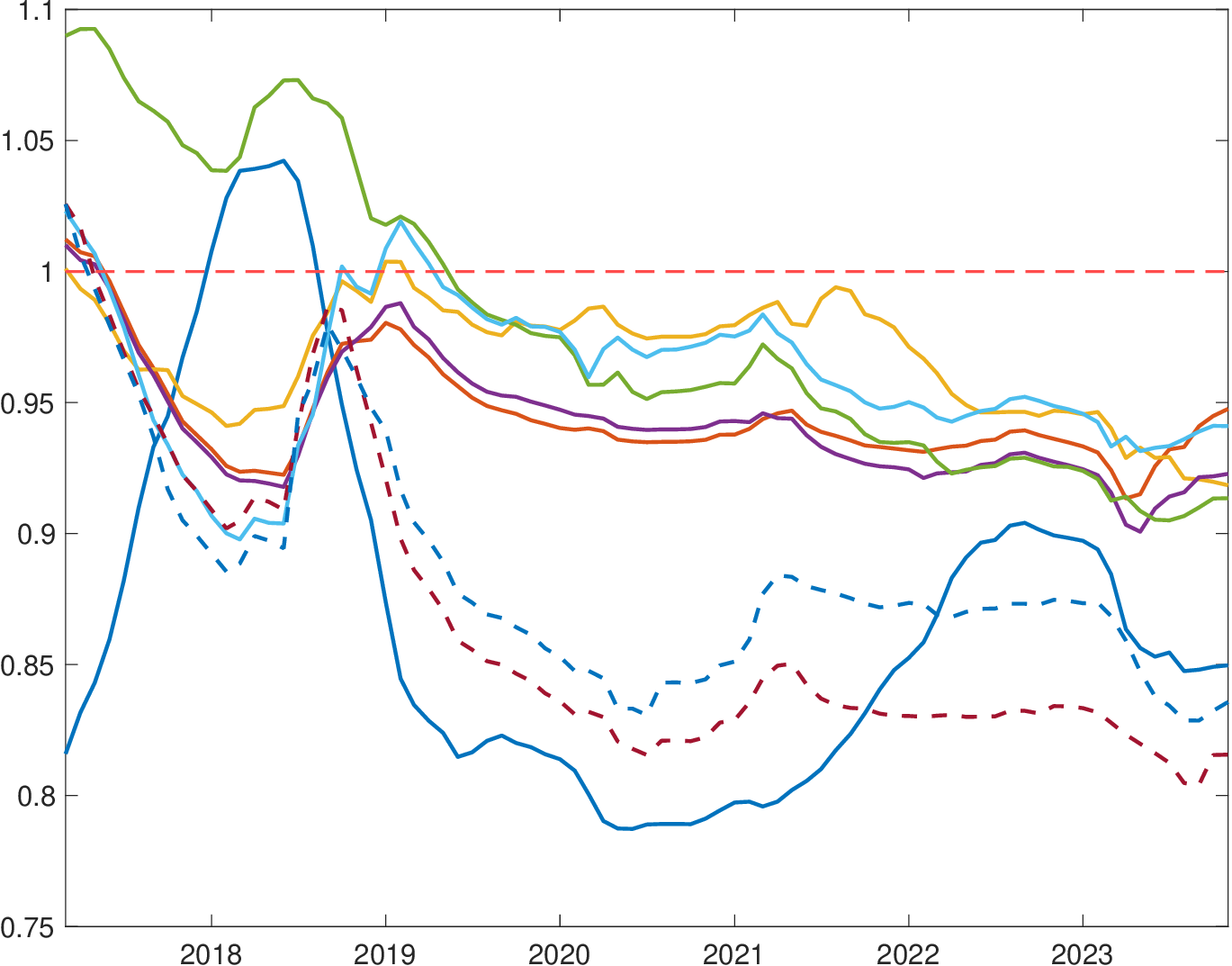}\\
\end{tabular}
\begin{tabular}{c}
\includegraphics[trim=2cm 2cm 2cm 2cm, clip, width=1\textwidth]{Figures/ratios_cum_legend.eps}
\end{tabular}

\label{fig:RMSPE_cum_12month}
\end{figure}

\begin{figure}[!ht]
\centering
\setlength{\tabcolsep}{.008\textwidth}
\caption*{}
\caption{Cumulative RMSPE ratios for different models (colored lines) and metals (panel) at horizon 18.}
\begin{tabular}{cc}
\begin{footnotesize} {Copper}\end{footnotesize} &\begin{footnotesize} {Aluminum}\end{footnotesize}\\
\includegraphics[trim=0cm 0cm 0cm 0cm, clip, width=0.48\textwidth]{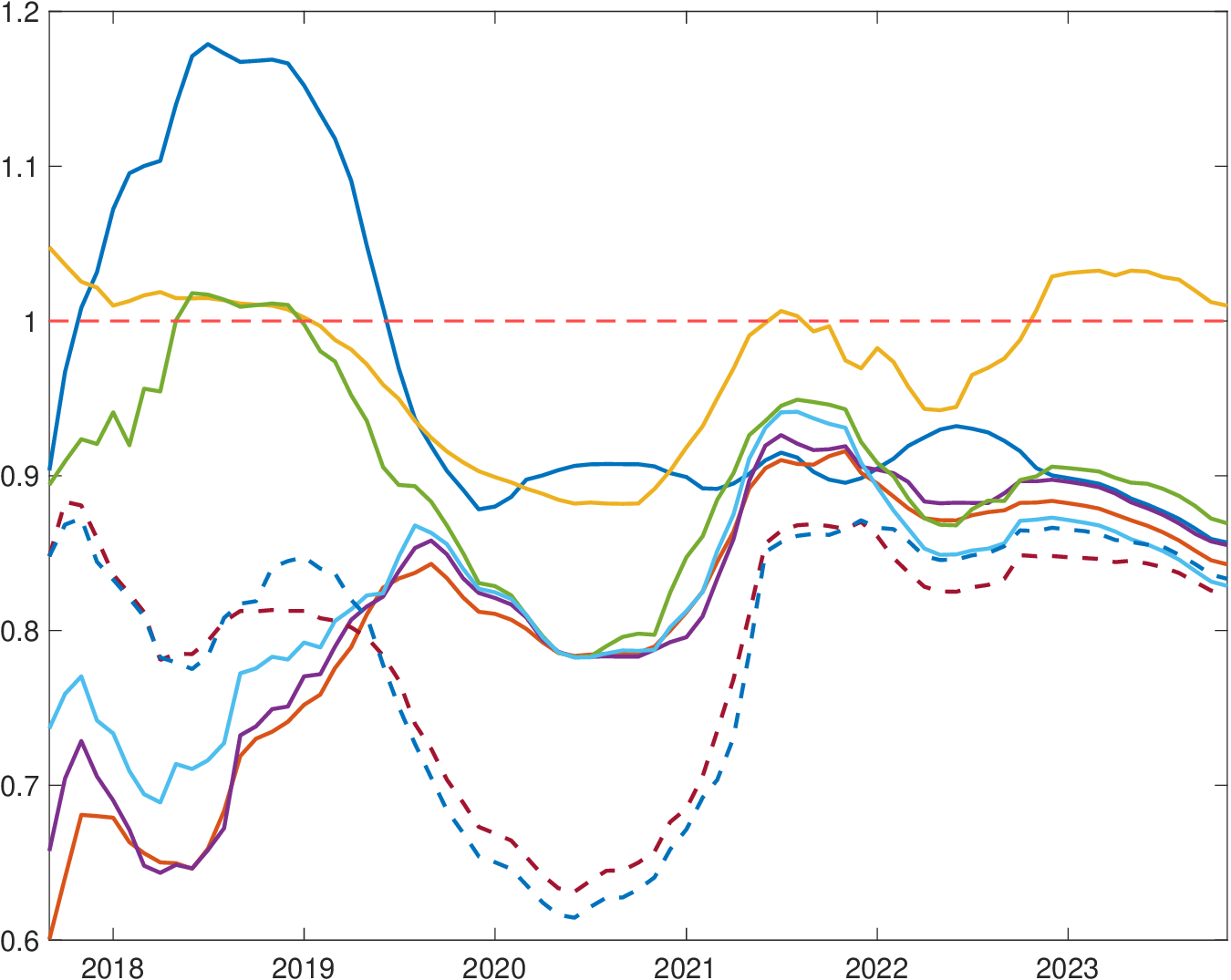} &
\includegraphics[trim=0cm 0cm 0cm 0cm, clip, width=0.48\textwidth]{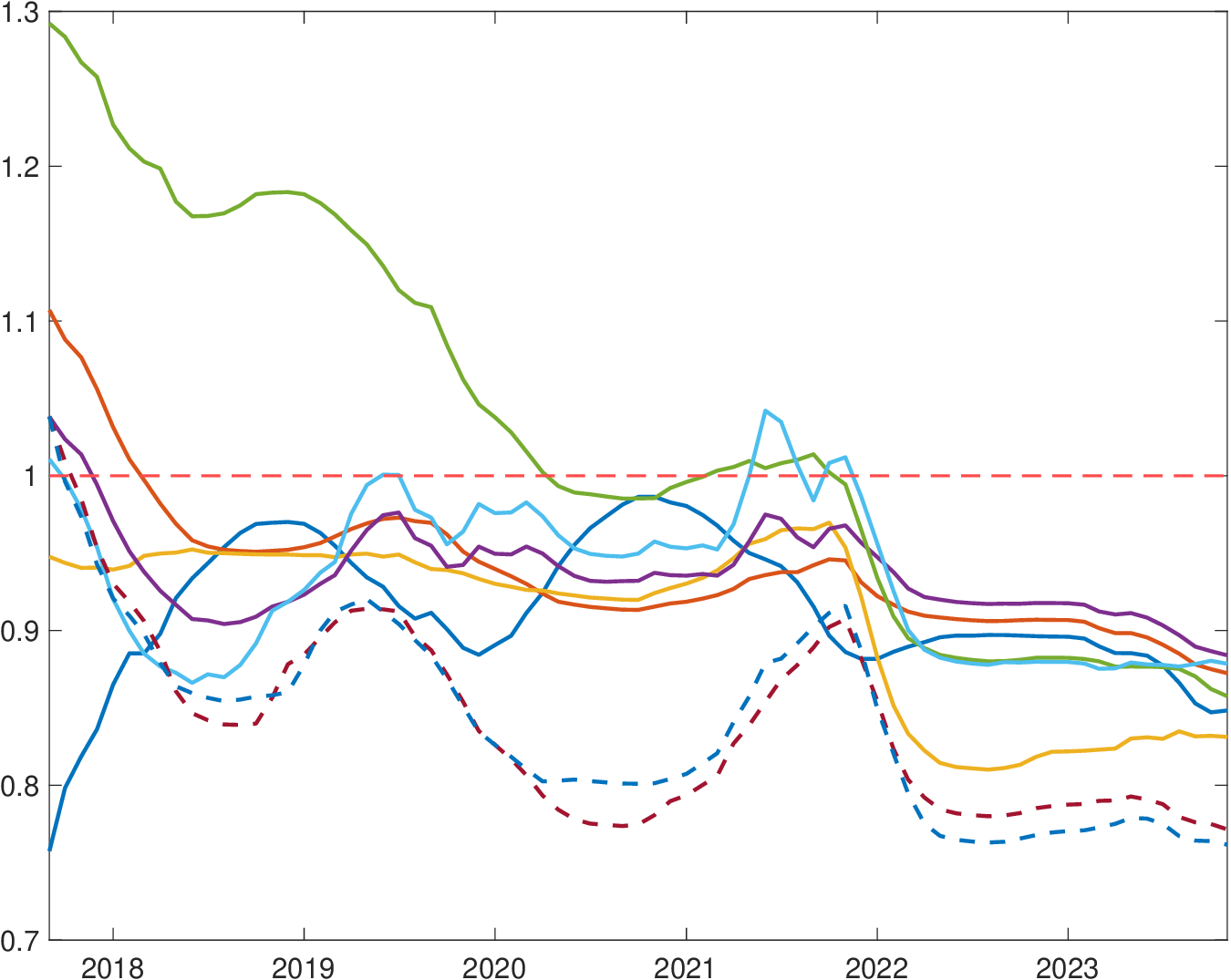}\\
\end{tabular}
\begin{tabular}{cc}\begin{footnotesize}\centering{ {Nickel}}\end{footnotesize} &\begin{footnotesize} {Zinc}\end{footnotesize}\\
\includegraphics[trim=0cm 0cm 0cm 0cm, clip, width=0.48\textwidth]{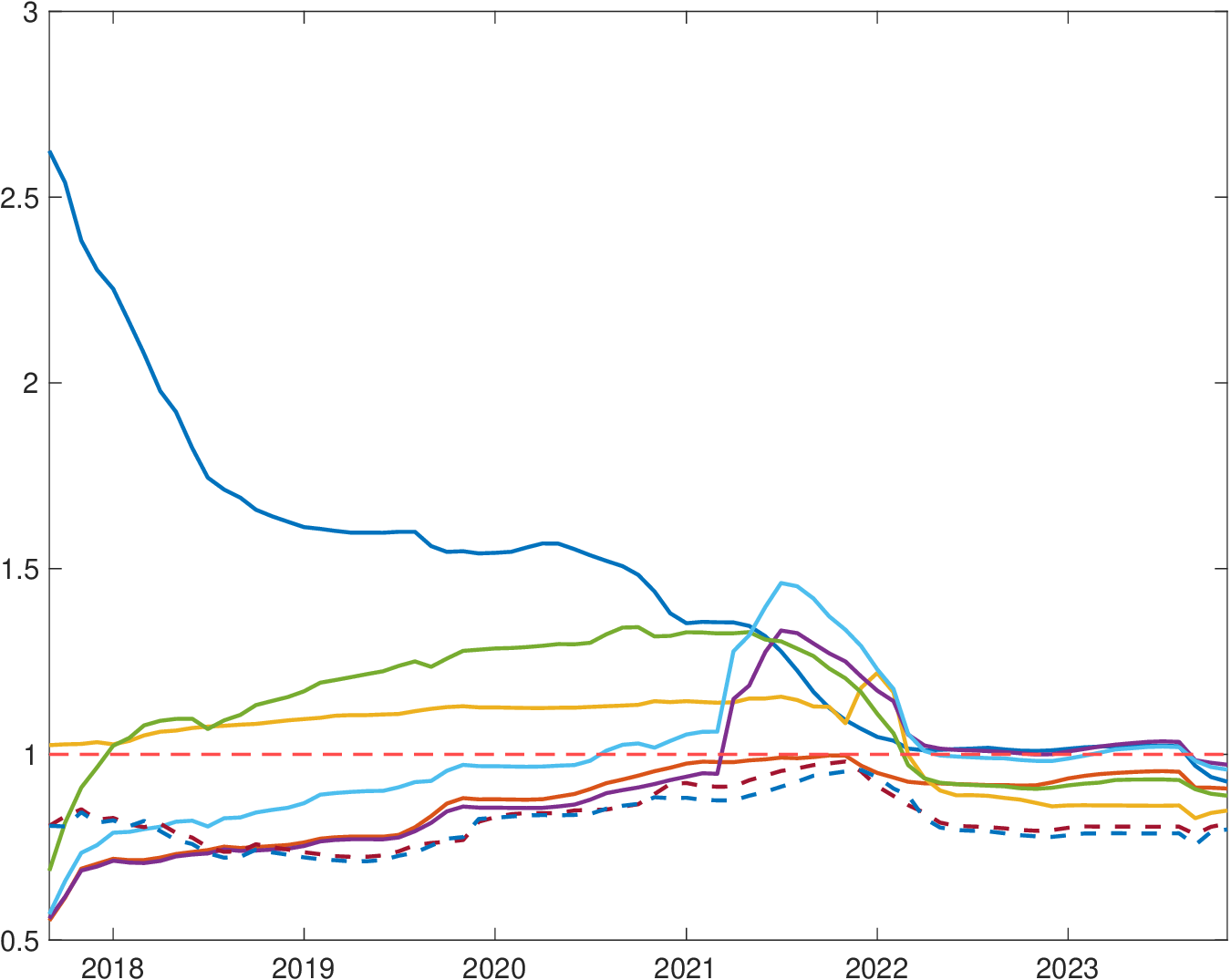} &
\includegraphics[trim=0cm 0cm 0cm 0cm, clip, width=0.48\textwidth]{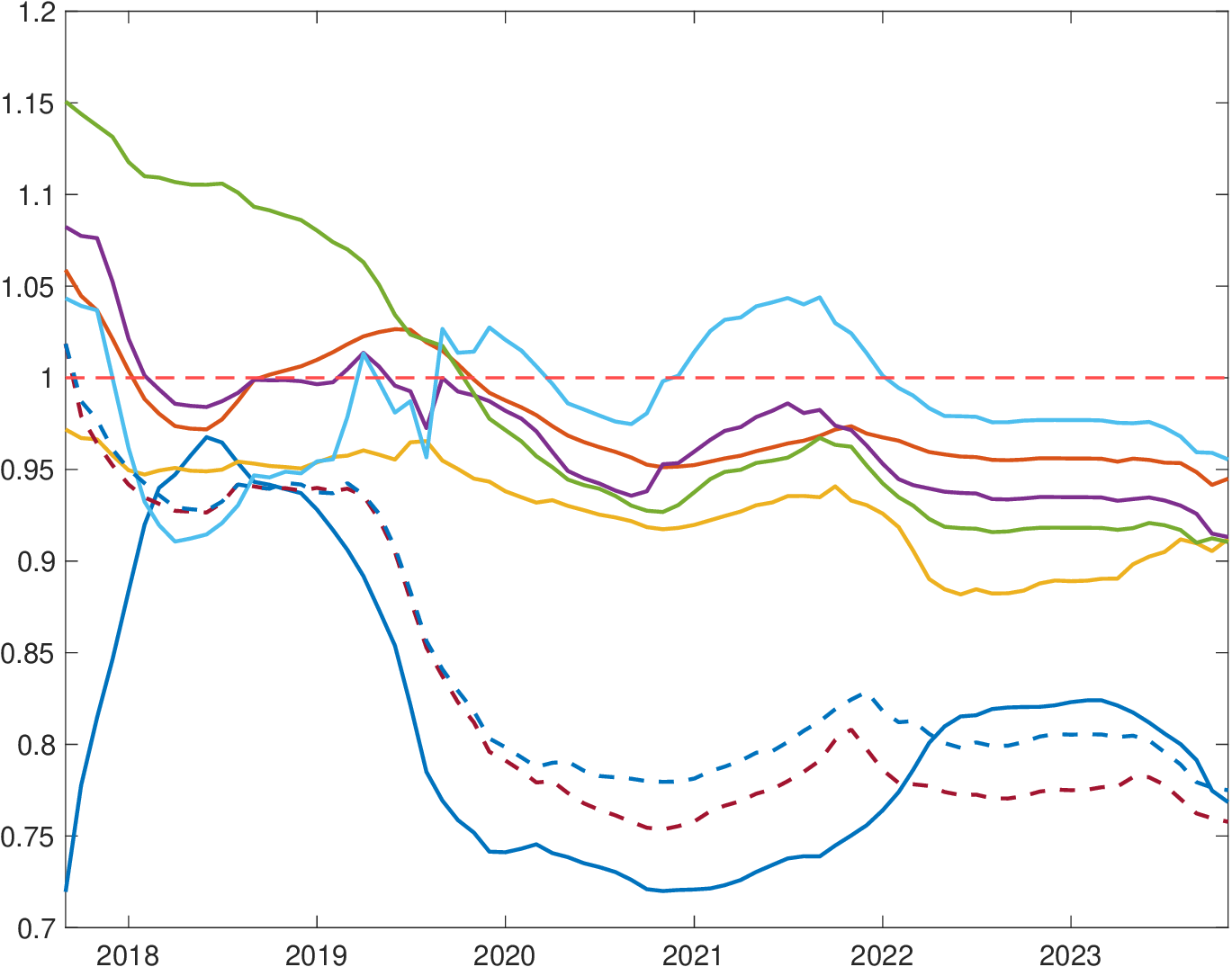}\\
\end{tabular}
\begin{tabular}{c}
\includegraphics[trim=2cm 2cm 2cm 2cm, clip, width=1\textwidth]{Figures/ratios_cum_legend.eps}
\end{tabular}

\label{fig:RMSPE_cum_18month}
\end{figure}

All forecasting statistics analyzed so far evaluate model performance by averaging squared predictive errors over the entire out-of-sample period. However, the presence of structural breaks can significantly impact model performance over time, leading different models to perform better in different subperiods. To account them, Figures \ref{fig:RMSPE_cum_6month}, \ref{fig:RMSPE_cum_12month}, and \ref{fig:RMSPE_cum_18month}\footnote{We provide evidence for the other horizons where model-free forecasts are available in the Supplement.} illustrate the time evolution of the RMSPE ratios of various models relative to the RW-D benchmark at a 6-, 12- and 18-month horizon, respectively\footnote{To allow the RMSPE ratio to stabilize, we skip the first 12 forecast periods.}. 
A downward (upward) sloping line indicates that a given model is performing better (worse) than the benchmark at that point in time, while a flat line suggests comparable performance between the model and the RW-D. The value of the ratio at any given time indicates the average forecasting performance of the model up to that point.

In the short term (6 horizons), we notice that forecasting combination does not perform the other models along time for all the metals considered. For copper, models that include new orders as a single variable or as a factor outperform the random walk and improve after the COVID-19 outbreak. 
The same happens for aluminum, where the model with capacity utilization outperforms the other models and is in line with the forecasting combination methods after the COVID-19 outbreak. F
For nickel and zinc (bottom panels), the situation is chaotic with models closed to the benchmark model across time and no big differences after 2020--2021.

Moving to the medium horizon (see Figure~\ref{fig:RMSPE_cum_12month}), we notice strong co-movements after 2019 between Factor augmented, VAR, and ARDI with new orders models. Similarly, the two forecasting combination models appear to react similarly to shocks. 
For aluminum, the forecasting combination shows important results along time appearing the best model with similarity with ARDI models with capacity utilization. 
These findings highlight the importance of including different models for those metals since they capture business cycle momentum and expected future industrial demand. 
In the bottom panel, nickel remains unpredictable and it seems that using a simple model, such as the RW-D, is providing good results. However, zinc seems to improve moderately when forecasting combination is considered, although professional forecasts seems to outperform all models.

In the long horizon (see Figure~\ref{fig:RMSPE_cum_18month}), we notice that forecasting combinations increasingly outperform individual models for all the metals. 
Indeed, for copper (top left panel), forecast pooling strategies exhibit markedly improved performance, with cumulative ratios lower than those of any individual model for most of the sample, and in particular, they capture strongly the post-COVID price surge.
Similar results are available for aluminum (top right), where  forecast pooling -- particularly when based on the two best-performing MCS models -- outperforms all other forecasting strategies.

For nickel (bottom left), no model or group of models clearly dominates across the entire evaluation period. Our models exhibit strong performance in the early part of the sample but perform worse than the benchmark thereafter, ultimately converging towards a ratio close to, but still below, one. On the contrary, professional forecasts show poor performance relative to the RW-D benchmark at the beginning of the evaluation period but improve over time -- as reflected by a declining cumulative ratio. A notable exception to these patterns is the forecast pooling strategies, which maintains a cumulative ratio below one for most of the sample.
For zinc (bottom right), at the 18-month horizon, the best-performing strategies are the forecast pooling approaches, which achieve forecast error reductions of approximately 20\% relative to individual models over the whole sample. Professional forecasters deliver forecasting performance comparable to that of forecast pooling over the entire evaluation period except the 2022--2023 period.

Overall, this time-varying forecast evaluation leads to two key takeaways. First, forecast pooling is significantly more effective at longer horizons than at shorter ones, although it still yields cumulative ratios well below one even at shorter horizons. Second, at longer horizons, forecast pooling consistently emerges as the best forecasting strategy across the entire evaluation period, with the partial exception of copper.

%====================%

\section{Conclusion}\label{sec:conclusion}
This paper develops a real-time forecasting framework for monthly real prices of key metals, addressing a gap in the forecasting literature where metal markets have received considerably less systematic attention than energy markets. By constructing a real-time database of macroeconomic predictors and evaluating a broad set of forecasting models under true real-time information constraints, the paper clarifies which predictors and model classes offer genuine forecasting gains for aluminum, copper, nickel, and zinc. Our results deliver three main messages.

First, real-time macroeconomic information substantially improves forecasting accuracy at medium-term horizons, whereas short-term dynamics remain extremely difficult to predict. The dominant predictors are demand-side indicators from the primary metals manufacturing sector--especially new orders and capacity utilisation -- which consistently outperform other macro-financial variables. This provides clear evidence that medium-term metal price dynamics reflect broader industrial activity rather than idiosyncratic noise.

Second, forecast performance is heterogeneous across metals. Copper and aluminum exhibit robust medium-term predictability; zinc displays moderate improvements driven by inventories and energy-transition variables; and nickel remains intrinsically difficult to forecast, with only modest gains from factor-augmented models. These differences highlight structural variation across metal markets that forecasting models must accommodate.

Third, while individual models based on metals-sector indicators often perform best at specific horizons, model uncertainty is pervasive. Forecast pooling, implemented via the Model Confidence Set, delivers accuracy gains for nearly all metals, particularly over longer horizons, where no single model dominates. A notable exception arises for copper, where the predictive performance of new orders is sufficiently strong that pooling may on the contrary not be necessary.

These results contribute to the forecasting literature in two ways. Methodologically, the paper demonstrates the value of constructing forecasting models using real-time data, an approach rarely applied to metal markets. Substantively, it provides evidence that medium-term metal prices are forecastable using publicly available macroeconomic indicators, offering a transparent alternative to survey and futures-based expectations, both of which perform consistently worse in our evaluation.

The analysis also highlights several limitations. The availability of metal-specific monthly indicators remains limited; the forecasting environment is strongly affected by geopolitical disruptions and structural breaks; and interdependencies across metals are only partially captured by the models considered.

\pagebreak
\onehalfspacing
\bibliography{biblio}

\end{document}

%%%%%%%%%%%%%%%%%%%%%%%%%%%%%%%%%%%%%%%%%%%%%%%%%%%%%%%%%%%%%%%%%%%%%%%%%%
% Title PAGE
%%%%%%%%%%%%%%%%%%%%%%%%%%%%%%%%%%%%%%%%%%%%%%%%%%%%%%%%%%%%%%%%%%%%%%%%%%
\setcounter{page}{0}
\pagenumbering{Alph}

\doublespacing
\begin{center}
\LARGE{\textbf{Supplement to A Real-Time Framework for Forecasting Metal Prices}}\\

\Large{%
\begin{tabular}{ccc}
Andrea Bastianin$^{a,b}$
& %
Luca Rossini$^{a,b}$
& % 
Lorenzo Tonni$^{a,\ast}$
\end{tabular}
}

\vspace{1cm}
\normalsize{\today}\\\vspace{.75cm}
\end{center}

\footnotesize{\textbf{Abstract.} Section~\ref{app::data_construction} described the details for the dataset construction. In Section~\ref{app::data} the data description with the original ID and the transformations is provided. Section~\ref{app:factors} shows the estimated factors for the model considered. In Section~\ref{app:ext_for} further forecasted results are displayed, while Section~\ref{app::model_comparison} emphasizes the results on the model confidence sets. In conclusion, Section~\ref{app:ext_cum} highlights further cumulative ratios for different horizons.

\vspace{.25cm}

\small{\noindent\textbf{Key Words:} First-Release Data; Energy Transition; Forecasting; Metals; Critical Raw materials.\\
\textbf{JEL Codes:} C32; Q02; Q41; Q43; Q48.}\\\vspace{\fill}

\footnotesize{%
\noindent $^{(a)}$ Department of Economics, Management, and Quantitative Methods, University of Milan, Milan, Italy.\\
\noindent $^{(b)}$ Fondazione Eni Enrico Mattei, Milan, Italy.\\
\noindent $^{(\ast)}$ \textit{Corresponding author}: Lorenzo Tonni, Department of Economics, Management, and Quantitative Methods, University of Milan, Via Conservatorio, 7, 20122, Milan, IT. Email: \url{lorenzo.tonni@unimi.it}.}
\vfill
\thispagestyle{empty}
\newpage
\setcounter{footnote}{0}
\pagenumbering{arabic}
\doublespacing
\normalsize

% ========= || Appendix || ========= %

\newgeometry{left=1.5cm,right=1.5cm,top=1cm,bottom=1.5cm}

%\appendix

\renewcommand{\thesection}{S.\arabic{section}}
\renewcommand{\theequation}{S.\arabic{equation}}
\renewcommand{\thefigure}{S.\arabic{figure}}
\renewcommand{\thetable}{S.\arabic{table}}
\setcounter{section}{0}
\setcounter{table}{0}
\setcounter{figure}{0}
\setcounter{equation}{0}
\setcounter{page}{1}

\section{Predictors dataset construction}
\label{app::data_construction}

Macroeconomic time series are routinely subject to revisions and typically released with a delay relative to the period they cover. As a result, the information available to real-time forecasters is affected by measurement error and by missing observations at the end of the sample. To replicate the constraints faced by real-time forecasters, we proceed as follows: (i) we collect only first-release data--that is, observations as initially published, prior to any subsequent revisions; and (ii) we nowcast the missing end-of-sample observations for each vintage.

Table \ref{tab:data_structure} illustrates the data structure for Industrial Production (IP). Rows indicate the periods being measured, while columns correspond to publication dates. Vintages differ in two aspects. First, each vintage includes one newly released observation, shown in bold along the diagonal. Because we use only first-release data, these values remain unchanged in all subsequent vintages. Second, with each new observation added to the dataset, we use the updated information set to nowcast the missing values at the end of the series.

For series reported in index form that have undergone changes in base year over time, we harmonize the base year using the following procedure: (i) log-difference each vintage, and (ii) set the first observation as the common base year for all vintages before cumulatively summing the resulting growth rates.

\begin{table}[htbp]
  \centering
  \caption{First Release dataset, Industrial Production (IP)}
    \begin{tabular}{cccccc}
    \toprule
    \toprule
      & \multicolumn{5}{c}{Vintages date} \\
    \midrule
    \multicolumn{1}{c|}{Observation date} & Jan 2012 & Feb 2012 &  Mar 2012 & Apr 2012 & … \\
    \midrule
    \multicolumn{1}{c|}{Jan 1992} & 100.00 & 100.00 & 100.00 & 100.00 & … \\
    \multicolumn{1}{c|}{Feb 1992} & 100.60 & 100.60 & 100.60 & 100.60 & … \\
    \multicolumn{1}{c|}{Mar 1992} & 100.90 & 100.90 & 100.90 & 100.90 & … \\
    \multicolumn{1}{c|}{\vdots} & \vdots & \vdots & \vdots & \vdots &  \\
    \multicolumn{1}{c|}{Sep 2011} & 155.28 & 155.28 & 155.28 & 155.28 & … \\
    \multicolumn{1}{c|}{Oct 2011} & 156.28 & 156.28 & 156.28 & 156.28 & … \\
    \multicolumn{1}{c|}{Nov 2011} & \textbf{155.94} & 155.94 & 155.94 & 155.94 & … \\
    \multicolumn{1}{c|}{Dec 2011} & Nowcast & \textbf{156.59} & 156.59 & 156.59 & … \\
    \multicolumn{1}{c|}{Jan 2012} & Nowcast & Nowcast & \textbf{156.63} & 156.63 & … \\
    \multicolumn{1}{c|}{Feb 2012} & NA & Nowcast & Nowcast & \textbf{156.66} & … \\
    \multicolumn{1}{c|}{Mar 2012} & NA & NA & Nowcast & Nowcast & … \\
    \multicolumn{1}{c|}{Apr 2012} & NA & NA & NA & Nowcast & … \\
    \multicolumn{1}{c|}{May 2012} & NA & NA & NA & NA & … \\
    \vdots & \vdots & \vdots & \vdots & \vdots & $\ddots$ \\
    \bottomrule
    \end{tabular}%
  \label{tab:data_structure}%
\end{table}%

\newpage
\section{Data description}
\label{app::data}

% ========= || macro predictors || =========== %

\begin{table}[!ht]
\caption{Predictors and Raw Materials Prices Data Description}
\begin{footnotesize}
%\begin{threeparttable}
\resizebox{7.0in}{!}{%
\begin{tabular}{  c | l | l |c | l | c | c | c}
\hline
\hline
\textbf{N} & \textbf{ID} & \textbf{Original ID}& Source &	\textbf{Series} &  \textbf{T} & \textbf{Missing obs} & \textbf{First Vintage}\\ 
\hline
\hline
\multicolumn{7}{c}{(1) \textbf{Macroeconomic Predictors}}\\
\hline
1  & IP & INDPRO & FRED   & Industrial Production & $\Delta\log(x_t)$ & 2 & Jul 2011\\ 
2  & CPI & CPIAUCSL & FRED  & Consumer Price Index for All Urban Consumers: All Items &  $\Delta^{2}\log(x_t)$ & 2  &  Jul 1972\\ 

3 & NO-M & A31SNO  & FRED  & New Orders: Primary Metals  								   &  $\log(x_t)$ & 2 &  Jul 2011\\ 
4 & NO-A & AANMNO  & FRED  & New Orders: Aluminum and Nonferrous Metal Products  								   &  $\log(x_t)$ & 2 &  Jul 2011\\

5 & HEM & AWHMAN & FRED   & Average Weekly Hours of Production and Non-supervisory Employees, Manufacturing 								   & $\log(x_t)$ & 2 &  Nov 1961\\ 
6 & MVS & A35DVS  & FRED  & Manufacturers’ Value of Shipments: Battery Manufacturing 								   &  $\log(x_t)$ & 3  & Jul 2011\\ 

7 & CU-P & CAPUTLB562A3CS & FRED & Capacity Utilization: Primary \& Semifinished Processing 								   &  $\log(x_t)$ & 2 &  Feb 2015\\ 
8 & CU-M & CAPUTLG331S & FRED & Capacity Utilization: Primary Metal   &  $\log(x_t)$ & 2  & Feb 2015\\ 

9 & PPI-M & PCU331331  & FRED  & Producer Price Indeces: Primary Metal Manufacturing  & $\Delta\log(x_t)$ & 2 & Apr 2015\\ 
10 & PPI-B & PCU335911335911  & FRED  & Producer Price Indeces: Battery Manufacturing  &  $\Delta\log(x_t)$ & 2 & Apr 2015\\ 

11 & AUS & RBAUBIS  & FRED  & Real Broad Effective Exchange Rate for Australia 		  &  $\Delta\log(x_t)$ & 2 &  Feb 2014\\ 
12 & CHL & RBCLBIS  & FRED  & Real Broad Effective Exchange Rate for Chile 		  & $\Delta\log(x_t)$ & 2 & Feb 2014\\ 
13 & CHN & RBCNBIS  & FRED  & Real Broad Effective Exchange Rate for China 		  & $\Delta\log(x_t)$ & 2 &  Feb 2014\\ 
14 & IDN & RBIDBIS  & FRED  & Real Broad Effective Exchange Rate for Indonesia 		  & $\Delta\log(x_t)$ & 2 & Feb 2014\\ 
15 & PER & RBPEBIS  & FRED  & Real Broad Effective Exchange Rate for Peru &  $\Delta\log(x_t)$ & 2 & Feb 2014\\ 
16 & PHL & RBPHBIS  & FRED  & Real Broad Effective Exchange Rate for Philippines		  &  $\Delta\log(x_t)$ & 2 & Feb 2014 \\ 
17 & RUS & RBRUBIS   & FRED & Real Broad Effective Exchange Rate for Russia 		  &  $\Delta\log(x_t)$ & 2 & Feb 2014
\\ 
18 & ALU-V & LAHWARE  & LSEG  & Aluminum Inventory Volume 		  & $\Delta\log(x_t)$ & - & - \\ 
19 & COP-V & LCPWARE  & LSEG  & Copper Inventory Volume &  $\Delta\log(x_t)$ & - & - \\ 
20 & NIC-V & LNIWARE & LSEG  & Nickel Inventory Volume &  $\Delta\log(x_t)$ & - & - \\ 
21 & ZNC-V & LZZWARE  & LSEG & Zinc Inventory Volume		  &  $\Delta\log(x_t)$ & -  & - \\

\hline
\multicolumn{7}{c}{(2) \textbf{Raw Materials Prices}}\\
\hline

1  & Aluminum & LAHCASH  & LSEG & Aluminum Spot Price & $\Delta\log(x_t)$ & - & - \\ 
2  & Copper & LCPCASH  & LSEG & Copper Spot Price &  $\Delta\log(x_t)$ & -   & - \\ 
3  & Nickel & LNICASH  & LSEG & Nickel Spot Price &  $\Delta\log(x_t)$ & -  & - \\ 
4 & Zinc & LZZCASH &  LSEG & Zinc Spot Price	 &  $\Delta\log(x_t)$ & - & - \\ 

\hline
\end{tabular}%
}

%\end{threeparttable}
\end{footnotesize}
\label{tab::predictors_detailed_description}
\end{table}

\newpage
\section{Estimated Factors}
\label{app:factors}

\normalsize{The figures below show the contribution of various predictors—grouped by category—to the first two factors as the rolling window shifts. The estimates refer to the smoothed factors extracted for the ARDI and FAVAR models used in forecasting $h = 12$. Since the price of the forecasted metal is excluded from the set of variables used for factor extraction in the ARDI specification—and the entire vector of endogenous variables is excluded in the FAVAR specification—the estimated factors differ slightly across metals and models.}

\begin{figure}[!ht]
    \centering
\begin{tabular}{cc}
 \includegraphics[width=0.5\textwidth]{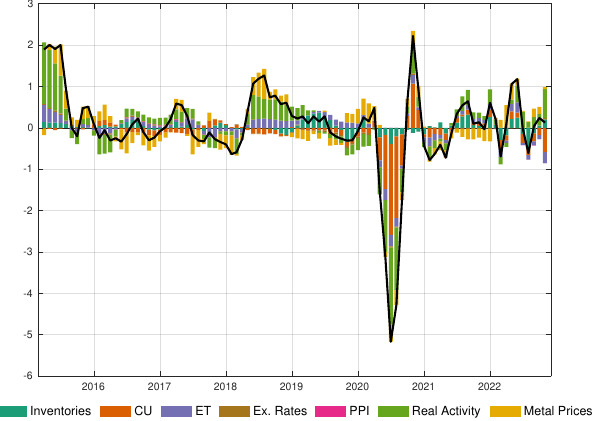}  &   \includegraphics[width=0.5\textwidth]{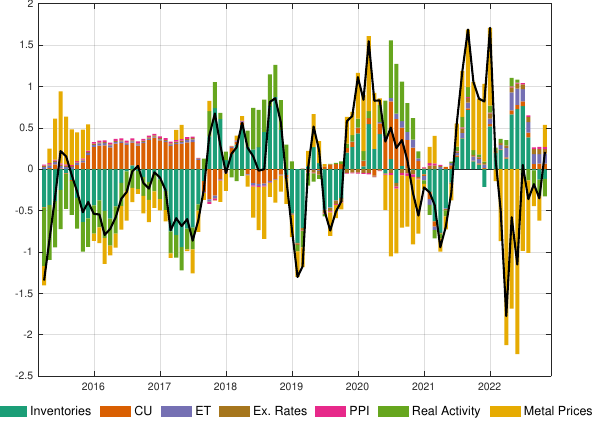}
\end{tabular}
        \caption{First (left) and second (right) estimated factor from the ARDI model with factors for the real price of Copper.}
        \label{fig:factor_ARDI_Copper_Supp}
\end{figure}

%================================%

\begin{figure}[!ht]
    \centering
\begin{tabular}{cc}
 \includegraphics[width=0.5\textwidth]{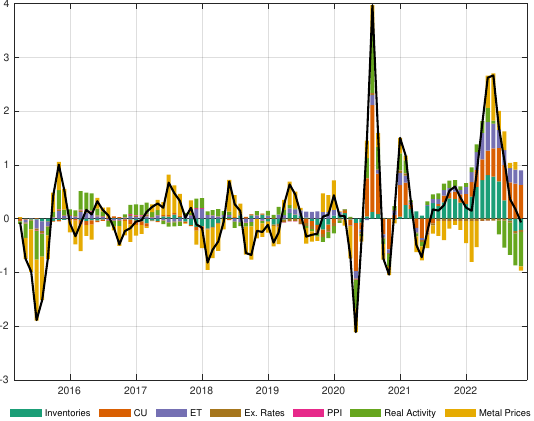}  &   \includegraphics[width=0.5\textwidth]{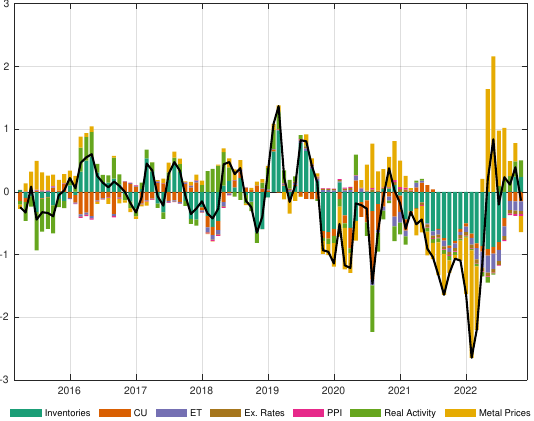}
\end{tabular}
        \caption{First (left) and second (right) estimated factor from the FAVAR model with factors for the real price of Copper.}
        \label{fig:factor_FAVAR_Copper_Supp}
\end{figure}

%================================%

\begin{figure}[!ht]
    \centering
\begin{tabular}{cc}
 \includegraphics[width=0.5\textwidth]{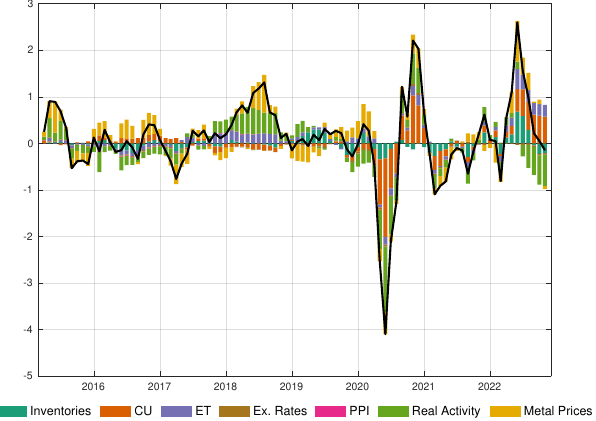}  &   \includegraphics[width=0.5\textwidth]{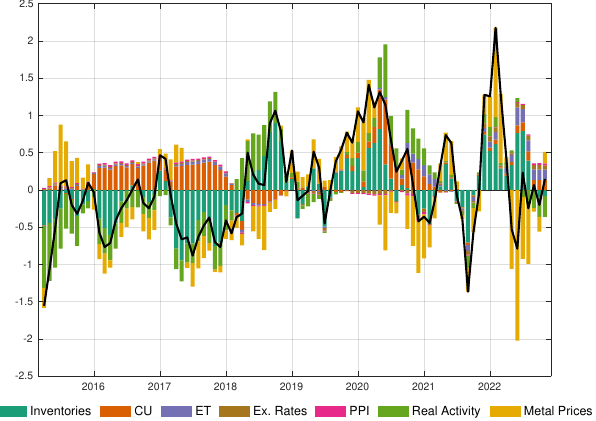}
\end{tabular}
        \caption{First (left) and second (right) estimated factor from the ARDI model with factors for the real price of Aluminum.}
        \label{fig:factor_ARDI_Aluminum_Supp}
\end{figure}

%================================%

\begin{figure}[!ht]
    \centering
\begin{tabular}{cc}
 \includegraphics[width=0.5\textwidth]{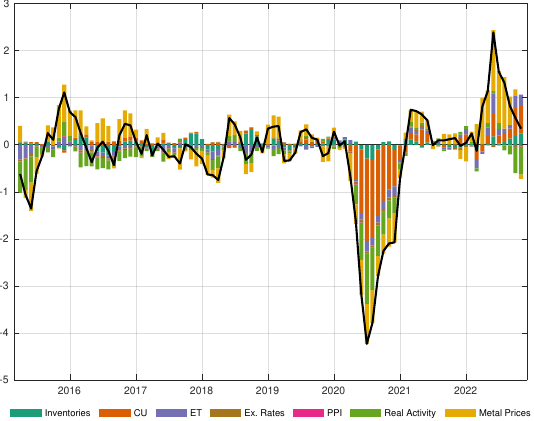}  &   \includegraphics[width=0.5\textwidth]{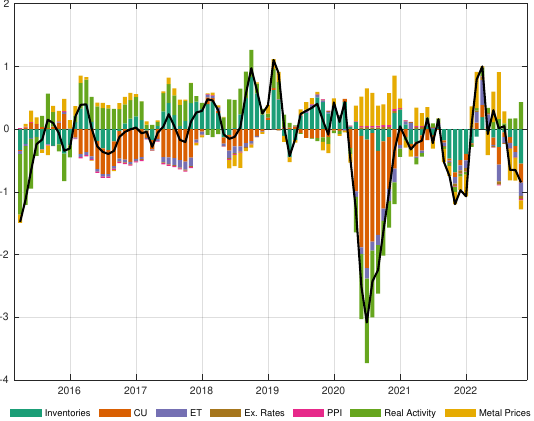}
\end{tabular}
        \caption{First (left) and second (right) estimated factor from the FAVAR model with factors for the real price of Aluminum.}
        \label{fig:factor_FAVAR_Aluminum_Supp}
\end{figure}

%================================%
\begin{figure}[!ht]
    \centering
\begin{tabular}{cc}
 \includegraphics[width=0.5\textwidth]{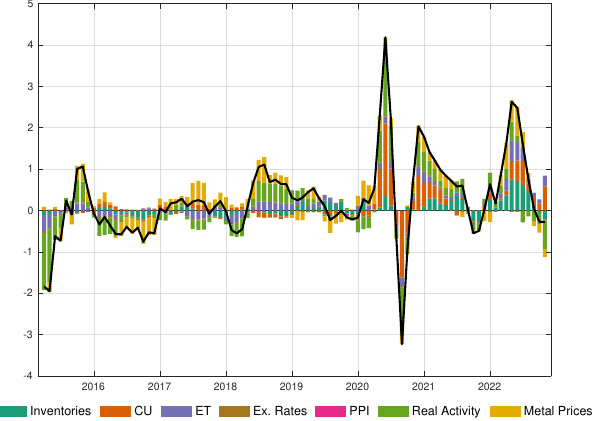}  &   \includegraphics[width=0.5\textwidth]{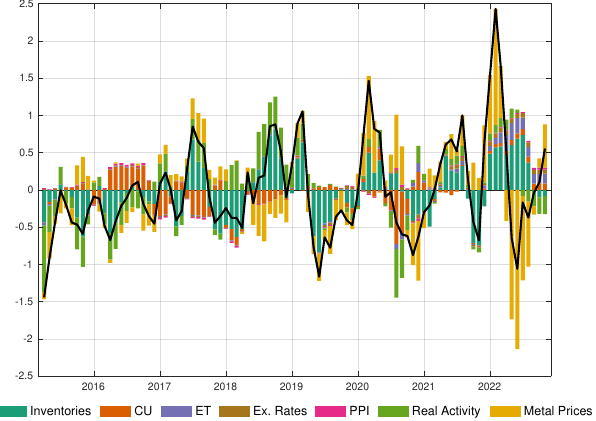}
\end{tabular}
        \caption{First (left) and second (right) estimated factor from the ARDI model with factors for the real price of Zinc.}
        \label{fig:factor_ARDI_Zinc_Supp}
\end{figure}

%================================%

\begin{figure}[!ht]
    \centering
\begin{tabular}{cc}
 \includegraphics[width=0.5\textwidth]{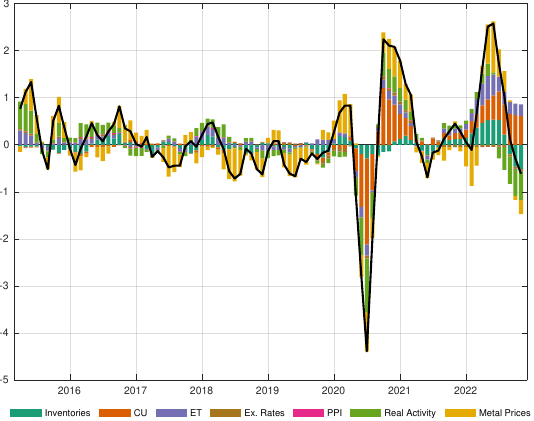}  &   \includegraphics[width=0.5\textwidth]{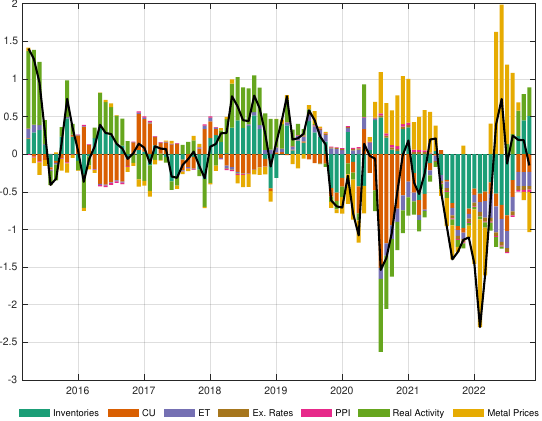}
\end{tabular}
        \caption{First (left) and second (right) estimated factor from the FAVAR model with factors for the real price of Zinc.}
        \label{fig:factor_FAVAR_Zinc_Supp}
\end{figure}

%================================%
\begin{figure}[!ht]
    \centering
\begin{tabular}{cc}
 \includegraphics[width=0.5\textwidth]{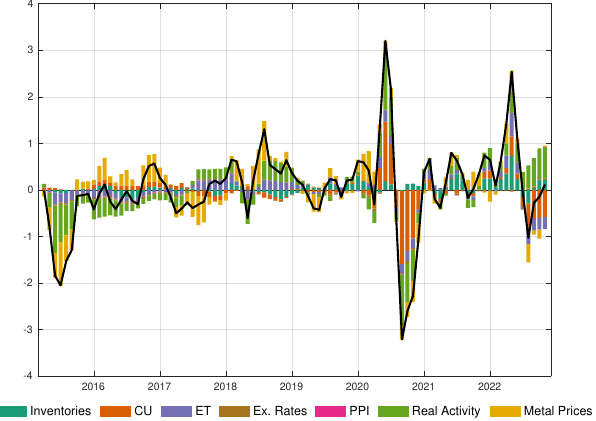}  &   \includegraphics[width=0.5\textwidth]{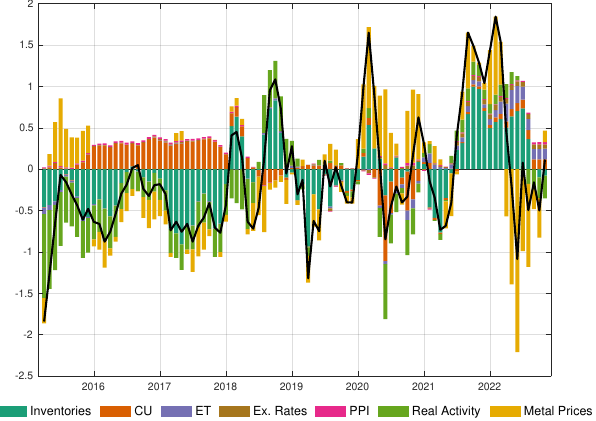}
\end{tabular}
        \caption{First (left) and second (right) estimated factor from the ARDI model with factors for the real price of Nickel.}
        \label{fig:factor_ARDI_Nickel_Supp}
\end{figure}

%================================%

\begin{figure}[!ht]
    \centering
\begin{tabular}{cc}
 \includegraphics[width=0.5\textwidth]{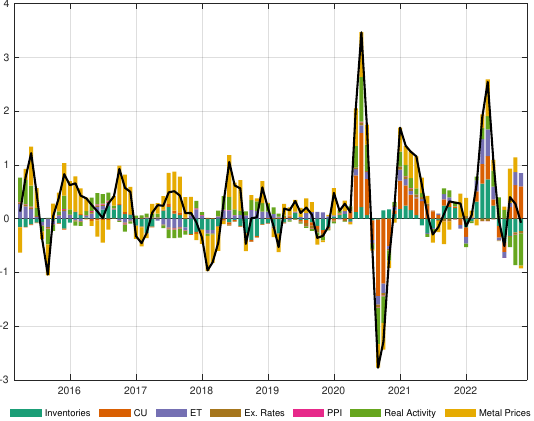}  &   \includegraphics[width=0.5\textwidth]{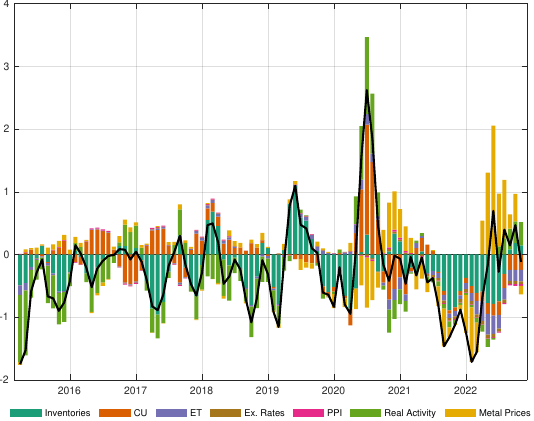}
\end{tabular}
        \caption{First (left) and second (right) estimated factor from the FAVAR model with factors for the real price of Nickel.}
        \label{fig:factor_FAVAR_Nickel_Supp}
\end{figure}

\clearpage

\newpage
\section{Extended Forecasting Exercise}
\label{app:ext_for}

\tiny{
\begin{center} 
\begin{longtable}{l|l|l|l|l|l|l|l|l|l|l}
\caption{Copper real prices point forecasts based on different models and horizons.} \label{tab:copper_supp} \\
    \toprule
    Model & Horizon 1 & Horizon 3 & Horizon 6 & Horizon 9 & Horizon 12 & Horizon 15 & Horizon 18 & Horizon 21 & Horizon 24 \\
    \endfirsthead

    \multicolumn{10}{c}{\tablename\ \thetable\ -- \textit{Continued from previous page}} \\  
    \toprule
    Model & Horizon 1 & Horizon 3 & Horizon 6 & Horizon 9 & Horizon 12 & Horizon 15 & Horizon 18 & Horizon 21 & Horizon 24 \\
    \midrule
    \endhead

    \hline \multicolumn{10}{r}{\textit{Continued on next page}} \\
    \endfoot

    \endlastfoot
    \bottomrule

    % ---- Contenuto della tabella ----
    \midrule
    \multicolumn{10}{c}{RMSPE} \\
    \midrule
%    \midrule
    RW-D  & 292.255 & 610.339 & 918.097 & 1138.998 & 1363.143 & 1527.964 & 1628.309 & 1634.843 & 1625.215 \\
    \midrule
 %   \midrule
    \multicolumn{10}{c}{RMSPE Ratios} \\
    \midrule
 %   \midrule
    \multicolumn{10}{c}{Model-Free} \\
    \midrule
    Futures & -     & 1.145*** & -     & -     & -     & 1.004*** & -     & -     & - \\
    Professional Forecasters & -     & -     & 1.069*** & 1.000*** & \textbf{0.942} & \textbf{0.898} & \textbf{0.857**} & -     & - \\
    \midrule
 %   \midrule
    \multicolumn{10}{c}{Univariate} \\
    \midrule
    AR(1) & \textbf{0.993} & \textbf{0.985} & 1.004*** & 1.031*** & 1.027*** & 1.046*** & 1.064*** & 1.077*** & 1.077*** \\
    AR(AIC) & \textbf{0.996*} & 1.028*** & 1.046*** & 1.052*** & 1.042*** & 1.047*** & 1.066*** & 1.029*** & 1.016*** \\
    \midrule
    \multicolumn{10}{c}{ARDL} \\
    \midrule
    ARDL(1,1) - IP & \textbf{0.993} & \textbf{0.985} & 1.005*** & 1.030*** & 1.023*** & 1.041*** & 1.061*** & 1.078*** & 1.087*** \\
    ARDL(AIC) - IP & \textbf{0.998**} & 1.027*** & 1.060*** & 1.202*** & 1.382*** & 1.858*** & 2.346*** & 3.807*** & 3.442*** \\
    ARDL(1,1) - NO-M & \textbf{0.973} & \textbf{0.888*} & \textbf{0.788**} & \textbf{0.742***} & \textbf{0.744***} & \textbf{0.795***} & \textbf{0.843***} & \textbf{0.923**} & \textbf{0.995} \\
    ARDL(AIC) - NO-M & \textbf{0.957} & \textbf{0.957} & \textbf{0.891} & \textbf{0.808**} & \textbf{0.773***} & \textbf{0.795***} & \textbf{0.832***} & \textbf{0.876**} & \textbf{0.933} \\
    ARDL(1,1) - HEM & \textbf{0.982} & \textbf{0.978} & \textbf{0.990*} & 1.013*** & 1.033*** & 1.112*** & 1.196*** & 1.284*** & 1.424*** \\
    ARDL(AIC) - HEM & \textbf{0.987} & 1.024*** & 1.041*** & 1.049*** & 1.071*** & 1.132*** & 1.196*** & 1.286*** & 1.426*** \\
    ARDL(1,1) - MVS & \textbf{0.977} & \textbf{0.934} & \textbf{0.935} & \textbf{0.938} & \textbf{0.939} & \textbf{0.964} & 1.016*** & 1.128*** & 1.204*** \\
    ARDL(AIC) - MVS & \textbf{0.986} & 1.025*** & 1.057*** & \textbf{0.967} & \textbf{0.934} & \textbf{0.957} & 1.018*** & 1.087*** & 1.158*** \\
    ARDL(1,1) - PPI-M & 1.001*** & \textbf{0.999**} & 1.022*** & 1.020*** & 1.007*** & 1.026*** & 1.049*** & 1.080*** & 1.109*** \\
    ARDL(AIC) - PPI-M & 1.067*** & 1.371*** & 1.303*** & 1.094*** & 1.023*** & \textbf{0.981} & \textbf{0.996*} & 1.002*** & \textbf{0.965} \\
    ARDL(1,1) - PPI-B & \textbf{0.992} & \textbf{0.984} & 1.003*** & 1.029*** & 1.023*** & 1.041*** & 1.053*** & 1.063*** & 1.065*** \\
    ARDL(AIC) - PPI-B & 1.031*** & 1.019*** & 1.075*** & 1.032*** & \textbf{0.969} & \textbf{0.990} & \textbf{0.999**} & \textbf{0.966} & 1.001*** \\
    ARDL(1,1) - CHL & \textbf{0.985} & \textbf{0.979} & 1.006*** & 1.038*** & 1.030*** & 1.046*** & 1.065*** & 1.077*** & 1.073*** \\
    ARDL(AIC) - CHL & \textbf{0.999**} & 1.028*** & 1.092*** & 1.062*** & 1.119*** & 1.084*** & 1.066*** & 1.029*** & 1.017*** \\
    ARDL(1,1) - AUS & \textbf{0.979} & \textbf{0.984} & 1.008*** & 1.044*** & 1.038*** & 1.058*** & 1.069*** & 1.077*** & 1.080*** \\
    ARDL(AIC) - AUS & \textbf{0.986} & 1.085*** & 1.058*** & 1.064*** & 1.054*** & 1.059*** & 1.071*** & 1.039*** & 1.034*** \\
    ARDL(1,1) - CHN & \textbf{0.990} & \textbf{0.984} & 1.008*** & 1.035*** & 1.031*** & 1.049*** & 1.064*** & 1.070*** & 1.065*** \\
    ARDL(AIC) - CHN & \textbf{0.997*} & 1.058*** & 1.051*** & 1.058*** & 1.049*** & 1.056*** & 1.056*** & 1.007*** & 1.000*** \\
    ARDL(1,1) - IDN & \textbf{0.993} & \textbf{0.986} & 1.004*** & 1.030*** & 1.026*** & 1.044*** & 1.059*** & 1.072*** & 1.075*** \\
    ARDL(AIC) - IDN & 1.003*** & 1.023*** & 1.043*** & 1.053*** & 1.039*** & 1.076*** & 1.077*** & 1.038*** & 1.027*** \\
    ARDL(1,1) - PER & \textbf{0.990} & \textbf{0.984} & 1.004*** & 1.032*** & 1.027*** & 1.046*** & 1.064*** & 1.078*** & 1.079*** \\
    ARDL(AIC) - PER & \textbf{0.984} & 1.027*** & 1.045*** & 1.057*** & 1.043*** & 1.047*** & 1.063*** & 1.034*** & 1.015*** \\
    ARDL(1,1) - PHL & \textbf{0.993*} & \textbf{0.987} & 1.004*** & 1.027*** & 1.025*** & 1.044*** & 1.063*** & 1.075*** & 1.074*** \\
    ARDL(AIC) - PHL & \textbf{0.999**} & 1.034*** & 1.041*** & 1.046*** & 1.041*** & 1.081*** & 1.115*** & 1.030*** & 1.013*** \\
    ARDL(1,1) - RUS & \textbf{0.993} & \textbf{0.986} & 1.005*** & 1.031*** & 1.025*** & 1.045*** & 1.064*** & 1.077*** & 1.079*** \\
    ARDL(AIC) - RUS & 1.046*** & 1.025*** & 1.041*** & 1.054*** & 1.041*** & 1.048*** & 1.070*** & 1.027*** & 1.017*** \\
    ARDL(1,1) - Aluminum & \textbf{0.998**} & \textbf{0.988} & 1.011*** & 1.058*** & 1.068*** & 1.059*** & 1.069*** & 1.059*** & 1.079*** \\
    ARDL(AIC) - Aluminum & 1.009*** & 1.040*** & 1.096*** & 1.094*** & 1.033*** & 1.034*** & 1.028*** & \textbf{0.934} & \textbf{0.951} \\
    ARDL(1,1) - Nickel & \textbf{0.999**} & 1.003*** & 1.132*** & 1.110*** & 1.085*** & 1.082*** & 1.074*** & 1.074*** & 1.074*** \\
    ARDL(AIC) - Nickel & 1.063*** & 1.217*** & 1.186*** & 1.166*** & 1.080*** & 1.078*** & 1.059*** & 1.003*** & 1.094*** \\
    ARDL(1,1) - Zinc & \textbf{0.988} & \textbf{0.990} & 1.064*** & 1.114*** & 1.094*** & 1.099*** & 1.094*** & 1.099*** & 1.092*** \\
    ARDL(AIC) - Zinc & \textbf{0.988} & 1.119*** & 1.180*** & 1.201*** & 1.168*** & 1.124*** & 1.102*** & 1.071*** & 1.046*** \\
    ARDL(1,1) - ZNC-V & \textbf{0.996**} & \textbf{0.989} & \textbf{0.997*} & 1.019*** & 1.004*** & \textbf{0.975} & \textbf{0.998**} & 1.052*** & 1.148*** \\
    ARDL(AIC) - ZNC-V & \textbf{0.997**} & 1.079*** & 1.027*** & 1.044*** & 1.017*** & \textbf{0.957} & \textbf{0.960} & 1.086*** & 1.252*** \\
    ARDL(1,1) - ALU-V & \textbf{0.960} & \textbf{0.990} & 1.085*** & 1.079*** & 1.055*** & 1.039*** & 1.046*** & 1.104*** & 1.205*** \\
    ARDL(AIC) - ALU-V & \textbf{0.971} & 1.062*** & 1.205*** & 1.207*** & \textbf{0.998**} & 1.094*** & 1.015*** & \textbf{0.973} & 1.230*** \\
    ARDL(1,1) - NIC-V & \textbf{0.999**} & \textbf{0.992} & 1.010*** & 1.054*** & 1.041*** & 1.074*** & 1.099*** & 1.107*** & 1.106*** \\
    ARDL(AIC) - NIC-V & 1.022*** & 1.033*** & 1.074*** & 1.071*** & 1.046*** & 1.099*** & 1.128*** & 1.061*** & 1.025*** \\
    ARDL(1,1) - COP-V & \textbf{0.998**} & \textbf{0.995} & \textbf{0.986} & 1.005*** & 1.006*** & 1.070*** & 1.063*** & 1.066*** & 1.059*** \\
    ARDL(AIC) - COP-V & \textbf{0.988} & 1.003*** & \textbf{0.995*} & 1.029*** & 1.005*** & 1.032*** & 1.074*** & 1.026*** & 1.073*** \\
    ARDL(1,1) - CU-M & \textbf{0.983} & \textbf{0.966} & \textbf{0.905} & \textbf{0.874} & \textbf{0.997**} & 1.065*** & 1.119*** & 1.252*** & 1.338*** \\
    ARDL(AIC) - CU-M & 1.000*** & \textbf{0.995*} & \textbf{0.859} & \textbf{0.844} & \textbf{0.954} & 1.042*** & 1.061*** & 1.120*** & 1.189*** \\
    ARDL(1,1) - CU-P & \textbf{0.980} & \textbf{0.948} & \textbf{0.890} & \textbf{0.822} & \textbf{0.871} & \textbf{0.941} & 1.010*** & 1.107*** & 1.196*** \\
    ARDL(AIC) - CU-P & \textbf{0.975} & \textbf{0.959} & \textbf{0.862*} & \textbf{0.777*} & \textbf{0.826*} & \textbf{0.904*} & \textbf{0.998**} & 1.630*** & 1.897*** \\
    ARDL(1,1) - NO-A & \textbf{0.968} & \textbf{0.914} & \textbf{0.897} & \textbf{0.907} & \textbf{0.870**} & \textbf{0.882**} & \textbf{0.912**} & \textbf{0.983} & 1.020*** \\
    ARDL(AIC) - NO-A & \textbf{0.965} & \textbf{0.982} & \textbf{0.993*} & \textbf{0.969} & \textbf{0.904} & \textbf{0.869**} & \textbf{0.882***} & \textbf{0.907*} & \textbf{0.897} \\
    ARDL(1) - Ec. Activity & 1.018*** & 1.029*** & 1.006*** & \textbf{0.941} & \textbf{0.860*} & \textbf{0.883} & \textbf{0.926} & \textbf{0.991*} & 1.058*** \\
    ARDL(1) - Ex. Rates & \textbf{0.975} & \textbf{0.980*} & 1.011*** & 1.045*** & 1.039*** & 1.052*** & 1.062*** & 1.069*** & 1.063*** \\
    ARDL(1) - ETM Prices & \textbf{0.984} & 1.011*** & 1.171*** & 1.197*** & 1.190*** & 1.141*** & 1.105*** & 1.077*** & 1.099*** \\
    ARDL(1) - ET & \textbf{0.973} & \textbf{0.926} & \textbf{0.931} & \textbf{0.933} & \textbf{0.929} & \textbf{0.950} & 1.000*** & 1.115*** & 1.199*** \\
    ARDL(1) - CU & 1.007*** & 1.068*** & 1.167*** & 1.199*** & 1.272*** & 1.280*** & 1.308*** & 1.445*** & 1.530*** \\
    ARDL(1) - Inventories & \textbf{0.976} & 1.007*** & 1.079*** & 1.106*** & 1.085*** & 1.149*** & 1.136*** & 1.167*** & 1.271*** \\
    \midrule
    \multicolumn{10}{c}{VAR} \\
    \midrule
    VAR(1) & \textbf{0.985} & \textbf{0.893*} & \textbf{0.796**} & \textbf{0.772**} & \textbf{0.750***} & \textbf{0.797***} & \textbf{0.855***} & \textbf{0.939*} & \textbf{0.997*} \\
    \midrule
    \multicolumn{10}{c}{Factor} \\
    \midrule
    ARDI(1,1) - 1 Factor & \textbf{0.942} & \textbf{0.856**} & \textbf{0.779**} & \textbf{0.717**} & \textbf{0.765**} & \textbf{0.832**} & \textbf{0.894**} & 1.004*** & 1.122*** \\
    ARDI(AIC) - 1 Factor & \textbf{0.941} & \textbf{0.884**} & \textbf{0.859**} & \textbf{0.703**} & \textbf{0.723***} & \textbf{0.774***} & \textbf{0.841***} & \textbf{0.948} & 1.048*** \\
    ARDI(1,1,1) - 2 Factors & \textbf{0.943} & \textbf{0.861**} & \textbf{0.786**} & \textbf{0.691**} & \textbf{0.733**} & \textbf{0.815**} & \textbf{0.869**} & 1.002*** & 1.152*** \\
    ARDI(AIC) - 2 Factors & \textbf{0.941} & \textbf{0.919} & \textbf{0.863*} & \textbf{0.713**} & \textbf{0.782**} & \textbf{0.817***} & \textbf{0.829***} & \textbf{0.943} & 1.102*** \\
    FAVAR(1) - 1 Factor & \textbf{0.956} & \textbf{0.859**} & \textbf{0.784**} & \textbf{0.760**} & \textbf{0.743***} & \textbf{0.780***} & \textbf{0.841***} & \textbf{0.946*} & 1.035*** \\
    FAVAR(1) - 2 Factors & \textbf{0.957} & \textbf{0.868**} & \textbf{0.789**} & \textbf{0.734**} & \textbf{0.709***} & \textbf{0.766***} & \textbf{0.829***} & \textbf{0.965} & 1.097*** \\
%    \midrule
 %   \multicolumn{1}{c}{LASSO} &       &       &       &       &       &       &       &       &  \\
 %   \midrule
 %   LASSO - 1  & \textbf{0.976} & \textbf{0.955*} & \textbf{0.939} & \textbf{0.872} & \textbf{0.830*} & \textbf{0.917} & \textbf{0.947} & 1.093*** & 1.179*** \\
 %   LASSO - 12 & \textbf{0.984} & 1.434*** & 1.367*** & 1.273*** & \textbf{0.932} & 1.018*** & 1.021*** & 1.169*** & 1.278*** \\
    \bottomrule
\end{longtable}
\begin{scriptsize}
      \begin{tablenotes}
        \item \parbox[t]{6.8in}{%
        \textbf{\textit{Notes:}} For the RW-D model, we report the RMSPE, while for all other models, we present the RMSPE ratio relative to the RW-D model. RMSPE refers to the real price in USD per metric ton, with February 2015 CPI as base year. A ratio below one (in bold) indicates that the model outperforms the RW-D model. Asterisks denote statistical significance at the 1\% (***), 5\% (**) and 10\% (*) based on the \cite{diebold1995comparing} test.
        }
      \end{tablenotes}
    \end{scriptsize}
\end{center}
}

\newpage
\tiny{
\begin{center} 
\begin{longtable}{l|l|l|l|l|l|l|l|l|l|l}
\caption{Aluminum real prices point forecasts based on different models and horizons.} \label{tab:aluminum_supp} \\
    \toprule
    Model & Horizon 1 & Horizon 3 & Horizon 6 & Horizon 9 & Horizon 12 & Horizon 15 & Horizon 18 & Horizon 21 & Horizon 24 \\
    \endfirsthead

    \multicolumn{10}{c}{\tablename\ \thetable\ -- \textit{Continued from previous page}} \\  
    \toprule
    Model & Horizon 1 & Horizon 3 & Horizon 6 & Horizon 9 & Horizon 12 & Horizon 15 & Horizon 18 & Horizon 21 & Horizon 24 \\
    \midrule
    \endhead

    \hline \multicolumn{10}{r}{\textit{Continued on next page}} \\
    \endfoot

    \endlastfoot
    \bottomrule

    % ---- Contenuto della tabella ----
    \midrule
    \multicolumn{10}{c}{RMSPE} \\
    \midrule
%    \midrule
    RW-D  & 89.9137 & 192.3754 & 276.6448 & 350.6078 & 424.5103 & 483.2490 & 525.4635 & 529.1453 & 523.4011 \\
    \midrule
%    \midrule
    \multicolumn{10}{c}{RMSPE Ratios} \\
    \midrule
%    \midrule
    \multicolumn{10}{c}{Model-Free} \\
    \midrule
    Futures & -     & 1.803*** & -     & -     & -     & 1.093*** & -     & -     & - \\
    Professional Forecasters & -     & -     & 1.151*** & 1.054*** & \textbf{0.978} & \textbf{0.918} & \textbf{0.848**} & -     & - \\
    \midrule
 %   \midrule
    \multicolumn{10}{c}{Univariate} \\
    \midrule
    AR(1) & \textbf{0.929} & 1.016*** & 1.011*** & 1.019*** & 1.023*** & 1.025*** & 1.038*** & 1.032*** & 1.022*** \\
\bottomrule
\end{longtable}
\begin{scriptsize}
      \begin{tablenotes}
        \item \parbox[t]{6.8in}{%
        See Notes in Table~\ref{tab:copper_supp} 
        }
      \end{tablenotes}
    \end{scriptsize}
\end{center}
}

\newpage
\tiny{
\begin{center} 
\begin{longtable}{l|l|l|l|l|l|l|l|l|l|l}
\caption{Zinc real prices point forecasts based on different models and horizons.} \label{tab:zinc_supp} \\
    \toprule
    Model & Horizon 1 & Horizon 3 & Horizon 6 & Horizon 9 & Horizon 12 & Horizon 15 & Horizon 18 & Horizon 21 & Horizon 24 \\
    \endfirsthead

    \multicolumn{10}{c}{\tablename\ \thetable\ -- \textit{Continued from previous page}} \\  
    \toprule
    Model & Horizon 1 & Horizon 3 & Horizon 6 & Horizon 9 & Horizon 12 & Horizon 15 & Horizon 18 & Horizon 21 & Horizon 24 \\
    \midrule
    \endhead

    \hline \multicolumn{10}{r}{\textit{Continued on next page}} \\
    \endfoot

    \endlastfoot
    \bottomrule

    % ---- Contenuto della tabella ----
    \midrule
    \multicolumn{10}{c}{RMSPE} \\
    \midrule
%    \midrule
    RW-D  & 153.583 & 297.878 & 448.554 & 548.229 & 633.019 & 714.203 & 782.877 & 817.696 & 850.193 \\
    \midrule
 %   \midrule
    \multicolumn{10}{c}{RMSPE Ratios} \\
    \midrule
 %   \midrule
    \multicolumn{10}{c}{Model-Free} \\
    \midrule
    Futures & -     & \textbf{0.973} & -     & -     & -     & \textbf{0.863} & -     & -     & - \\
    Professional Forecasters & -     & -     & 1.017*** & \textbf{0.917} & \textbf{0.850} & \textbf{0.800*} & \textbf{0.769**} & -     & - \\
    \midrule
%    \midrule
    \multicolumn{10}{c}{Univariate} \\
    \midrule
    AR(1) & 1.007*** & 1.012*** & 1.026*** & 1.040*** & 1.051*** & 1.043*** & 1.021*** & \textbf{0.977} & \textbf{0.937} \\
\bottomrule
\end{longtable}
\begin{scriptsize}
      \begin{tablenotes}
        \item \parbox[t]{6.8in}{%
        See Notes in Table~\ref{tab:copper_supp} 
        }
      \end{tablenotes}
    \end{scriptsize}
    \end{center}
}

\newpage
\tiny{
\begin{center} 
\begin{longtable}{l|l|l|l|l|l|l|l|l|l|l}
\caption{Nickel real prices point forecasts based on different models and horizons.} \label{tab:nickel_supp} \\
    \toprule
    Model & Horizon 1 & Horizon 3 & Horizon 6 & Horizon 9 & Horizon 12 & Horizon 15 & Horizon 18 & Horizon 21 & Horizon 24 \\
    \endfirsthead

    \multicolumn{10}{c}{\tablename\ \thetable\ -- \textit{Continued from previous page}} \\  
    \toprule
    Model & Horizon 1 & Horizon 3 & Horizon 6 & Horizon 9 & Horizon 12 & Horizon 15 & Horizon 18 & Horizon 21 & Horizon 24 \\
    \midrule
    \endhead

    \hline \multicolumn{10}{r}{\textit{Continued on next page}} \\
    \endfoot

    \endlastfoot
    \bottomrule

    % ---- Contenuto della tabella ----
    \midrule
    \multicolumn{10}{c}{RMSPE} \\
%    \midrule
    \midrule
    RW-D  & 1557.230 & 2913.438 & 3526.039 & 3610.095 & 3936.267 & 4100.563 & 4675.748 & 4858.198 & 5283.199 \\
%    \midrule
    \midrule
    \multicolumn{10}{c}{RMSPE Ratios} \\
%    \midrule
    \midrule
    \multicolumn{10}{c}{Model-Free} \\
    \midrule
    Futures & -     & \textbf{0.929} & -     & -     & -     & \textbf{0.960} & -     & -     & - \\
    Professional Forecasters & -     & -     & \textbf{0.961} & 1.004*** & \textbf{0.955} & \textbf{0.968} & \textbf{0.927} & -     & - \\
%    \midrule
    \midrule
    \multicolumn{10}{c}{Univariate} \\
    \midrule
    AR(1) & 1.072*** & 1.032*** & 1.139*** & 1.075*** & 1.081*** & 1.106*** & 1.105*** & 1.076*** & 1.081*** \\
\bottomrule
\end{longtable}
\begin{scriptsize}
      \begin{tablenotes}
        \item \parbox[t]{6.8in}{%
        See Notes in Table~\ref{tab:copper_supp} 
        }
      \end{tablenotes}
    \end{scriptsize}
\end{center}
}

\vspace{-1.8cm}

\newpage
\section{Extended Model Confidence Set Results}
\label{app::model_comparison}

\tiny{
\begin{center} 
\begin{longtable}{l|llllllllll}
\caption{Model Confidence Sets (p-values) for Copper real price Forecasts} \label{tab:copper_MCS_supp} \\
    \toprule
    Model & Horizon 1 & Horizon 3 & Horizon 6 & Horizon 9 & Horizon 12 & Horizon 15 & Horizon 18 & Horizon 21 & Horizon 24 \\
    \endfirsthead

    \multicolumn{10}{c}{\tablename\ \thetable\ -- \textit{Continued from previous page}} \\  
    \toprule
    Model & Horizon 1 & Horizon 3 & Horizon 6 & Horizon 9 & Horizon 12 & Horizon 15 & Horizon 18 & Horizon 21 & Horizon 24 \\
    \midrule
    \endhead

    \hline \multicolumn{10}{r}{\textit{Continued on next page}} \\
    \endfoot

    \endlastfoot
    \bottomrule

    % ---- Contenuto della tabella ----
\midrule
    \multicolumn{10}{c}{Model-Free} \\
    \midrule
    Futures & -     & 0.044 & -     & -     & -     & 0.028 & -     & -     & - \\
    Professional Forecasts & -     & -     & 0.076 & 0.095 & 0.083 & 0.377** & \color{red}{0.952}** & -     & - \\
    \midrule
    \multicolumn{10}{c}{Univariate} \\
    \midrule
    RW    & 0.130* & 0.051 & 0.076 & 0.065 & 0.027 & 0.039 & 0.062 & 0.578** & \color{red}{0.706}** \\
    AR(1) & 0.083 & 0.051 & 0.076 & 0.065 & 0.027 & 0.028 & 0.055 & 0.123* & 0.113* \\
    AR(AIC) & 0.096 & 0.051 & 0.076 & 0.065 & 0.027 & 0.028 & 0.055 & 0.396** & 0.461** \\
    \midrule
    \multicolumn{10}{c}{ARDL} \\
\end{longtable}
 \begin{scriptsize}
      \begin{tablenotes}
        \item \parbox[t]{6in}{%
          \textbf{\textit{Notes:}} MCS $p$-values computed with 10.000 block bootstrap replications using a block size of $6$. Asterisks indicate models included in the Set of Superior Models (SSM) at $\alpha = 0.25$ (**) and at $\alpha = 0.1$ (*) with the MCS procedure. Red indicates the top $5$ according to the MCS ranking.
        }
      \end{tablenotes}
    \end{scriptsize}
\end{center}

% Table generated by Excel2LaTeX from sheet 'MCS - New'
\tiny{
\begin{center} 
\begin{longtable}{l|llllllllll}
\caption{Model Confidence Sets (p-values) for Aluminum real price Forecasts} \label{tab:aluminum_MCS_supp} \\
    \toprule
    Model & Horizon 1 & Horizon 3 & Horizon 6 & Horizon 9 & Horizon 12 & Horizon 15 & Horizon 18 & Horizon 21 & Horizon 24 \\
    \endfirsthead

    \multicolumn{10}{c}{\tablename\ \thetable\ -- \textit{Continued from previous page}} \\  
    \toprule
    Model & Horizon 1 & Horizon 3 & Horizon 6 & Horizon 9 & Horizon 12 & Horizon 15 & Horizon 18 & Horizon 21 & Horizon 24 \\
    \midrule
    \endhead

    \hline \multicolumn{10}{r}{\textit{Continued on next page}} \\
    \endfoot

    \endlastfoot
    \midrule
    \multicolumn{10}{c}{Model-Free} \\
    \midrule
    Futures & -     & 0.074 & -     & -     & -     & 0.406** & -     & -     & - \\
    Professional Forecasts & -     & -     & 0.095 & 0.116* & 0.153* & 0.496** & \color{red}{0.989}** & -     & - \\
    \midrule
    \multicolumn{10}{c}{Univariate} \\
    \midrule
    RW    & 0.166* & \color{red}{0.388}** & 0.288** & 0.175* & 0.091 & 0.114* & 0.138* & 0.485** & 0.546** \\
    AR(1) & 0.536** & 0.384** & 0.275** & 0.153* & 0.084 & 0.102* & 0.109* & 0.119* & 0.172* \\
    AR(AIC) & 0.239* & 0.171* & 0.042 & 0.080 & 0.063 & 0.135* & 0.109* & 0.657** & 0.708** \\
    \midrule
    \multicolumn{10}{c}{ARDL} \\
\bottomrule
\end{longtable}
 \begin{scriptsize}
      \begin{tablenotes}
        \item \parbox[t]{6in}{%
          See Notes in Table~\ref{tab:copper_MCS_supp}
        }
      \end{tablenotes}
    \end{scriptsize}
\end{center}
}

% Table generated by Excel2LaTeX from sheet 'MCS - New'
\tiny{
\begin{center} 
\begin{longtable}{l|llllllllll}
\caption{Model Confidence Sets (p-values) for Zinc real price Forecasts} \label{tab:nickel_MCS_supp} \\
    \toprule
    Model & Horizon 1 & Horizon 3 & Horizon 6 & Horizon 9 & Horizon 12 & Horizon 15 & Horizon 18 & Horizon 21 & Horizon 24 \\
    \endfirsthead

    \multicolumn{10}{c}{\tablename\ \thetable\ -- \textit{Continued from previous page}} \\  
    \toprule
    Model & Horizon 1 & Horizon 3 & Horizon 6 & Horizon 9 & Horizon 12 & Horizon 15 & Horizon 18 & Horizon 21 & Horizon 24 \\
    \midrule
    \endhead

    \hline \multicolumn{10}{r}{\textit{Continued on next page}} \\
    \endfoot

    \endlastfoot
    \midrule
    \multicolumn{10}{c}{Model-Free} \\
    \midrule
    Futures & -     & \color{red}{0.969}** & -     & -     & -     & \color{red}{0.944}** & -     & -     & - \\
    Professional Forecasts & -     & -     & \color{red}{0.427}** & \color{red}{0.946}** & \color{red}{1.000}** & \color{red}{1.000}** & \color{red}{1.000}** & -     & - \\
    \midrule
    \multicolumn{10}{c}{Univariate} \\
    \midrule
    RW    & 0.346** & 0.500** & 0.368** & 0.399** & 0.218* & 0.218* & 0.429** & 0.426** & 0.123* \\
    AR(1) & 0.346** & 0.451** & 0.317** & 0.190* & 0.096 & 0.143* & 0.321** & 0.314** & 0.122* \\
    AR(AIC) & 0.346** & 0.451** & 0.178* & 0.137* & 0.087 & 0.160* & 0.279** & 0.718** & 0.678** \\
    \midrule
    \multicolumn{10}{c}{ARDL} \\
\bottomrule
\end{longtable}
 \begin{scriptsize}
      \begin{tablenotes}
        \item \parbox[t]{6in}{%
          See Notes in Table~\ref{tab:copper_MCS_supp}
        }
      \end{tablenotes}
    \end{scriptsize}
\end{center}
}

% Table generated by Excel2LaTeX from sheet 'MCS - New'
\tiny{
\begin{center} 
\begin{longtable}{l|llllllllll}
\caption{Model Confidence Sets (p-values) for Nickel real price Forecasts} \label{tab:zinc_MCS_supp} \\
    \toprule
    Model & Horizon 1 & Horizon 3 & Horizon 6 & Horizon 9 & Horizon 12 & Horizon 15 & Horizon 18 & Horizon 21 & Horizon 24 \\
    \endfirsthead

    \multicolumn{10}{c}{\tablename\ \thetable\ -- \textit{Continued from previous page}} \\  
    \toprule
    Model & Horizon 1 & Horizon 3 & Horizon 6 & Horizon 9 & Horizon 12 & Horizon 15 & Horizon 18 & Horizon 21 & Horizon 24 \\
    \midrule
    \endhead

    \hline \multicolumn{10}{r}{\textit{Continued on next page}} \\
    \endfoot

    \endlastfoot
    \bottomrule
    \multicolumn{10}{c}{Model-Free} \\
    \midrule
    Futures & -     & \color{red}{0.995}** & -     & -     & -     & \color{red}{0.966}** & -     & -     & - \\
    Professional Forecasts & -     & -     & \color{red}{0.787}** & \color{red}{0.982}** & \color{red}{1.000}** & \color{red}{1.000}** & \color{red}{1.000}** & -     & - \\
    \midrule
    \multicolumn{10}{c}{Univariate} \\
    \midrule
    RW    & 0.139* & 0.562** & 0.501** & 0.515** & 0.356** & 0.389** & 0.273** & 0.229* & 0.289** \\
    AR(1) & \color{red}{0.941}** & 0.898** & 0.599** & 0.530** & 0.259** & 0.536** & 0.607** & 0.219* & 0.114* \\
    AR(AIC) & \color{red}{0.941}** & 0.898** & 0.571** & 0.421** & 0.176* & 0.416** & 0.579** & 0.722** & 0.837** \\
    \midrule
    \multicolumn{10}{c}{ARDL} \\
\bottomrule
\end{longtable}
 \begin{scriptsize}
      \begin{tablenotes}
        \item \parbox[t]{6in}{%
          See Notes in Table~\ref{tab:copper_MCS_supp}
        }
      \end{tablenotes}
    \end{scriptsize}
\end{center}
}

\newpage

\section{Extended Cumulative Ratios over time}
\label{app:ext_cum}

\begin{figure}[!ht]
\centering
\setlength{\tabcolsep}{.008\textwidth}
\caption*{}
\begin{tabular}{cc}
\begin{footnotesize} {Copper}\end{footnotesize} &\begin{footnotesize} {Aluminum}\end{footnotesize}\\
\includegraphics[trim=0cm 0cm 0cm 0cm, clip, width=0.48\textwidth]{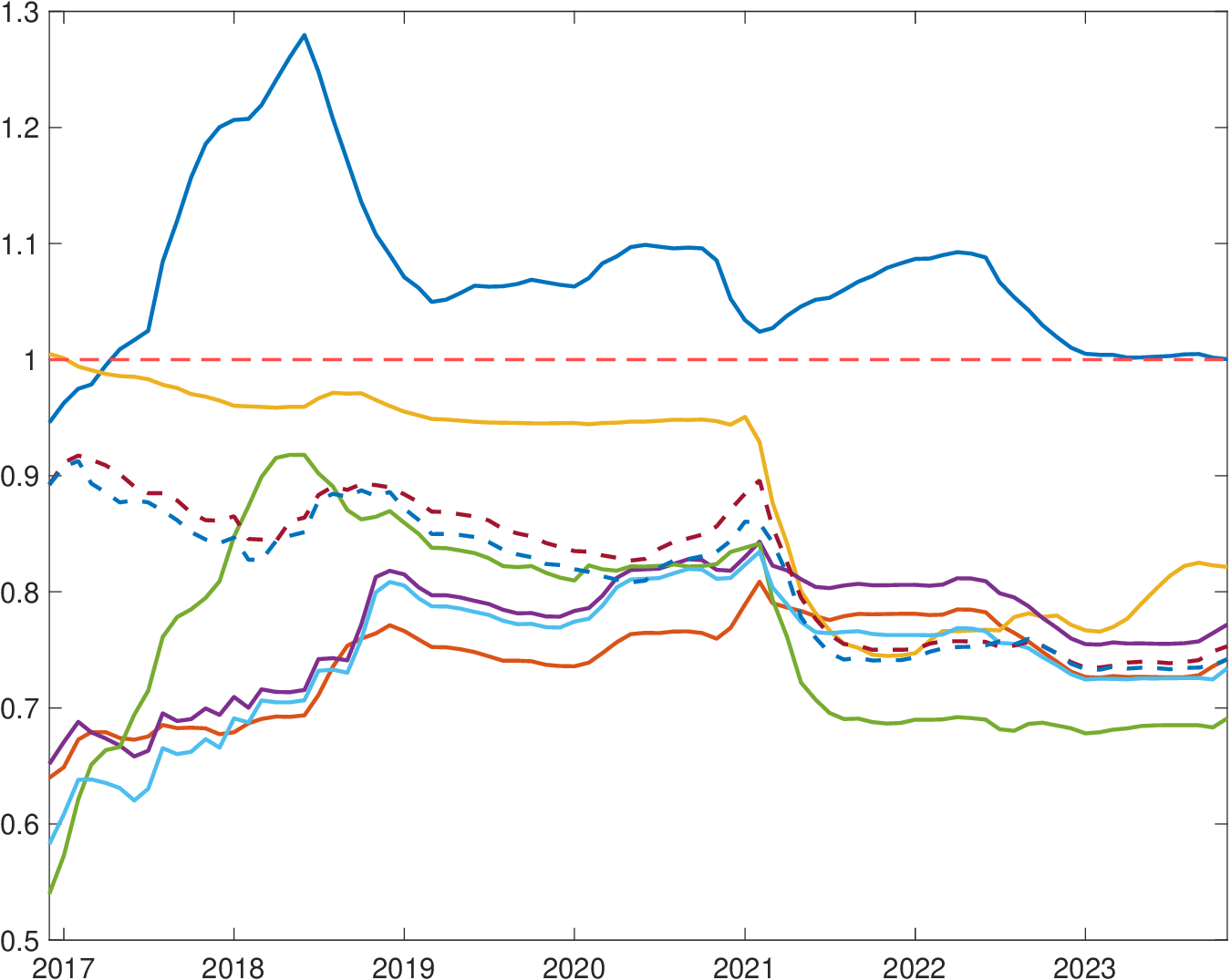} &
\includegraphics[trim=0cm 0cm 0cm 0cm, clip, width=0.48\textwidth]{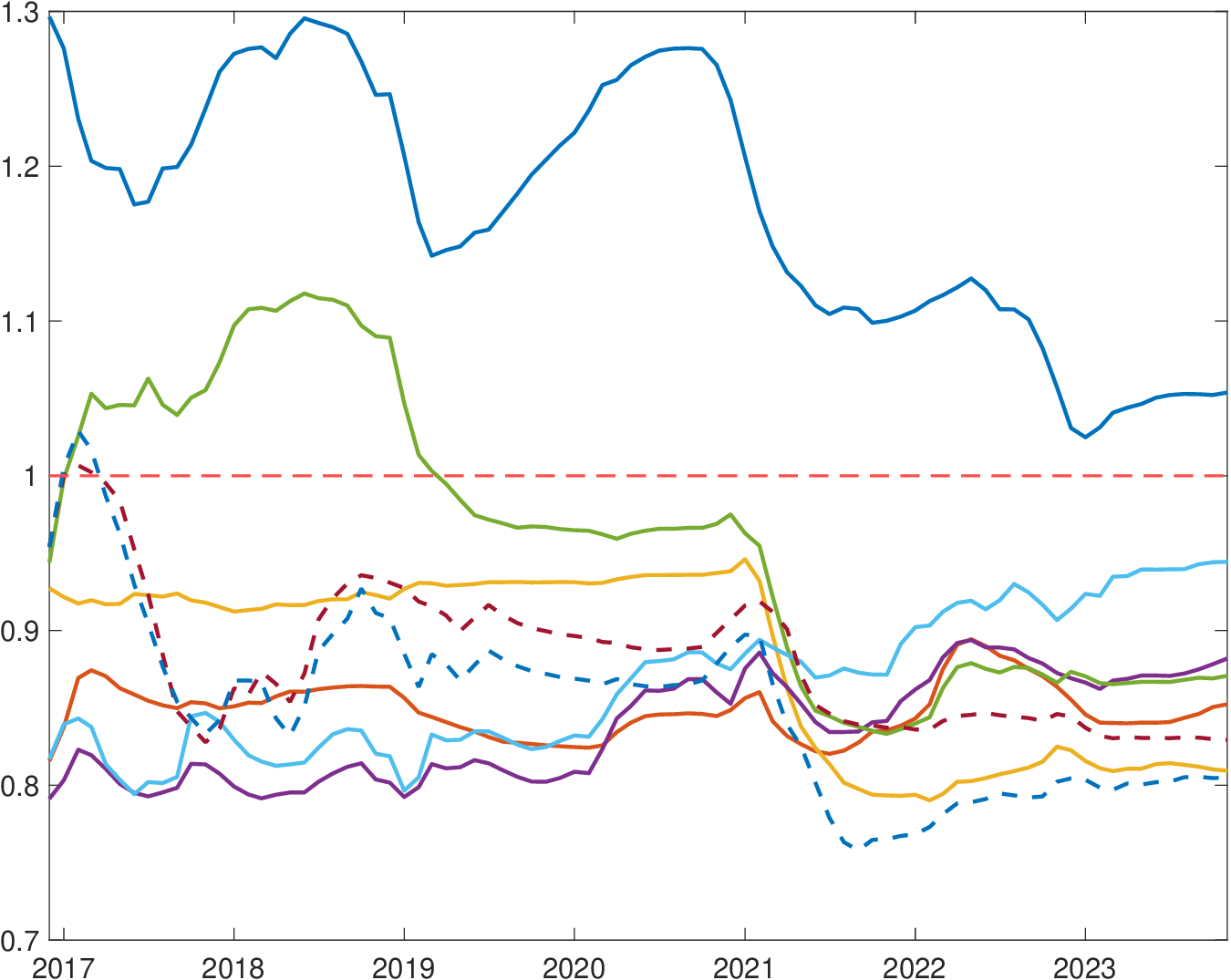}\\
\end{tabular}
\begin{tabular}{cc}\begin{footnotesize}\centering{ {Nickel}}\end{footnotesize} &\begin{footnotesize} {Zinc}\end{footnotesize}\\
\includegraphics[trim=0cm 0cm 0cm 0cm, clip, width=0.48\textwidth]{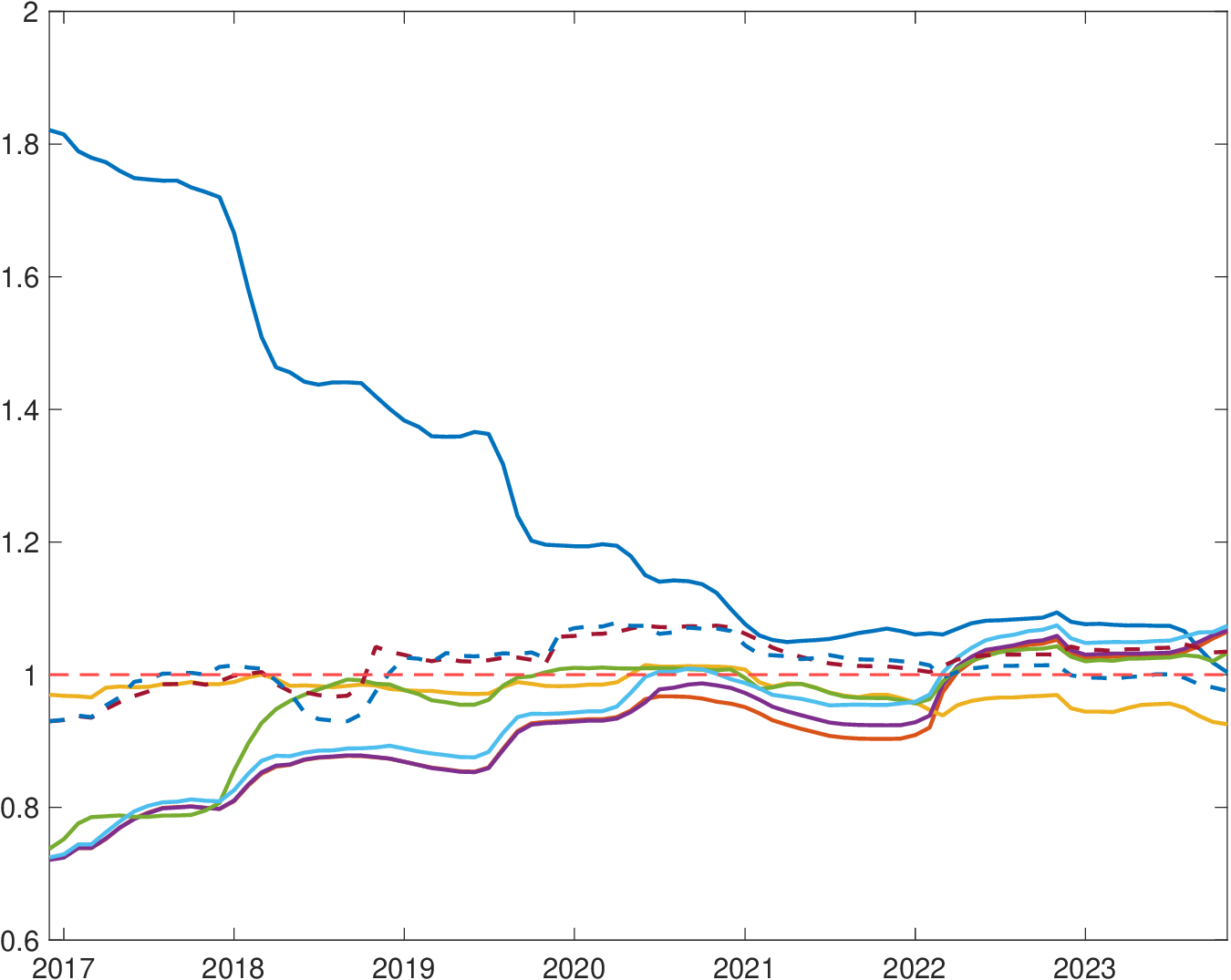} &
\includegraphics[trim=0cm 0cm 0cm 0cm, clip, width=0.48\textwidth]{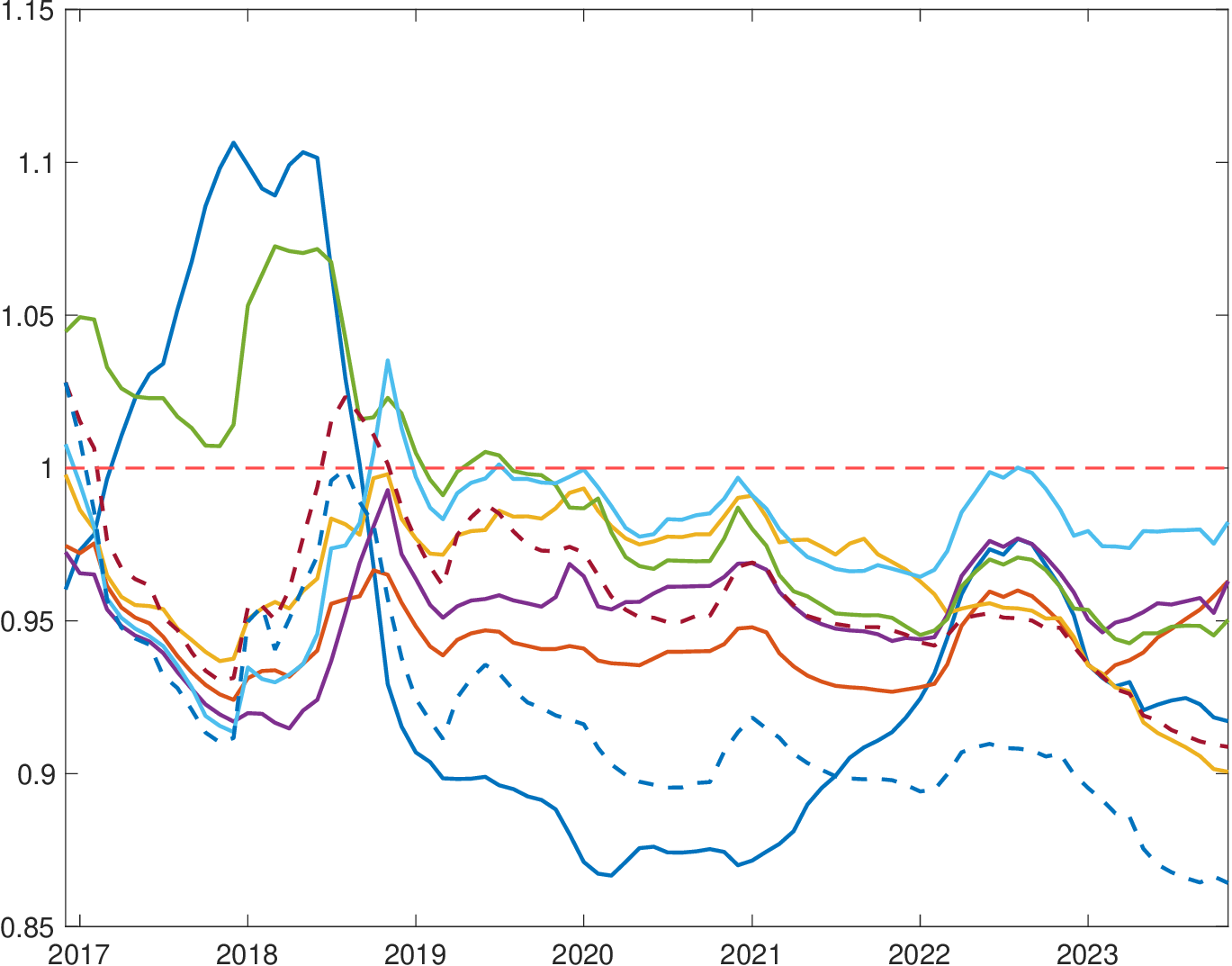}\\
\end{tabular}
\begin{tabular}{c}
\includegraphics[trim=2cm 2cm 2cm 2cm, clip, width=1\textwidth]{Figures/ratios_cum_legend.eps}
\end{tabular}
\caption{Cumulative RMSPE ratios at horizon 9 for different models and metals.}
\label{fig:RMSPE_cum_9month}
\end{figure}

\begin{figure}[!ht]
\centering
\setlength{\tabcolsep}{.008\textwidth}
\caption*{}
\begin{tabular}{cc}
\begin{footnotesize} {Copper}\end{footnotesize} &\begin{footnotesize} {Aluminum}\end{footnotesize}\\
\includegraphics[trim=0cm 0cm 0cm 0cm, clip, width=0.48\textwidth]{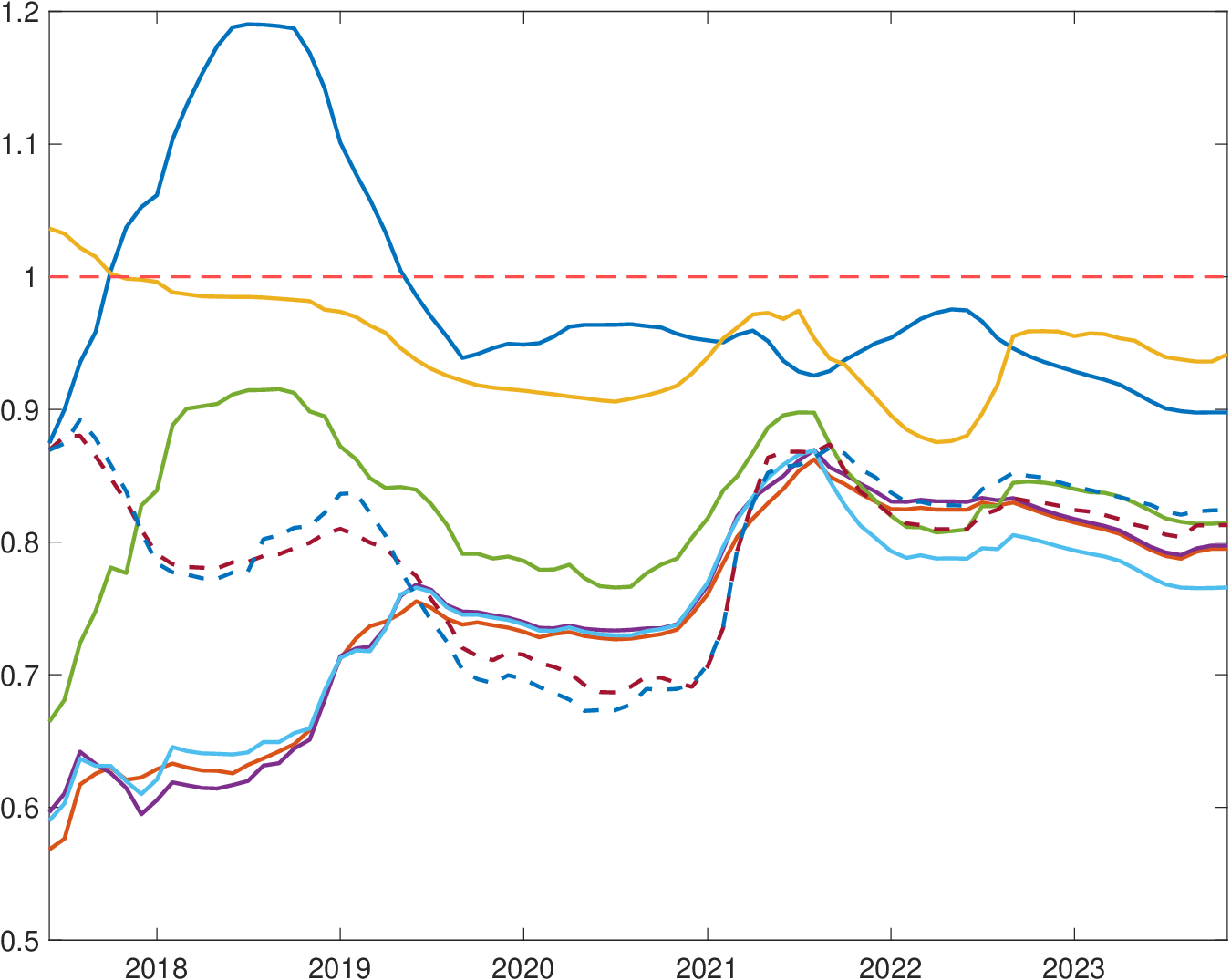} &
\includegraphics[trim=0cm 0cm 0cm 0cm, clip, width=0.48\textwidth]{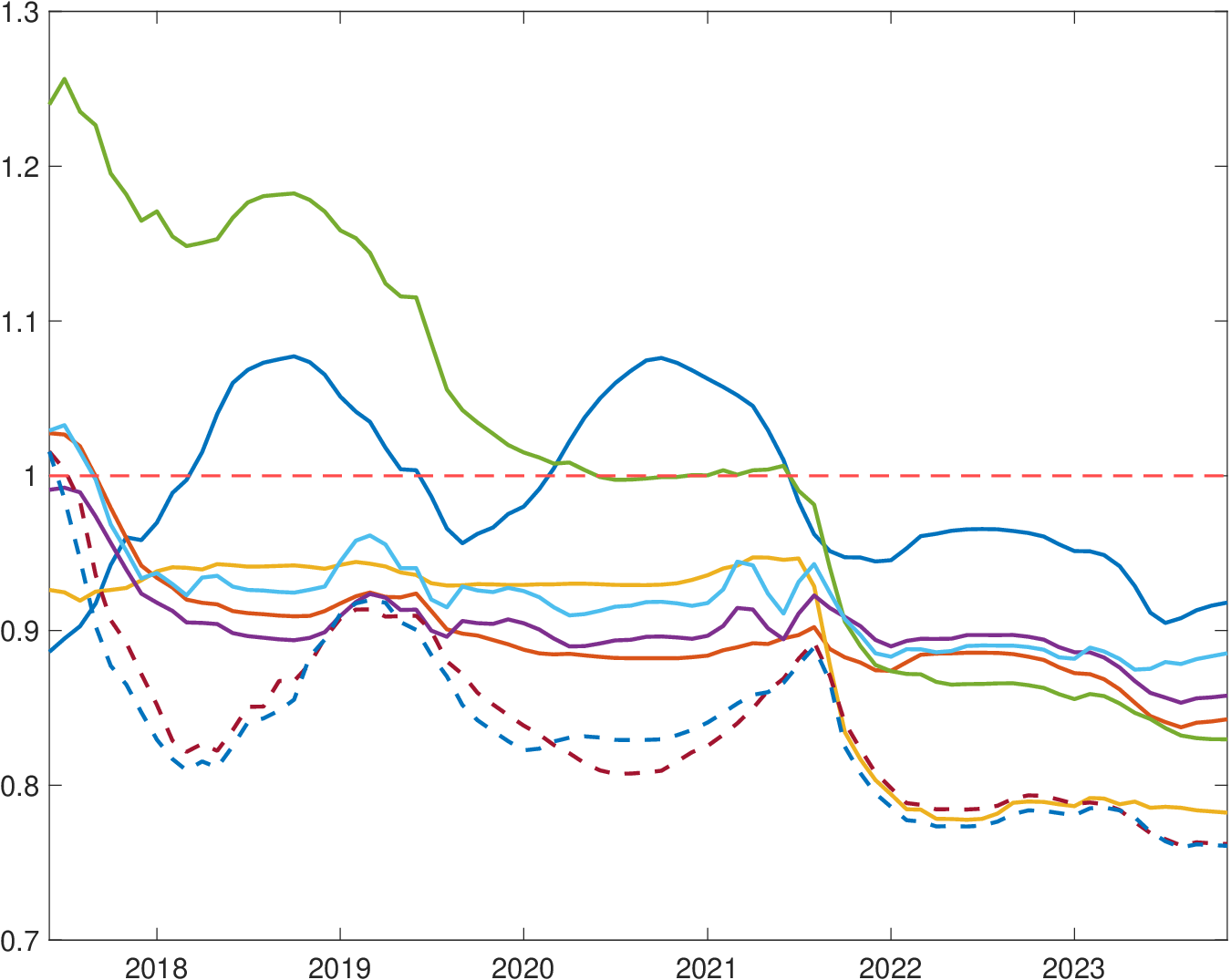}\\
\end{tabular}
\begin{tabular}{cc}\begin{footnotesize}\centering{ {Nickel}}\end{footnotesize} &\begin{footnotesize} {Zinc}\end{footnotesize}\\
\includegraphics[trim=0cm 0cm 0cm 0cm, clip, width=0.48\textwidth]{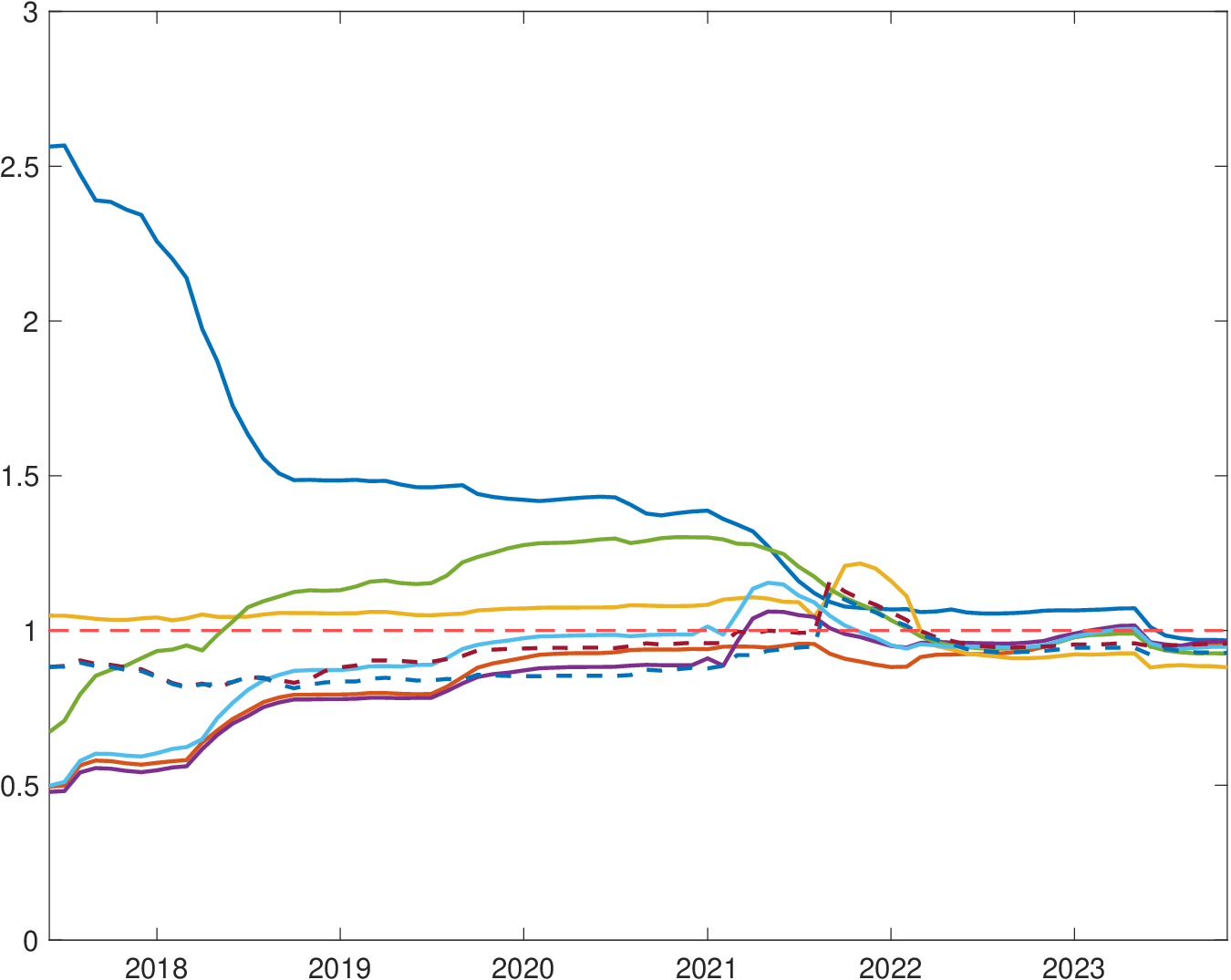} &
\includegraphics[trim=0cm 0cm 0cm 0cm, clip, width=0.48\textwidth]{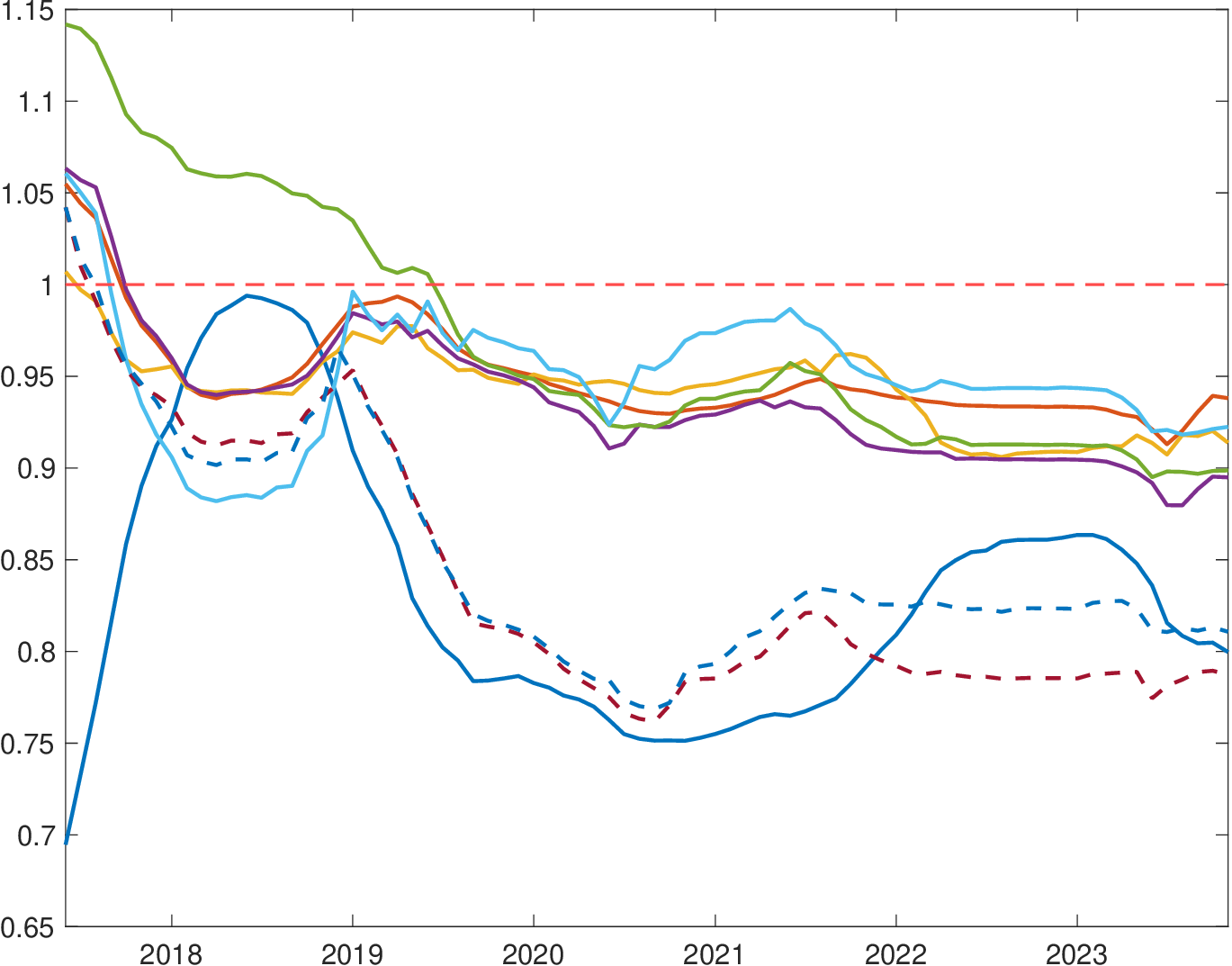}\\
\end{tabular}
\begin{tabular}{c}
\includegraphics[trim=2cm 2cm 2cm 2cm, clip, width=1\textwidth]{Figures/ratios_cum_legend.eps}
\end{tabular}
\caption{Cumulative RMSPE ratios at horizon 15 for different models and metals.}
\label{fig:RMSPE_cum_15month}
\end{figure}

\bibliography{biblio}